\documentclass[%
 reprint,
superscriptaddress,
nofootinbib,
 amsmath,amssymb,
 aps,
 prx,
]{revtex4-2}

\usepackage{amssymb}
\usepackage{placeins}
\usepackage{graphicx}% Include figure files
\usepackage{dcolumn}% Align table columns on decimal point
\usepackage{bm}% bold math
\usepackage[colorlinks=true,allcolors=blue]{hyperref}
\usepackage{xcolor}
\usepackage{amsmath}
\usepackage{mathtools}

\usepackage[english]{babel}
\usepackage{blindtext}
\usepackage{setspace}

\newcommand{\PmSp}{\mathrm{Pm}_{\mathrm{Sp}}}
\newcommand{\rhot}{\tilde{\rho}}
\newcommand{\pt}{\tilde{p}}
\newcommand{\bBt}{\tilde{\bB}}
\newcommand{\bAt}{\tilde{\bA}}
\newcommand{\Bt}{\tilde{B}}
\newcommand{\ut}{\tilde{u}}
\newcommand{\but}{\tilde{\bu}}
\newcommand{\nut}{\tilde{\nu}}
\newcommand{\etat}{\tilde{\eta}}

\newcommand{\dd}{\mathrm{d}}
\newcommand{\br}{\boldsymbol{r}}
\newcommand{\bcdot}{\boldsymbol{\cdot}}
\newcommand{\bx}{\boldsymbol{x}}

\newcommand{\bu}{\boldsymbol{u}}

\newcommand{\bB}{\boldsymbol{B}}

\newcommand{\bA}{\boldsymbol{A}}

\newcommand{\bL}{\boldsymbol{L}}

\newcommand{\bnabla}{\boldsymbol{\nabla}}
\newcommand{\p}{\partial}
\newcommand{\mcE}{\mathcal{E}}

\newcommand{\const}{\mathrm{const}}

\newcommand{\blue}[1]{\textcolor{black}{#1}}

\makeatletter 
    
\renewcommand\onecolumngrid{% <<<<<<
\do@columngrid{one}{\@ne}%
\def\set@footnotewidth{\onecolumngrid}% <<<<<<<<<<<<<<<<
\def\footnoterule{\kern-6pt\hrule width 1.5in\kern6pt}%
}

\renewcommand\twocolumngrid{% <<<<<<
        \def\footnoterule{% restore rule
        \dimen@\skip\footins\divide\dimen@\thr@@
        \kern-\dimen@\hrule width.5in\kern\dimen@}
        \do@columngrid{mlt}{\tw@}
}%

\makeatother    
\begin{document}

\preprint{APS/123-QED}

\title{Cosmic-void observations reconciled with primordial magnetogenesis}% Force line breaks with \\

\author{David N. Hosking}
 \email{david.hosking@physics.ox.ac.uk}
 \affiliation{Oxford Astrophysics, Denys Wilkinson Building, Keble Road, Oxford OX1 3RH, UK}%Lines break automatically or can be forced with \\
 \affiliation{Merton College, Merton Street, Oxford, OX1 4JD, UK
}
\author{Alexander A. Schekochihin}%

\affiliation{Merton College, Merton Street, Oxford, OX1 4JD, UK
}
\affiliation{The Rudolf Peierls Centre for Theoretical Physics, University of Oxford, Clarendon Laboratory, Parks Road, Oxford, OX1 3PU, UK}%

\date{\today}% It is always \today, today,

\begin{abstract}
It has been suggested that the weak magnetic field hosted by the intergalactic medium in cosmic voids could be a relic from the early Universe. However, accepted models of turbulent magnetohydrodynamic decay predict that the present-day strength of fields originally generated at the electroweak phase transition (EWPT) \blue{without parity violation} would be too low to explain the observed scattering of $\gamma$-rays from TeV blazars. Here, we propose that the decay is mediated by magnetic reconnection and conserves the mean square fluctuation level of magnetic helicity. We find that the relic fields would be stronger by several orders of magnitude under this theory than was indicated by previous treatments, which restores the consistency of the EWPT-relic hypothesis with the observational constraints. \blue{Moreover, efficient EWPT magnetogenesis would produce relics at the strength required to resolve the Hubble tension via magnetic effects at recombination and seed galaxy-cluster fields close to their present-day strength.}
\end{abstract}

\maketitle

It is widely believed that cosmic voids host magnetic fields. Evidence for this comes chiefly from $\gamma$-ray observations of blazars~(\cite{NeronovVovk10, Tavecchio10, Taylor11, Dermer11, Dolag11, Essey11, Huan11, Tavecchio11, Takahashi12, Arlen14, Finke15, Archambault17}; see~\cite{DurrerNeronov13,Subramanian16, Vachaspati21} for reviews): extragalactic magnetic fields (EGMFs) in voids would, if present, scatter the electrons produced in electromagnetic cascades of TeV \mbox{$\gamma$-rays} emitted by blazars, thus suppressing the number of secondary (GeV) $\gamma$-rays received at Earth. Such suppression is indeed observed, and can be used to constrain the root-mean-square strength $B\equiv\langle \bB^2 \rangle^{1/2}$ and \blue{energy-containing} scale $\lambda_B$ of the magnetic fields. Using spectra measured by the \textit{Fermi} telescope, Ref.~\cite{Taylor11, Archambault17} estimate that
\begin{equation}
    B \gtrsim 10^{-17} \,\mathrm{G}\, \left(\frac{\lambda_B}{1\,\mathrm{Mpc}}\right)^{-1/2},\label{constraint_delay}
\end{equation}where $10^{-17} \,\mathrm{G}$ can increase to $\sim10^{-15} \,\mathrm{G}$ depending on modelling assumptions, \blue{including the effect of time delay due to the larger distance traveled by scattered electrons~\cite{Taylor11, Ackermann18}. Eq.~\eqref{constraint_delay} may also be subject to some modification due to the cooling of cascade electrons by plasma instabilities~\cite{Broderick12, Broderick18, Batista19, PerryLyubarsky21} --- what effect, if any, this has on the constraint~\eqref{constraint_delay} is poorly understood --- see~\cite{Addazi22, BatistaSaveliev21} for recent discussions.}

Where might fields in voids come from? A popular idea (although not the only one, see~\cite{Beck13}) is that they could be relics of primordial magnetic fields (PMFs) generated in the early Universe~\cite{BanerjeeJedamzik04}, including, prominently, at the electroweak phase transition (EWPT)~\cite{Vachaspati91}. If so, the physics of the early Universe could be constrained by observations of the fields in voids --- a remarkable possibility --- provided the magnetohydrodynamic (MHD) decay of the PMFs between their genesis and the present day were understood. However, conventional theory of the decay (\cite{BanerjeeJedamzik04}; see~\cite{DurrerNeronov13,Subramanian16, Vachaspati21} for reviews) appears inconsistent with the EWPT-relic hypothesis: Ref. \cite{WagstaffBanerjee16} argue that the lower bound~\eqref{constraint_delay} on $B$ is too high to be consistent with PMFs generated at the EWPT without magnetic helicity (a topological quantity that quantifies the number of twists and linkages in the field, which is conserved even as energy decays~\cite{Taylor86}). Furthermore, they show that the amount of magnetic helicity required for consistency with Eq.~\eqref{constraint_delay} is greater than can be generated by baryon asymmetry at the EWPT, as estimated by Ref.~\cite{Vachaspati01}. In principle, other mechanisms of magnetic-helicity generation may have been present in the early Universe; one idea is \textit{chiral MHD} (see \cite{Boyarksy21} and references therein). Whether enough net helicity can be generated via these mechanisms for PMFs to become maximally helical during their evolution remains an open question~\cite{Brandenburg17_cosmic, Brandenburg17_chiral}.

On the other hand, Ref.~\cite{WagstaffBanerjee16} note that their conclusions could be subject to modification by the contemporaneous discovery of ``inverse transfer'' of magnetic energy in simulations of non-helical MHD turbulence~\cite{Zrake14, Brandenburg15} (see~\cite{Kahniashvili13, Brandenburg17_cosmic, Ellis19, Mtchedlidze22} for schemes for doing so based on decay laws obtained numerically). Recently, the inverse transfer was explained as a consequence of \emph{local fluctuations} in the magnetic helicity, which are generically present even when the global helicity vanishes~\cite{HoskingSchekochihin20decay}. In this paper, we demonstrate how this insight, together with the other key result of Ref.~\cite{HoskingSchekochihin20decay}, and of Refs.~\cite{Zhou19, Zhou20, Bhat21}, that the decay timescale is the one on which magnetic fields reconnect, restores consistency of the hypothesis of a non-helical EWPT-generated PMF with~Eq.~\eqref{constraint_delay}. Intriguingly, we find that reasonably efficient magnetogenesis of non-helical magnetic field at the EWPT could produce relics with the $\sim 10^{-11}\,\mathrm{G}$ comoving strength that, it has been suggested, is sufficient to resolve the Hubble tension~\cite{JedamzikPogosian20, Galli22}. Relics of this strength would also constitute seed fields for galaxy clusters that would not require much amplification by turbulent dynamo after structure formation to reach their observed present-day strength~\cite{BanerjeeJedamzik03} (although dynamo would still be required to \textit{maintain} cluster fields at present levels).
 
\vspace{1.5mm}
\noindent\textbf{\large{Results}}
\vspace{.3mm}

We take the metric of the expanding Universe to be
\begin{equation}
    \dd s^2 = a^2(t)(-\dd t^2 + \dd x_i \,\dd x^i ),\label{metric}
\end{equation}where $a(t)$ is the scale factor, normalised to 1 at the present day, $t$ is conformal time (related to cosmic time~$\overline{t}$ by $a(t)\dd t = \dd \overline{t}$), and $x_i$ are comoving coordinates. The expanding-Universe MHD equations can be transformed to those for a static Universe by a simple rescaling~\cite{Brandenburg96}: the scaled variables
\begin{align}
    \rhot=a^4\rho,\quad& \pt= a^4p,\quad \bBt= a^2 \bB,\quad \but =\bu, \nonumber\\
    & \etat = \eta/a, \quad \nut=\nu/a,\label{transformation}
\end{align}[where $\rho$, $p$, $\bB$, $\bu$, $\eta$ and $\nu$ are the physical values of the total (matter + radiation) density, pressure, magnetic field, velocity, magnetic diffusivity and kinematic viscosity, respectively] evolve according to the MHD equations in Minkowski spacetime. As in previous work (see~\cite{DurrerNeronov13, Subramanian16, Vachaspati21}), we consider the dynamics of the ``tilded'' variables in Minkowski spacetime and transform the result to the spacetime~\eqref{metric} of the expanding Universe via Eq.~\eqref{transformation}.

\noindent\textbf{Selective decay of small-scale structure.}\\ Historically, it has been believed that statistically isotropic MHD turbulence decays while preserving the small-$k$ asymptotic of the magnetic-energy spectrum~$\mcE_M(k)$~(see~\cite{DurrerNeronov13, Subramanian16} and references therein). This idea, sometimes called ``selective decay of small-scale structure'', amounts to a statement of the invariance in time of the magnetic Loitsyansky integral,
\begin{equation}
    I_{\bL_M}\equiv-\int\dd^3 \br \,r^2 \langle \bBt(\bx)\bcdot\bBt(\bx+\br) \rangle,\label{magnetic_Loitsyansky}
\end{equation}which, for isotropic turbulence without long-range spatial correlations, is related to $\mcE_M(k)$ by
\begin{equation}
    \mcE_{M}(k\to 0) = \frac{I_{\bL_M} k^4}{24 \pi^2} + O(k^6). \label{EMexpansion}
\end{equation}
\begin{figure}
    \centering
    \includegraphics[width=\columnwidth]{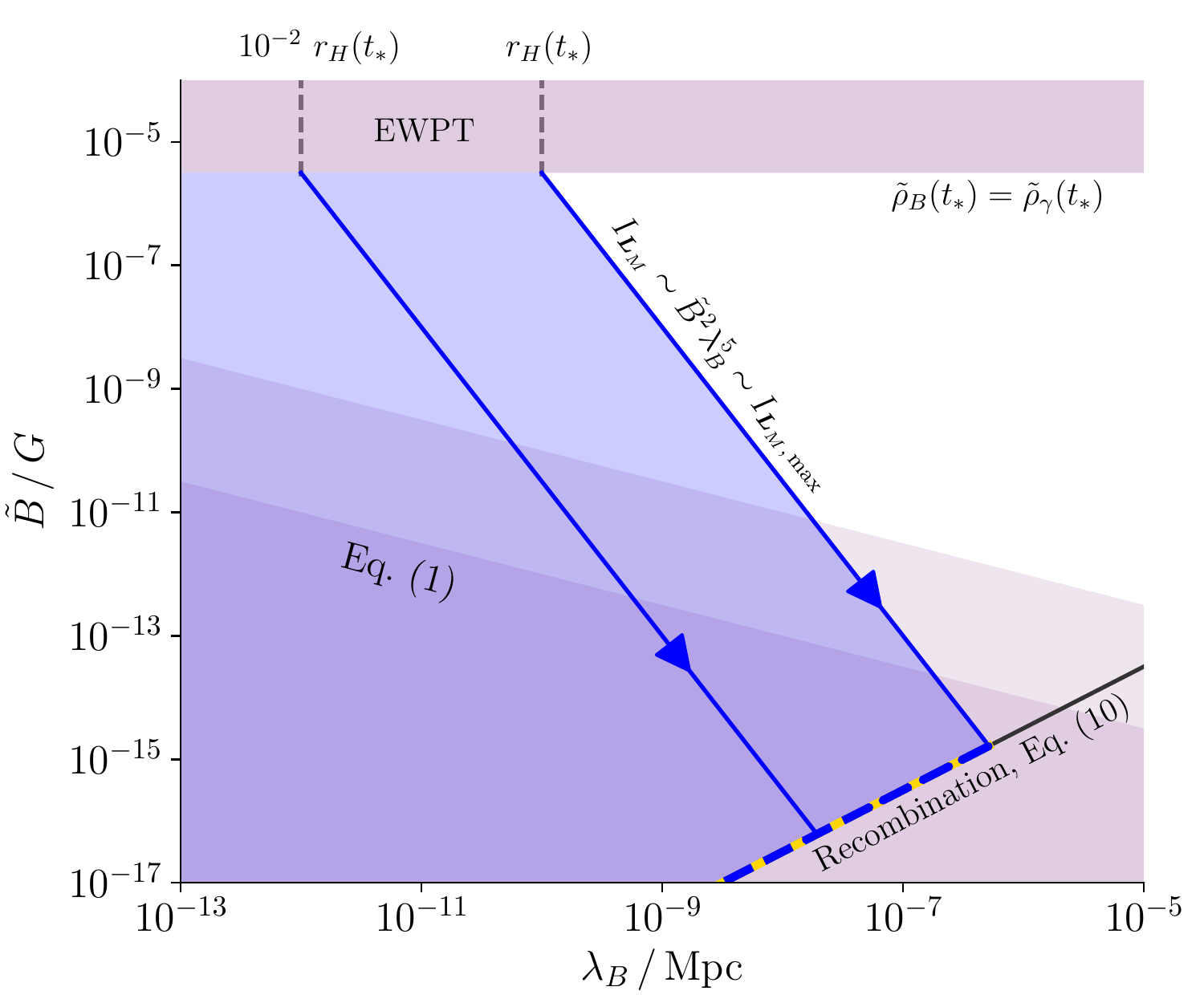}
    \caption{\textbf{Inconsistency of the decay theory based on Eqs.~\eqref{b2l5_const} and~\eqref{Alfvenic} with observational constraints for EWPT-generated PMFs.}\\
    % The Coulomb-gauge ($\bnabla \bcdot \bAt = 0$) magnetic-helicity density $\tilde{h}=\tilde{\bA}\bcdot \tilde{\bB}$,  , taken from a direct numerical simulation of non-helical MHD turbulence decaying from a magnetically dominated state.
    Purple regions denote values of $\Bt$ and $\lambda_B$ excluded on physical [$\tilde{\rho}_B(t)\lesssim \tilde{\rho}_{\gamma}(t_*)$] or observational [the two forms of the constraint~\eqref{constraint_delay}] grounds. Under decays that conserve $I_{\bL_M}$ [Eq.~\eqref{magnetic_Loitsyansky}], $\Bt$ and $\lambda_B$ evolve along lines parallel to the ones shown in blue. The predicted values of modern-day $\Bt$ and $\lambda_B$ are given by the intersection of these lines with Eq.~\eqref{ideal_law}. We see that even PMFs generated with $\tilde{\rho}_B(t_*)\sim \tilde{\rho}_{\gamma}(t_*)$ and $\lambda_B(t_*) \sim r_H(t_*)$ produce modern-day relics that are inconsistent with Eq.~\eqref{constraint_delay}.
    }
    \label{fig:b2l5}
\end{figure}Invariance of $I_{\bL_M}$ implies
\begin{equation}
    I_{\bL_M} \sim \Bt^2 \lambda_B^5 \sim \const. \label{b2l5_const}
\end{equation}\blue{In writing~\eqref{b2l5_const}, we have assumed that the magnetic-energy spectrum is sufficiently peaked around the energy-containing scale $\lambda_B$ for the latter to be equal to the correlation, or integral, scale of the field. This would not be the case for a scale-invariant magnetic field (often conjectured to be generated by inflationary mechanisms). We exclude such fields from our analysis in this paper, in which we consider ``causal'' fields --- the sort that could be generated at a phase transition --- exclusively.}

\noindent\textbf{Decay timescale.} Eq.~\eqref{b2l5_const} can be translated into a decay law for magnetic energy by a suitable assumption about how the energy-decay timescale,
\begin{equation}
    \tau \,(\Bt, \lambda_B, t) \equiv - \left(\frac{\dd \log \Bt^2}{\dd t}\right)^{-1},\label{rate}
\end{equation}depends on $\Bt$, $\lambda_B$ and $t$.
Regardless of this choice, Eqs.~\eqref{b2l5_const} and~\eqref{rate} have the following important property. Suppose that, after some intermediate time $t_c$, $\tau(B,\lambda_B, t)$ can be approximated by some particular product of powers of its arguments. Then, for all $t\gg \tau(t_c)$, $\tilde{B}^2$ decays as a power law: $\tilde{B}^2 \propto t^{-p}$, where $p$ is a number of order unity. Substituting this back into Eq.~\eqref{rate}, one finds
\begin{equation}
    \tau \,(\Bt, \lambda_B, t)\sim t,\label{eddyprocessing}
\end{equation}which is an implicit equation for $\Bt = \Bt(\lambda_B)$ that can be solved simultaneously with Eq.~\eqref{b2l5_const} for $\Bt(t)$ and $\lambda_B(t)$. Eq.~\eqref{eddyprocessing} was first suggested by Ref.~\cite{BanerjeeJedamzik04} on phenomenological grounds. Its great utility, which has perhaps not been spelled out explicitly, is that it implies that one need not know the functional form of $\tau(\Bt, \lambda_B, t)$ during the \textit{early} stages of the decay in order to compute $\Bt$ and $\lambda_B$ at later times. Thus, the effect of early-Universe physics (e.g., neutrino viscosity) on the decay dynamics can be safely neglected.

\noindent\textbf{Inconsistency with observations.} Assuming that the decay satisfies Eq.~\eqref{b2l5_const} and that its timescale is Alfv\'{e}nic, viz.,
\begin{equation}
    \tau\sim \frac{\lambda_B}{\tilde{v}_A}, \quad \tilde{v}_A = \frac{\tilde{B}}{\sqrt{4\pi \tilde{\rho}_b}},\label{Alfvenic}
\end{equation}when it terminates at the recombination time $t_{\mathrm{recomb}}$~\cite{Subramanian16} [Eq.~\eqref{t_recomb} in Methods], Eq.~\eqref{eddyprocessing} implies~\cite{BanerjeeJedamzik04}
\begin{equation}
    \Bt(t_{\mathrm{recomb}}) \sim 10^{-8.5} \mathrm{G}\, \frac{\lambda_{B}(t_\mathrm{recomb})}{1\,\mathrm{Mpc}}
    \label{ideal_law}
\end{equation}[see Eq.~\eqref{ideal_law2} in Methods]. In~\eqref{Alfvenic}, $\tilde{\rho}_b$ is the baryon density, which appears because photons do not contribute to the fluid inertia at scale $\lambda_B$ at the time of recombination~\cite{JedamzikSaveliev19} [see Eq.~\eqref{photon_mfp} in Methods]. An approximate upper bound, $I_{\boldsymbol{L}_M,\,\mathrm{max}}$, on $I_{\boldsymbol{L}_M}$ follows from assuming that the magnetic-energy density $\tilde{\rho}_B \equiv \tilde{B}^2/8\pi$ and the electromagnetic-radiation density $\tilde{\rho}_{\gamma}$ were equal at the time $t_*$ of the EWPT while $\lambda_B(t_*)$ was equal to the Hubble radius $r_H(t_*)$. This corresponds to~${\Bt(t_{*})\sim 10^{-5.5} \,\mathrm{G}}$ and ${\lambda_B(t_*) \sim r_H(t_*)\sim 10^{-10} \,\mathrm{Mpc}}$~\cite{DurrerNeronov13, WagstaffBanerjee16}. As is shown in Fig.~\ref{fig:b2l5}, these values and Eq.~\eqref{ideal_law} together lead to values of $\Bt$ and $\lambda_B$ at~$t_{\mathrm{recomb}}$ that violate the observational constraint~\eqref{constraint_delay}. Note that $\lambda_B(t_*)\sim 10^{-2}\, r_H(t_*)$ is, in fact, a more popular estimate, corresponding to the typical coalescence size of ``bubbles of new phase'' that form at the phase transition~\cite{Turok92}; for this initial correlation scale, the predicted value of $\tilde{B}$ is separated from the allowed values by around three orders of magnitude. A similar calculation led Ref.~\cite{WagstaffBanerjee16} to conclude that genesis of EGMFs at the EWPT was unlikely \blue{(although we note that significant modification of Eq.~\eqref{constraint_delay} by inclusion of the effects of plasma instabilities in the modelling of the electromagnetic cascade --- see the comment below Eq.~\eqref{constraint_delay} --- could alter this conclusion)}.

\begin{figure*}
    \centering
    \includegraphics[width=0.8\textwidth]{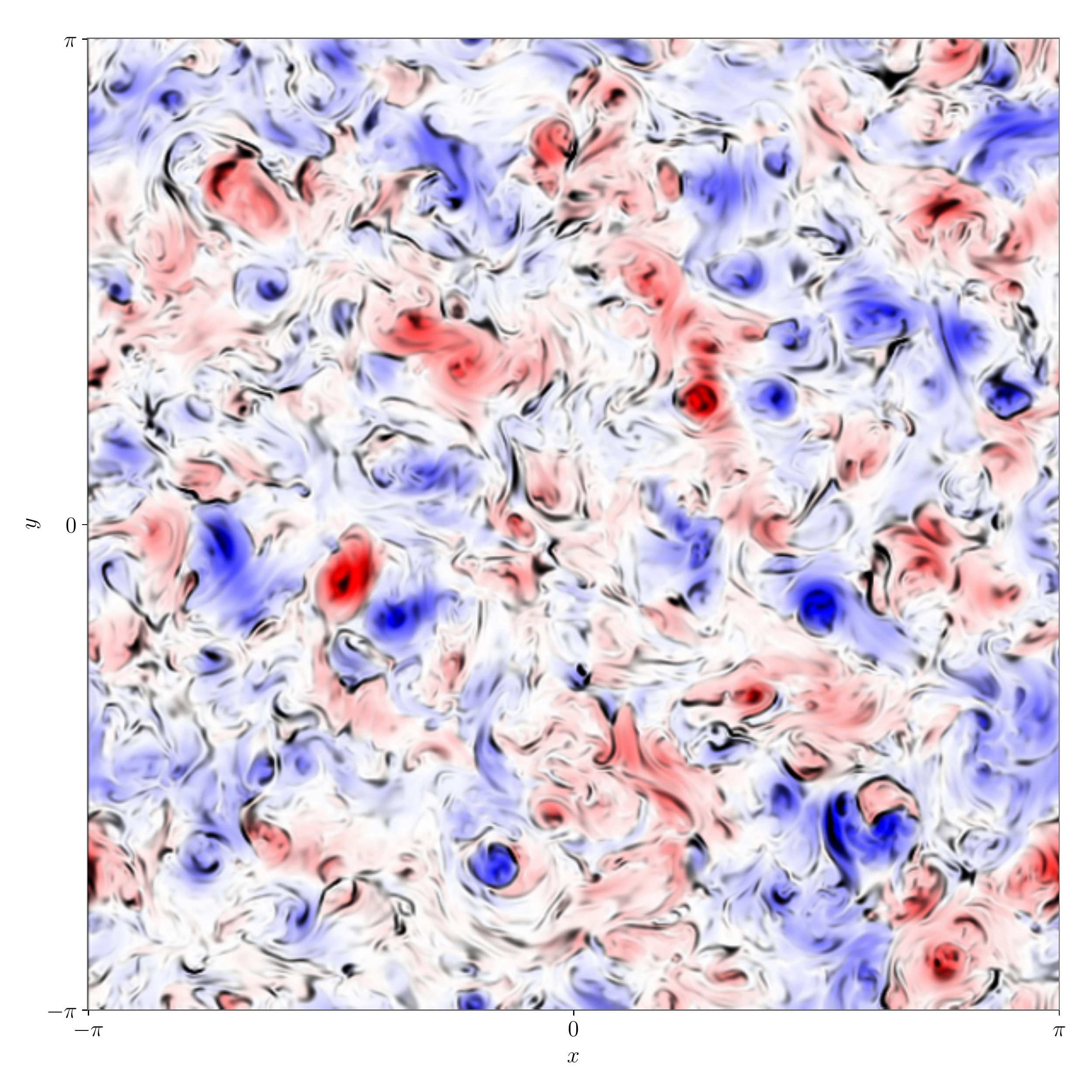}
    \caption{\textbf{Slice of magnetic-helicity density from a simulation of decaying non-helical MHD turbulence.}
    \\
    The turbulence breaks up into patches of positive and negative helicity $h$ (computed in the Coulomb gauge; $\bnabla \bcdot \bAt = 0$), shown in red and blue, respectively. The invariance of $I_H$ (see main text) is a manifestation of the conservation of the net magnetic-helicity fluctuation level arising in large volumes. Because of the complex magnetic-field topology, the rate-setting process for the decay is magnetic reconnection: reconnection sites, indicated in the figure by patches of large current density $|\bnabla \times \tilde{\bB}|$ (black; variable opacity scale), typically form between the helical structures. See Methods for details of the numerical setup.}
    \label{fig:slices}
\end{figure*}

\noindent\textbf{Saffman helicity invariant.} We argue that the theory outlined above requires revision. First, the idea of ``selective decay of small-scale structure'' is flawed. This is because the $k\lambda_B \ll 1$ tail of the magnetic-energy spectrum $\mcE_M(k)$ corresponds not to physical structures (as in the Richardson-cascade picture of inertial-range hydrodynamic turbulence) but to cumulative statistical properties of the structures of size~$\lambda_B$~\cite{Davidson15}. Absent a physical principle to support the invariance of $I_{\bL_M}$ (such as angular-momentum conservation for its hydrodynamic equivalent~\cite{LandauLifshitzFluids, Davidson15}), there is therefore no reason to suppose that the small-$k$ asymptotic of $\mcE_M(k)$ evolves on a longer timescale than the dynamical one of $\lambda_B$-scale structures (if this is long compared to the magnetic-diffusion timescale at scale $\lambda_B$, then selective decay \textit{is} valid, as the simulations of~\cite{BanerjeeJedamzik04, ReppinBanerjee17} confirm, but this is not the regime relevant to PMFs).

Instead, we propose that the decay of PMFs is controlled by a different integral invariant~\cite{HoskingSchekochihin20decay}:
\begin{equation}
    I_H = \int\dd^3\br \,\langle h(\bx)h(\bx+\br)\rangle,\label{I_H} 
\end{equation}where $h=\bAt\bcdot\bBt $ is the helicity density ($\bBt=\bnabla\times\bAt$). Eq.~\eqref{I_H} is equivalent to \begin{equation}
    I_H = \lim_{V\to\infty} \frac{1}{V}\left\langle\left[\int_V\dd^3\bx\, h(\bx)\right]^2 \right\rangle= \lim_{V\to\infty} \frac{\langle H_V^2\rangle}{V},\label{randomwalk}
\end{equation}where $H_V$ is the total magnetic helicity contained within the control volume $V$. The invariance of $I_H$ can therefore be understood intuitively as expressing the conservation of the net mean square \textit{fluctuation level} of magnetic helicity per unit volume that arises in any finite volume of non-helical MHD turbulence (see Fig.~\ref{fig:slices}; we refer the reader concerned about the existence of such fluctuations to Section B of the Supplementary Information). \blue{Numerical evidence supporting the invariance of $I_H$ has been presented by Ref.~\cite{HoskingSchekochihin20decay} and independently by Refs.~\cite{Zhou22_Hosking, Brandenburg22_Hosking}.}
From $I_H=\const$, we deduce
\begin{equation}
   I_H \sim \Bt^4 \lambda_B^5 \sim \const.\label{b4l5}
\end{equation}

We make two brief remarks. First, growth of $I_{{\bL}_M}$, and, therefore, the inverse-transfer effect discovered by Refs.~\cite{Kahniashvili10, Zrake14, Brandenburg15}, follows immediately from Eq.~\eqref{b4l5}. This is because $I_{{\bL}_M}\sim \Bt^2 \lambda_B^5 \sim I_H/\Bt^2$ under self-similar evolution, so that $\mcE_M(k\to0)\propto I_{{\bL}_M} k^4$ [see Eq.~\eqref{EMexpansion}] grows while $\Bt$ decays. Second, the value of the large-scale spectral exponent does not affect the late-time limit of the decay laws in our theory (see Section~C of the Supplementary Information), unlike in the ``selective-decay'' paradigm.

\noindent\textbf{Reconnection-controlled decay timescale.} The second revision that we propose to the existing theory is that the field's decay timescale $\tau$ should be identified not with the Alfv\'{e}nic timescale~\eqref{Alfvenic}, but with the magnetic-reconnection one. This is because relaxation of stochastic magnetic fields via the generation of Alfv\'{e}nic motions is prohibited by topological constraints, which can only be broken by reconnection. Refs.~\cite{HoskingSchekochihin20decay, Bhat21, Zhou22_Hosking} have presented numerical evidence for a reconnection-controlled timescale for decays that occur with a dominance of magnetic over kinetic energy (see~\cite{Zhou19, Zhou20} for the same in 2D). Magnetically dominated conditions are relevant to the decay of PMFs because (i) the large neutrino and photon viscosities in the early Universe favour them, and (ii) once established, they are maintained, as reconnection is typically slow compared with the Alfv\'{e}nic timescale. The identification of $\tau$ as the reconnection timescale implies that a number of different decay regimes are possible, as we now explain.

Under resistive-MHD theory, reconnecting structures in a fluid with large conductivity generate a hierarchy of current sheets at increasingly small scales via the plasmoid instability~\cite{Loureiro07}. The global reconnection timescale is the one associated with the smallest of these sheets (the ``critical sheet''), which is short enough to be marginally stable~(\cite{Uzdensky10, Bhattacharjee09}, see \cite{Schekochihin20} for a review). This timescale is
\begin{figure*}[t!]
    \centering
    \includegraphics[width=0.9\textwidth]{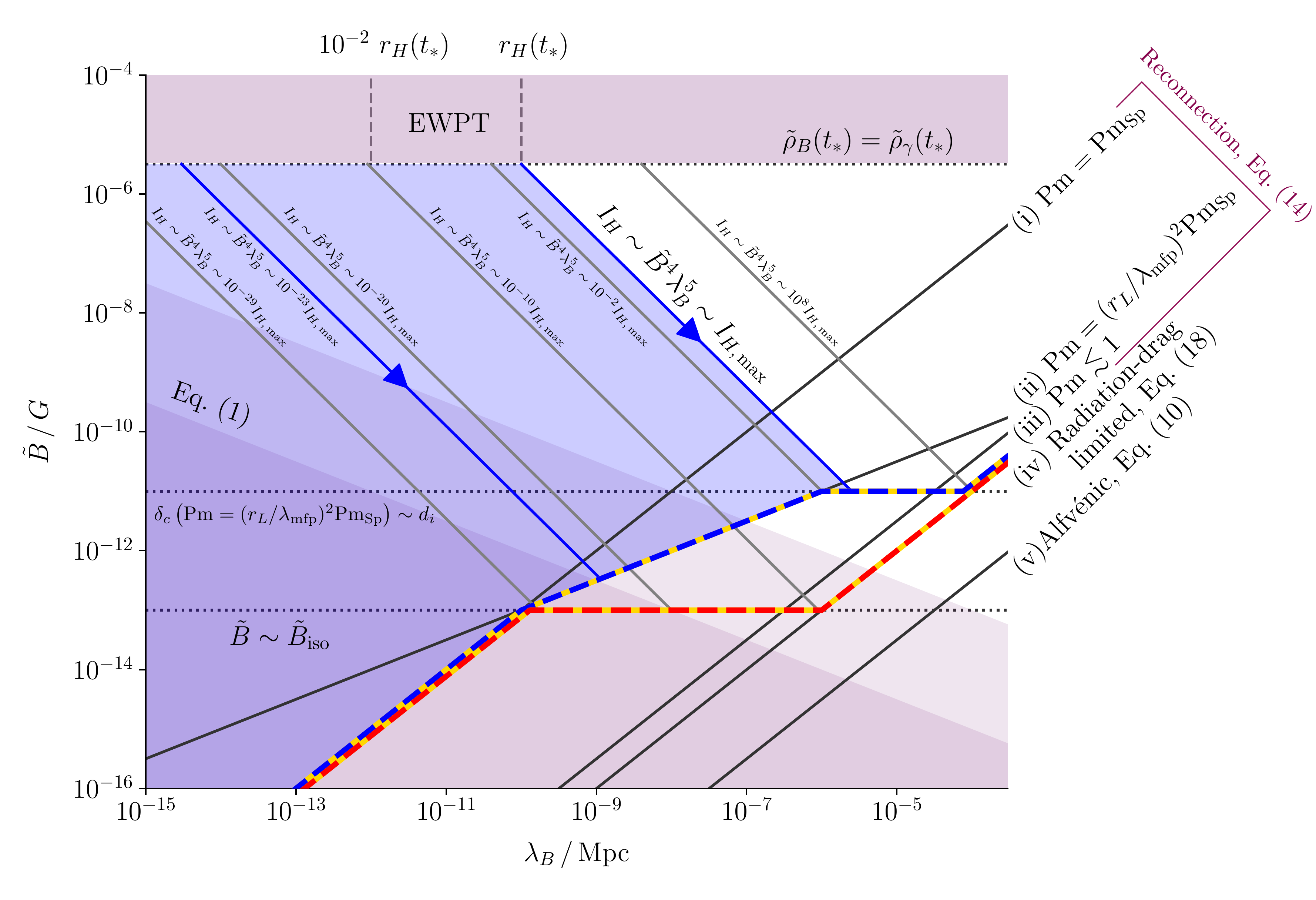}
    \caption{\textbf{Reconnection-controlled decay of non-helical PMFs.} As in Fig.~\ref{fig:b2l5}, purple regions denote values of $\Bt$ and $\lambda_B$ excluded on physical or observational grounds [Eq.~\eqref{constraint_delay}]. Under decays that conserve $I_{H}$ [Eq.~\eqref{I_H}], $\Bt$ and $\lambda_B$ evolve along lines parallel to the ones shown in blue. The predicted values of modern-day $\Bt$ and $\lambda_B$ are given by the intersection of these lines with Eq.~\eqref{eddyprocessing} evaluated at recombination [represented by lines~(i-v), which are derived in Methods], with $\tau$ the prevailing decay timescale. The blue-gold line shows the locus of possible present-day states resulting from reconnection-controlled decays on the timescales explained in the main text, assuming that the microscopic viscosity of the primordial plasma was controlled by collisions between protons. The effective value of $\mathrm{Pm}$ in Eq.~\eqref{MHD_rec_maintext} might have been heavily suppressed when $\Bt>\Bt_{\mathrm{iso}}$ if viscosity were then instead governed by plasma microinstabilities --- the red-gold line shows the locus of modern-day states corresponding to the extreme choice of $\mathrm{Pm}\lesssim 1$ for~$\Bt>\Bt_{\mathrm{iso}}$. In either case, we see that PMFs generated at the EWPT with a wide range of values of $I_H$ produce modern-day relics that are consistent with Eq.~\eqref{constraint_delay}, and even with the stronger version of this constraint [see text below Eq.~\eqref{constraint_delay}] which is indicated by the pale purple region.}
    \label{fig:b4l5}
\end{figure*}
\begin{equation}
    \tau_{\mathrm{rec}} = (1+\mathrm{Pm})^{1/2}\, \mathrm{min}\left\{ S^{1/2}, S_c^{1/2} \right\} \frac{\lambda_B}{\tilde{v}_{A}},\label{MHD_rec_maintext}
\end{equation}where $\mathrm{Pm} = \tilde{\nu}/\tilde{\eta}$ is the magnetic Prandtl number, which appears because viscosity can suppress the outflows that advect reconnected field away from the reconnection site,
\begin{equation}
    S = \frac{\tilde{v}_{A}\lambda_B}{\tilde{\eta}\,(1+\mathrm{Pm})^{1/2}}\label{Ssymbolic_maintext}
\end{equation}is the Lundquist number based on the reconnection outflow and $S_c\sim 10^4$ is the critical value of $S$ for the onset of the plasmoid instability. Eq.~\eqref{MHD_rec_maintext} is a straightforward theoretical generalisation~\cite{Schekochihin20} to arbitrary $\mathrm{Pm}$ of a prediction for  $\mathrm{Pm}=1$~\cite{Uzdensky10} that has been confirmed numerically~\cite{Bhattacharjee09, Loureiro12}. $\mathrm{Pm}$ is given by Spitzer's theory~\cite{Spitzer56} [$\mathrm{Pm}_{\mathrm{Sp}}\sim 10^7$ at recombination, see Eq.~\eqref{Pm} in Methods] if the plasma is collisional, i.e., if the Larmor radius of protons $r_L = m_i cv_{\mathrm{th},i}/aeB$ is large compared to their mean free path, $\lambda_{\mathrm{mfp}}$ ($m_i$ and $v_{\mathrm{th},i}\equiv \sqrt{2T/m_i}$ are the mass and thermal speed of protons respectively). If, on the other hand, $r_L< \lambda_{\mathrm{mfp}}$, which happens if $B> B_{\mathrm{iso}} \equiv  m_i cv_{\mathrm{th},i}/ea\lambda_{\mathrm{mfp}}$, then the components of the viscosity tensor perpendicular to the magnetic field are reduced by a factor $(r_L/\lambda_{\mathrm{mfp}})^2$, because protons' motions across $\bBt$ are inhibited by their Larmor gyration~\cite{Braginskii65}. These are the components that limit reconnection outflows because velocity gradients in reconnection sheets are perpendicular to the mean magnetic field. Therefore, $\mathrm{Pm} \to (r_L/\lambda_{\mathrm{mfp}})^2 \PmSp = (\Bt_{\mathrm{iso}}/\Bt)^2 \PmSp$ in Eq.~\eqref{MHD_rec_maintext} if~$\Bt> \Bt_{\mathrm{iso}}\equiv a^2 B_{\mathrm{iso}}$.

The validity of the resistive-MHD treatment that leads to Eq.~\eqref{MHD_rec_maintext} requires the fluid approximation to hold at the scale of the critical sheet: its width
\begin{equation}
    \delta_c \sim \frac{S_c^{1/2}}{S}\lambda_B,\label{delta_c}
\end{equation}
must be larger than either $r_L$ or the ion inertial length ${d_i=\sqrt{m_i c^2/4\pi e^2 n_i a^2}}$ ($n_i$ is the proton number density)~\cite{Uzdensky10, Ji22}. If $\delta_c < r_L, d_i$, then the physics of the critical sheet is kinetic, not fluid, and the reconnection timescale is
\begin{equation}
    \tau_{\mathrm{rec}} \sim 10 \frac{\lambda_B}{\tilde{v}_{A}},\label{kinetic_rec_maintext}
\end{equation}rather than~\eqref{MHD_rec_maintext}. Eq.~\eqref{kinetic_rec_maintext} is a robust numerical result whose theoretical explanation is an active research topic (see~\cite{Liu22} for a recent study,~\cite{ComissoBhattacharjee16, Cassak17} for reviews). \blue{We shall find in the next section that~\eqref{kinetic_rec_maintext} is not the limiting timescale at recombination for almost any choice of initial condition consistent with EWPT magnetogenesis; our conclusions therefore do not depend sensitively on the validity of~\eqref{kinetic_rec_maintext}.}

The decay timescale can also be limited by radiation drag due to photons~\cite{BanerjeeJedamzik04}; this imparts a force~$-\tilde{\alpha} \but$ per unit density of fluid [see Eq.~\eqref{alpha} in Methods]. The drag is subdominant to magnetic tension at sufficiently small scales (as it does not depend on gradients of $\but$), so does not contribute to $\mathrm{Pm}$ in Eq.~\eqref{MHD_rec_maintext}. However, it can inhibit \textit{inflows} to the reconnection layer. Balancing drag with magnetic tension at the integral scale $\lambda_B$, we find an inflow speed $\ut \sim \tilde{v}_A^2/\tilde{\alpha}\lambda_B$, so the timescale for magnetic flux to be processed by reconnection is
\begin{equation}
    \tau_{\alpha}\equiv \frac{\tilde{\alpha} \lambda_B^2}{\tilde{v}_A^2}.
\end{equation}The timescale for energy decay depends on whether large-scale drag or small-scale reconnection physics is most restrictive:
\begin{equation}
    \tau = \max\{\tau_{\mathrm{rec}}, \tau_{\alpha}\}.\label{alpha_vs_rec}
\end{equation}

\noindent\textbf{Comparison with observations.} The locus of possible PMF states for different values of $I_H\sim \Bt^4\lambda_B^5$ under the theory that we have described is represented by the blue-gold line in Fig.~\ref{fig:b4l5}. We denote the largest value of $I_{H}$ consistent with EWPT magnetogenesis by $I_{H,\,\mathrm{max}}$; this corresponds to $\rhot_{B}(t_*)=\rhot_{\gamma}(t_*)$ and $\lambda_B(t_*)= r_H(t_*)$. For $I_H \lesssim 10^{-29} I_{H,\,\mathrm{max}}$, decays terminate on line~(i) in Fig.~\ref{fig:b4l5} [Eq.~\eqref{MHD_processing} in  Methods], which represents Eq.~\eqref{eddyprocessing} with $\tau = \tau_\mathrm{rec}$ given by Eq.~\eqref{MHD_rec_maintext} and $\mathrm{Pm}=\PmSp$. Use of Eq.~\eqref{MHD_rec_maintext} is valid here because $\delta_c \gtrsim r_L,\,d_i$ [see Eqs.~\eqref{delta_c/rho_i} and \eqref{dcdi_PmSp} in Methods]. The Spitzer estimate of $\mathrm{Pm}$ is valid at recombination only if ${\tilde{B}\lesssim \tilde{B}_{\mathrm{iso}}\sim 10^{-13}\,\mathrm{G}}$ [Eq.~\eqref{B_iso} in Methods], so decays with $I_H \gtrsim 10^{-29} I_{H,\,\mathrm{max}}$ have a shorter timescale at recombination --- they terminate on line~(ii) [Eq.~\eqref{line(ii)} in Methods], which represents Eq.~\eqref{eddyprocessing} with $\tau = \tau_\mathrm{rec}$ given by Eq.~\eqref{MHD_rec_maintext} and $\mathrm{Pm}\sim (r_L/\lambda_{\mathrm{mfp}})^2 \PmSp$. For $I_H\gtrsim 10^{-2} I_{H,\,\mathrm{max}}$, the states on line~(ii) have $\delta_c < d_i,\,r_L$ [see Eqs.~\eqref{dcdi_Brag} and \eqref{dcrL_Brag} in Methods], so Eq.~\eqref{MHD_rec_maintext} is invalid for them. These decays pass through line (ii) at some time before recombination with timescale given by Eq.~\eqref{kinetic_rec_maintext}. However, they do access the domain of validity of Eq.~\eqref{MHD_rec_maintext} if, before $t_{\mathrm{recomb}}$, $\tilde{B}$ becomes small enough for $\delta_c$ to be comparable with relevant kinetic scales. When that happens, their timescale becomes much larger than~$t_{\mathrm{recomb}}$ so further decay is prohibited --- these decays all terminate with $\Bt\sim10^{-11}\mathrm{G}$, which corresponds to~$\delta_c \sim d_i$ at $t_{\mathrm{recomb}}$ [see Eq.~\eqref{dcdi_Brag} in Methods]. Decays with $I_H \gtrsim 10^8 I_{H,\,\mathrm{max}}$ are radiation-drag limited at recombination [line~(iv); Eq.~\eqref{MHD_processing_alpha} in Methods] --- such decays are inconsistent with EWPT magnetogenesis, but could originate from magnetogenesis at the quantum-chromodynamic (QCD) phase transition, when $r_H\sim 10^{-6}\,\mathrm{Mpc}$~\cite{DurrerNeronov13, WagstaffBanerjee16}.

The EGMF parameters represented by the blue-gold line are consistent with Eq.~\eqref{constraint_delay} for~${I_H \gtrsim 10^{-23} I_{H,\,\mathrm{max}}}$,~i.e.,
\begin{equation}
    \bigg[\frac{\Bt(t_*)}{10^{-5.5}\,\mathrm{G}}\bigg]^4\left[\frac{\lambda_B(t_*)}{ 10^{-10}\,\mathrm{Mpc}}\right]^5\gtrsim 10^{-23}.
\end{equation}The relic of a field with $\lambda_{B}(t_*)\sim 10^{-2}\,r_H(t_*)$ $\sim 10^{-10}\,\mathrm{Mpc}$ at the EWPT would therefore be consistent with Eq.~\eqref{constraint_delay} \blue{--- modulo any modifications for plasma instabilities in voids~\cite{Broderick12, Broderick18, Batista19, PerryLyubarsky21, Addazi22, BatistaSaveliev21} ---} if $\tilde{\rho}_B(t_*) \gtrsim 10^{-6.5}{\tilde{\rho}}_{\gamma}(t_*)$. This confirms the assertion in the title of this paper. Intriguingly, if instead $\tilde{\rho}_B(t_*) \sim \tilde{\rho}_{\gamma}(t_*)$ and $\lambda_B(t_*)\gtrsim 10^{-2} r_H(t_*)$, then we find $\Bt\sim 10^{-11}\,\mathrm{G}$ at recombination. PMFs of this strength would provide a seed for magnetic fields in galaxy clusters that would not require significant amplification by turbulent dynamo after structure formation to reach their present day strength of~$\sim \mu \mathrm{G}$~\cite{BanerjeeJedamzik03}, \blue{although dynamo would still be required to \textit{maintain} cluster fields at present levels. We emphasise that a cluster field so maintained by dynamo need not (and, in all likelihood, would not) retain memory of its primordial seed.} We also note that PMFs of $10^{-11}\,\mathrm{G}$ strength are considered a promising candidate to resolve the Hubble tension, by modifying the local rate of recombination~\cite{JedamzikPogosian20, Galli22}.

As an aside, we note that the relevance of reconnection physics is not restricted to non-helical decay~\cite{HoskingSchekochihin20decay}. Some analogues for maximally helical PMFs of the results of this section (relevant for magnetogenesis mechanisms capable of parity violation) are presented in Section~A of the Supplementary Information.

\noindent\textbf{Role of plasma microinstabilities.} Finally, we note that, for $\Bt>\Bt_{\mathrm{iso}}$, the effective values of $\tilde{\nu}$ and $\tilde{\eta}$ might be dictated by plasma ``microinstabilities'' rather than by collisions between protons~\cite{Schekochihin10} (this is conjectured to happen in galaxy clusters~\cite{Schekochihin05}). In Methods, we show that the decay of the integral-scale magnetic energy is too slow to excite the ``firehose'' instability that is important in the cluster context [see Eq.~\eqref{beta_Delta}]. Nonetheless, we cannot rule out other microinstabilities --- for example, the excitation of the ``mirror'' instability by reconnection has been studied recently by Ref.~\cite{Winarto22}, although its effect on the rate of reconnection remains unclear. The most dramatic effect that microinstabilities in general could plausibly have would be to reduce the effective value of $\mathrm{Pm}$ to $\lesssim 1$ if $\Bt>\Bt_{\mathrm{iso}}$ (see~\cite{St-OngeKunz18, Kunz16}). This corresponds to the red-gold line in Fig.~\ref{fig:b4l5}, which remains consistent with Eq.~\eqref{constraint_delay} for $I_H\gtrsim 10^{-20} I_{H,\,\mathrm{max}}$. Compatibility between the EWPT-magnetogenesis scenario and the observational constraints on EGMFs therefore appears robust.

\section*{Methods}
\small

\noindent\textbf{Post-recombination evolution.} In the matter-dominated Universe after recombination, the transformation that maps Minkowski-spacetime MHD onto its expanding-Universe equivalent is not~Eq.~\eqref{transformation}, but~\cite{BanerjeeJedamzik04}
\begin{align}
    \rhot=a^3\rho,\quad& \pt= a^4p,\quad \bBt= a^2 \bB,\quad \but =a^{1/2}\bu, \nonumber\\
    & \etat = \eta/a^{1/2}, \quad \nut=\nu/a^{1/2},\quad \dd\tilde{t} =\dd t / a^{1/2}.\label{transformation2}
\end{align}As $a \propto t^{2}$ in the matter-dominated Universe, $\tilde{t}\propto \log t$, so a power-law decay in rescaled variables corresponds to only a logarithmic decay in comoving variables~\cite{Subramanian16}. Thus, in computing the expected present-day strength of EGMFs, one may assume the decay of $\Bt$ to terminate at recombination with negligible error.

\noindent\textbf{Derivation of Eq.~\eqref{ideal_law}.} In order to apply Eq.~\eqref{eddyprocessing}, we require an expression for the conformal time at recombination,~$t_{\mathrm{recomb}}$. From the Friedmann equation,
\begin{equation}
    \frac{1}{a^4}\left(\frac{\dd a}{\dd t}\right)^2=\frac{8\pi G \rho}{3},
\end{equation}where $G$ is the gravitational constant, the ``entropy equation''
\begin{equation}
    g T^3 a^3 = \const,\label{entropy}
\end{equation}where $g$ is the number of degrees of freedom of the radiation field and $T$ is the temperature, and Stefan's law for the radiation density
\begin{equation}
    \rho = 3\chi g T^4,
\end{equation}where $\chi = \pi^2 / 90 c^5 \hbar^3$ (we work in ``energy units'' for temperature, with Boltzmann constant $k_B=1$), it can be shown that
\begin{equation}
    \left(\frac{\dd T}{\dd t}\right)^2 = 8\pi G g_0 \chi T^4 T_0^2 \left(\frac{g}{g_0}\right)^{1/3},\label{dT/dt}
\end{equation}where the subscript 0 refers to quantities evaluated at the present day. Because $(g/g_0)^{1/6}\simeq 1$, one may solve Eq.~\eqref{dT/dt} to give an expression for the cosmic temperature as a function of conformal time,
\begin{equation}
    T = \frac{1}{t T_0} \sqrt{\frac{1}{8\pi G g_0\chi }}.
\end{equation}With $g_0=2$ (for the two photon-polarisation states), one obtains
\begin{equation}
    t \sim  10^{16.5} \mathrm{s} \left(\frac{T}{0.3\,\mathrm{eV}}\right)^{-1}. \label{t_recomb}
\end{equation}Therefore, Eq.~\eqref{eddyprocessing} becomes
\begin{equation}
    \tau \sim 10^{16.5} \mathrm{s} \left(\frac{T}{0.3\,\mathrm{eV}}\right)^{-1}.\label{tau=t_numerical}
\end{equation}Thus, $t_{\mathrm{recomb}}\sim 10^{16.5} \mathrm{s}$. Eq.~\eqref{tau=t_numerical} can be used to relate $\Bt$ and $\lambda_B$ under the assumption that the decay occurs on the Alfv\'{e}nic timescale $\tau\sim \lambda_B/\tilde{v}_A$ [Eq.~\eqref{Alfvenic}]. As noted in the main text, $\tilde{v}_A$ should be computed using the baryon density~$\tilde{\rho}_b$, because the photon mean free path~\cite{DurrerNeronov13}
\begin{equation}
     \lambda_{\mathrm{mfp},\,\gamma}= \frac{1}{a\sigma_T n_e}\sim 1 \mathrm{Mpc} \left(\frac{T}{0.3\,\mathrm{eV}}\right)^{-2}\label{photon_mfp}
\end{equation}(where $\sigma_T$ is the Thompson-scattering cross-section) is large compared with $\lambda_B$ at the time of recombination, indicating that photons are not strongly coupled to the fluid~\cite{JedamzikSaveliev19}. However, because $\tilde{\rho}_b\simeq \tilde{\rho}_{\gamma}$ at the time of recombination, the decoupling of photons does not affect Eq.~\eqref{ideal_law}. The Alfv\'{e}n speed~is
\begin{equation}
    \tilde{v}_A = \frac{\tilde{B}}{\sqrt{4\pi \tilde{\rho}_b}} \simeq 10^{16} \mathrm{cm\,s^{-1}}\frac{\tilde{B}}{1G} \left(\frac{T}{0.3\,\mathrm{MeV}}\right)^{1/2}, \label{vA}
\end{equation}where we have used $\tilde{\rho}_{b}=a^4 \rho_{b} \simeq a^4 m_i n_b$, with $m_i$ the proton mass and $n_b$ the WMAP value for the baryon number density $n_b\simeq 2.5 \times 10^{-7}\,\mathrm{cm}^{-3} a^{-3}$~\cite{Bennett03}, and taken ${a\simeq T_0/T}$ [Eq.~\eqref{entropy}]. Comparing Eq.~\eqref{Alfvenic} and Eq.~\eqref{tau=t_numerical}, and substituting Eq.~\eqref{vA}, we have
\begin{equation}
    \Bt \sim  10^{-8.5}\, \mathrm{G}\, \left(\frac{\lambda_{B}}{1\,\mathrm{Mpc}}\right) \left(\frac{T}{0.3\,\mathrm{eV}}\right)^{1/2}.
    \label{ideal_law2}
\end{equation}Evaluated at $T=T(t_{\mathrm{recomb}})=0.3\,\mathrm{eV}$, this is Eq.~\eqref{ideal_law}.

\noindent\textbf{Derivation of line~(i) of Fig.~\ref{fig:b4l5}.}
Line~(i) represents Eq.~\eqref{MHD_rec_maintext} evaluated at the time of recombination $t_{\mathrm{recomb}}$, with $\mathrm{Pm}=\PmSp\equiv \nut_{\mathrm{Sp}}/\etat_{\mathrm{Sp}}$, where $\nut_{\mathrm{Sp}}$ and $\etat_{\mathrm{Sp}}$ are the comoving Spitzer values of kinematic viscosity and magnetic diffusivity respectively~\cite{Spitzer56}. We first evaluate $\PmSp$. 

Under Spitzer theory, the dominant component of the plasma viscosity at the scale of the rate-determining current sheet is due to ion-ion (i.e., proton-proton) collisions. The collision frequency is~\cite{Spitzer56}
\begin{equation}
    \nu_{ii} \sim \frac{e^4 n_i \ln{\Lambda_{ii}}}{m_i^{1/2}{T_i}^{3/2}},\label{nu_ii}
\end{equation}where $e$ is the elementary charge, $n_i$ the ion number density, $m_i$ the ion mass, $T_i$ the ion temperature, and $\ln{\Lambda_{ii}}$ the Coulomb logarithm for ion-ion collisions. Neglecting any anisotropising effect of the magnetic field (see main text), the comoving isotropic kinematic viscosity is~\cite{Parra19}
\begin{equation}
        \nut_{\mathrm{Sp}} \sim \frac{v_{\mathrm{th},i}^2}{a\nu_{ii}} \sim  \frac{T_i^{5/2}}{a m_i^{1/2} e^4 n_i \ln{\Lambda_{ii}}}\sim 10^{18} \mathrm{cm^2 s^{-1}} \left(\frac{T}{0.3\,\mathrm{eV}}\right)^{1/2},\label{nu}
\end{equation}where $v_{\mathrm{th},i}=\sqrt{2T_i/m_i}$ is the thermal speed of ions, and we have assumed $T_i \simeq T$, used ${a\simeq T_0/T}$ [Eq.~\eqref{entropy}], taken $n_i$ to be equal to the WMAP value for the baryon number density $n_b\simeq 2.5 \times 10^{-7}\,\mathrm{cm}^{-3} a^{-3}$~\cite{Bennett03}, and estimated the Coulomb logarithm $\ln{\Lambda_{ii}}$ by
\begin{equation}
    \ln{\Lambda_{ii}}\simeq \ln{\frac{T_i^{3/2}}{e^3 n_i^{1/2}}}\simeq 20.
\end{equation}Similarly, the electron-ion collision frequency is~\cite{Parra19}
\begin{equation}
    \nu_{ei} \sim \frac{e^4 n_e \ln{\Lambda_{ei}}}{m_e^{1/2}{T_e}^{3/2}},\label{nu_ei}
\end{equation}where $n_e\simeq n_i$ is the electron number density, $T_e$ the electron temperature, and $\ln{\Lambda_{ei}}$ the Coulomb logarithm for electron-ion collisions. Eq.~\eqref{nu_ei} leads to the Spitzer~\cite{Spitzer56} value for the magnetic diffusivity
\begin{equation}
    \etat_{\mathrm{Sp}} \sim \frac{\nu_{ei}m_e c^2}{4\pi n_e e^2 a}\sim  10^{10.5} \mathrm{cm^2 s^{-1}} \left(\frac{T}{0.3\,\mathrm{eV}}\right)^{-1/2},\label{eta}
\end{equation}where we have used $\ln{\Lambda_{ei}}\simeq \ln{\Lambda_{ii}}\simeq 20$, assumed the electron temperature $T_e\simeq T$, and again neglected any anisotropy resulting from the magnetic field. From Eqs.~\eqref{nu} and~\eqref{eta}, we have
\begin{equation}
    \PmSp=\frac{\nut_{\mathrm{Sp}}}{\etat_{\mathrm{Sp}}}\sim\frac{ T^4}{m_e^{1/2} m_i^{1/2} e^6 n_i \ln{\Lambda_{ii}} \ln{\Lambda_{ei}}}\sim 10^7
    \left(\frac{T}{0.3\,\mathrm{eV}}\right).\label{Pm}
\end{equation}

Let us now evaluate the Lundquist number, Eq.~\eqref{Ssymbolic_maintext}, in order to compare it with $S_c$, as Eq.~\eqref{MHD_rec_maintext} requires. Note that, as above, it is the Alfv\'{e}n speed based on baryon inertia that appears in Eq.~\eqref{Ssymbolic_maintext}; photons are even more weakly coupled to the cosmic fluid at reconnection scales than at scale $\lambda_B$ as the former are typically small compared with the latter. Using Eqs.~\eqref{b4l5},~\eqref{vA}, and~\eqref{Pm}, we find the Lundquist number
\begin{align}
        S &= \frac{1}{\sqrt{1+\PmSp}} \frac{\tilde{v}_A(t_*)\lambda_B(t_*)}{\etat}\left[\frac{\lambda_B(t_*)}{\lambda_B}\right]^{1/4} \nonumber \\&\sim 10^{9} \left[\frac{\Bt(t_*)}{10^{-5.5} \,\mathrm{G}}\right]\left[\frac{\lambda_B(t_*)}{10^{-12} \,\mathrm{Mpc}}\right]\nonumber \\&\phantom{3000000000}\times
        \left[\frac{T}{0.3\,\mathrm{eV}}\right]^{1/2} \left[\frac{\lambda_B(t_*)}{\lambda_B}\right]^{1/4}.\label{S_vab}
\end{align}
Eq.~\eqref{S_vab} shows that $S \gg S_c\sim 10^4$ [unless $\tilde{B}(t_*)$ or $\lambda_B(t_*)$ are very small, in which case their evolution is inconsistent with the observational constraint~\eqref{constraint_delay}, so we neglect this possibility for simplicity]. Substituting Eq.~\eqref{Pm}, we find that the decay timescale~\eqref{MHD_rec_maintext} is
\begin{equation}
    \tau \sim 10^{5.5} \left(\frac{T}{0.3\,\mathrm{eV}} \right)^{1/2} \frac{\lambda_B}{\tilde{v}_{A}}\label{MHD_rate}.
\end{equation}Comparing Eqs.~\eqref{tau=t_numerical} and \eqref{MHD_rate}, and again substituting Eq.~\eqref{vA}, we find
\begin{equation}
  \Bt \sim  10^{-3} \mathrm{G}\, \left(\frac{\lambda_{B}}{1\,\mathrm{Mpc}}\right) \left(\frac{T}{0.3\,\mathrm{eV}}\right).
    \label{MHD_processing}
\end{equation}
Evaluated at $T=T(t_{\mathrm{recomb}})=0.3\,\mathrm{eV}$, this is line~(i) of Fig.~\ref{fig:b4l5}.

Finally, we note that when reconnection occurs under large-$\mathrm{Pm}$ conditions with isotropic Spitzer viscosity, the ratio of $\delta_c$ [Eq.~\eqref{delta_c}] to $r_L$ [defined below Eq.~\eqref{Ssymbolic_maintext}] prior to recombination is independent of the magnetic-field strength, temperature and density:
\begin{equation}
    \frac{\delta_c}{r_L}\sim  S_c^{1/2}\left(\frac{m_e}{m_i}\right)^{1/4}\sim 10,\label{delta_c/rho_i}
\end{equation}where we have used Eqs~\eqref{eta}, \eqref{Pm} and \eqref{vA}. 
Thus, ${\delta_c>r_L}$ always. Furthermore, we find from Eqs.~\eqref{Ssymbolic_maintext}, \eqref{delta_c}, \eqref{vA}, \eqref{eta}, \eqref{Pm} and the definition of $d_i$ [see below Eq.~\eqref{delta_c}] that
\begin{equation}
    \frac{\delta_c}{d_i}\sim S_c^{1/2} \left(\frac{m_e}{m_i}\right)^{1/4} \frac{v_{\mathrm{th},i}}{\tilde{v}_A} \sim \left(\frac{\Bt}{ 10^{-9}\,\mathrm{G}}\right)^{-1}.\label{dcdi_PmSp}
\end{equation}Therefore, $\delta_c > d_i, r_L$ at recombination for all relevant field strengths, so we are justified in using fluid theory to describe decays with ${\Bt<\Bt_{\mathrm{iso}}}$ [evaluated in Eq.~\eqref{B_iso}].

As described in the main text, Eq.~\eqref{MHD_processing} is valid when $\Bt$ is small enough for the Larmor radius of ions $r_L$ to be larger than their mean free path
\begin{equation}
    \lambda_{\mathrm{mfp}} \sim \frac{v_{\mathrm{th},i}}{\nu_{ii}a}\sim 10^{12}\mathrm{cm}.
\end{equation}The critical magnetic field strength above which this condition is no longer satisfied is
\begin{equation}
    \Bt_{\mathrm{iso}} \sim \frac{m_i c \nu_{ii} a^2}{e} \sim 10^{-13}\,\mathrm{G} \left(\frac{T}{0.3\,\mathrm{eV}} \right)^{-1/2}.\label{B_iso}
\end{equation}

\noindent\textbf{Derivation of line~(ii) of Fig.~\ref{fig:b4l5}.}

Line~(ii) represents Eq.~\eqref{MHD_rec_maintext} evaluated at the time of recombination $t_{\mathrm{recomb}}$, with magnetic Prandtl number $\mathrm{Pm}\sim (r_L/\lambda_{\mathrm{mfp}})^2 \PmSp = (\Bt_{\mathrm{iso}}/\Bt)^2 \PmSp$. Note that this suppression of~$\mathrm{Pm}$ relative to~$\PmSp$ \emph{increases} the value of $S$ at any given $\tilde{v}_A$ and $\lambda_B$ relative to the value~\eqref{S_vab} of $S$ that corresponds to $\mathrm{Pm}=\PmSp$. We therefore expect this family of decays also to have $S\gg S_c \sim 10^4$. 

The inclusion of the factor of $(\Bt_{\mathrm{iso}}/\Bt)^2$ in $\mathrm{Pm}$ modifies Eq.~\eqref{MHD_processing} straightforwardly: it becomes
\begin{multline}
    \Bt \sim  10^{-3} \mathrm{G}\,\left(\frac{\Bt_{\mathrm{iso}}}{\Bt}\right) \left(\frac{\lambda_{B}}{1\,\mathrm{Mpc}}\right) \left(\frac{T}{0.3\,\mathrm{eV}}\right).\\ \implies \Bt \sim  10^{-8} \mathrm{G}\,\left(\frac{\lambda_{B}}{1\,\mathrm{Mpc}}\right)^{1/2} \left(\frac{T}{0.3\,\mathrm{eV}}\right)^{1/4}.\label{line(ii)}
\end{multline}
Evaluated at $T=T(t_{\mathrm{recomb}})=0.3\,\mathrm{eV}$, this is line~(iv) of Fig.~\ref{fig:b4l5}. 

The analogue of Eq.~\eqref{dcdi_PmSp} for $\mathrm{Pm}\sim (\Bt_{\mathrm{iso}}/\Bt)^2 \PmSp$ is
\begin{align}
    \frac{\delta_c}{d_i} & \sim S_c^{1/2} \left(\frac{m_e}{m_i}\right)^{1/4} \frac{v_{\mathrm{th},i}}{\tilde{v}_A} \frac{\Bt_{\mathrm{iso}}}{\Bt} \nonumber \\ & \sim \left\{\Bt\Bigg/ \left[10^{-11}\,\mathrm{G} \left(\frac{T}{0.3\,\mathrm{eV}}\right)^{-1/4}\right]\right\}^{-2},\label{dcdi_Brag}
\end{align}
while the corresponding analogue of Eq.~\eqref{delta_c/rho_i} is
\begin{align}
    \frac{\delta_c}{r_L} & \sim S_c^{1/2}\left(\frac{m_e}{m_i}\right)^{1/4} \frac{\Bt_{\mathrm{iso}}}{\Bt} \nonumber \\ & \sim \left\{\Bt\Bigg/ \left[10^{-12}\,\mathrm{G} \left(\frac{T}{0.3\,\mathrm{eV}}\right)^{-1/2}\right]\right\}^{-1}.\label{dcrL_Brag}
\end{align}Eq.~\eqref{dcdi_Brag} shows that $\delta_c \gtrsim d_i$ at $t_{\mathrm{recomb}}$ if ${\Bt \lesssim 10^{-11}\mathrm{G}}$, while Eq.~\eqref{dcrL_Brag} indicates that ${\delta_c \gtrsim r_L}$ if ${\Bt \lesssim 10^{-12}\mathrm{G}}$. Following the prescription described in~\cite{Uzdensky10}, we use the former condition on $\Bt$ as the domain of validity of Eq.~\eqref{MHD_rec_maintext} in Fig.~\ref{fig:b4l5}, though we note that our results do not depend strongly on this choice --- the order-of-magnitude difference between the two critical values of $\Bt$ is comparable to the degree of accuracy to which our scaling arguments are valid.

We also note that the temperature dependence of Eq.~\eqref{dcdi_Brag} means that a decaying field that developed $\delta_c \gtrsim d_i$ \emph{before} recombination would have done so at a field strength $\Bt< 10^{-11}\mathrm{G}$; strictly, therefore, the decay of primordial fields should terminate somewhere below the horizontal part of the blue-gold curve in Fig.~\ref{fig:b4l5}, not directly on it. However, the difference is order unity and thus negligible for the purposes of our order-of-magnitude estimates. This is because magnetic decay was strongly suppressed by radiative drag at early times [a consequence of the strong temperature dependence of Eq.~\eqref{MHD_processing_alpha}] --- i.e., when temperatures exceeded around $10^2 \times 0.3\,\mathrm{eV}$. For all relevant values of $I_H$, the magnetic-field strength would therefore have greatly exceeded the critical value required for $\delta_c \sim d_i$ until the time that corresponds to this temperature, and by that time the critical field strength indicated by Eq.~\eqref{dcdi_Brag} was already within a small factor of its value at recombination. 

\noindent\textbf{Derivation of line~(iii) of Fig.~\ref{fig:b4l5}.} Line~(iii) represents Eq.~\eqref{MHD_rec_maintext} at the time of recombination $t_{\mathrm{recomb}}$, with $\mathrm{Pm}\lesssim1$. With $\mathrm{Pm}\lesssim 1$, Eq.~\eqref{S_vab} should be replaced by 
\begin{multline}
        S \sim 10^{12.5} \left[\frac{\Bt(t_*)}{10^{-5.5} \,\mathrm{G}}\right]\left[\frac{\lambda_B(t_*)}{10^{-12} \,\mathrm{Mpc}}\right] \\ \times\left[\frac{T}{0.3\,\mathrm{eV}}\right] \left[\frac{\lambda_B(t_*)}{\lambda_B}\right]^{1/4},\label{S_vab_Pm1}
\end{multline}so that $S\gg S_c\sim 10^4$ for all decays of interest. The decay timescale~\eqref{MHD_rec_maintext} therefore becomes
\begin{equation}
    \tau \simeq 10^2 \frac{\lambda_B}{\tilde{v}_{A}}\label{MHD_rate_Pm1}.
\end{equation}Comparing Eqs.~\eqref{tau=t_numerical} and \eqref{MHD_rate}, and substituting Eq.~\eqref{vA}, we find
\begin{equation}
  \Bt \sim 10^{-6.5} \mathrm{G}\, \left(\frac{\lambda_{B}}{1\,\mathrm{Mpc}}\right) \left(\frac{T}{0.3\,\mathrm{eV}}\right)^{1/2}.
    \label{MHD_processing_Pm1}
\end{equation}
Evaluated at $T=T(t_{\mathrm{recomb}})=0.3\,\mathrm{eV}$, this is line~(iii) of Fig.~\ref{fig:b4l5}.

The analogues of Eqs.~\eqref{dcdi_PmSp} and \eqref{delta_c/rho_i} for $\mathrm{Pm}\lesssim 1$ (but ${\etat\sim \etat_{\mathrm{Sp}}}$) are
\begin{equation}
    \frac{\delta_c}{r_L}\sim S_c^{1/2} \frac{c}{v_{\mathrm{th},e}}\frac{\ln \Lambda_{ei}}{\Lambda_{ii}} \sim 10^{-2.5} \left(\frac{T}{0.3\,\mathrm{eV}}\right)^{-1/2},
\end{equation}and
\begin{multline}
    \frac{\delta_c}{d_i}\sim S_c^{1/2} \frac{c}{\tilde{v}_A} \left(\frac{m_e}{m_i}\right)^{1/2}\frac{\ln \Lambda_{ei}}{\Lambda_{ii}}  \\ \sim
    \left\{\Bt\Bigg/ \left[10^{-13}\,\mathrm{G} \left(\frac{T}{0.3\,\mathrm{eV}}\right)^{-1/2}\right]\right\}^{-1}.
\end{multline}Note that the field strength at which $\delta_c\sim d_i$ is approximately equal to $\Bt_{\mathrm{iso}}$ at recombination (both are $\sim 10^{-13}\,\mathrm{G}$), while $\delta_c \ll r_L$. The red-gold line in Fig.~\ref{fig:b4l5} therefore extends past line~(iii) to line~(iv) along the line $\Bt\sim \Bt_{\mathrm{iso}}$.

\noindent\textbf{Radiation drag and the derivation of line~(iv) of Fig.~\ref{fig:b4l5}.}
As well as by viscosity arising from collisions between ions, the kinetic energy of primordial-plasma flows (after neutrino decoupling) can be dissipated by electron-photon collisions (Thompson scattering). Around the time of recombination, the comoving mean free path of photons, Eq.~\eqref{photon_mfp}, is much larger than the anticipated correlation scale of the magnetic field (and, therefore, of any magnetically driven flows). Under these conditions, the effect of Thompson scattering is to induce a drag on electrons. Owing to the collisional coupling between ions and electrons, this drag can dissipate bulk plasma flows. 

The comoving drag force on the fluid per unit baryon density is
\begin{equation}
    \tilde{\boldsymbol{F}}_D = -\tilde{\alpha} \tilde{\bu},
\end{equation}where~\cite{BanerjeeJedamzik04}
\begin{equation}
    \tilde{\alpha} \sim  \frac{c}{\lambda_{\mathrm{mfp},\,\gamma}}\frac{\rho_{\gamma}}{\rho_{b}} \sim 10^{-13.5}\mathrm{s}^{-1} \left(\frac{T}{0.3\,\mathrm{eV}}\right)^{3}.\label{alpha}
\end{equation}
As explained in the main text, the effect of drag is most important at the scale $\lambda_B$ (it becomes increasingly subdominant to magnetic tension at smaller scales) where it inhibits inflows to the reconnection layer. When the timescale $\tau_{\alpha}  \equiv \tilde{\alpha} \lambda_B^2/\tilde{v}_A^2$ on which flux can be delivered to the layer by strongly dragged inflows is larger than the reconnection timescale of the critical sheet $\tau_{\mathrm{rec}}$ [see Eq.~\eqref{alpha_vs_rec}], $\tau_{\alpha}$ gives the timescale for energy decay. Eq.~\eqref{tau=t_numerical} with $\tau = \tau_{\alpha}$ yields, after substitution of Eqs.~\eqref{vA} and Eq.~\eqref{alpha} 
\begin{equation}
  \Bt \sim 10^{-7} \mathrm{G}\, \left(\frac{\lambda_{B}}{1\,\mathrm{Mpc}}\right) \left(\frac{T}{0.3\,\mathrm{eV}}\right)^{3/2}.
    \label{MHD_processing_alpha}
\end{equation}Evaluated at $T=T(t_{\mathrm{recomb}})=0.3\,\mathrm{eV}$, this is line~(iv) of Fig.~\ref{fig:b4l5}. 

\noindent\textbf{Non-excitation of the firehose instability.}
Plasma with an anisotropic viscosity tensor can, in principle, be unstable to a variety of instabilities that develop at kinetic scales. For a decaying magnetic field, an instability of particular importance is the ``firehose'', which can generate the growth of small-scale magnetic fields in response to the decay of large-scale ones~\cite{Schekochihin10, Melville16}. This happens if the size of the (negative) pressure anisotropy $\Delta$ exceeds a critical value:
\begin{equation}
    \Delta \equiv \frac{p_{\perp} - p_{\|}}{p_{\|}}\leq -\frac{2}{\beta_i} \label{instability_condition}
\end{equation}where $p_{\|}$ and $p_{\perp}$ are the thermal pressures parallel and perpendicular to the magnetic field, and
\begin{equation}
    \beta_i \equiv \frac{p_{\|}}{B^2/8\pi}
\end{equation}is the ``plasma beta''. $\Delta$ can be estimated as~\cite{Schekochihin10}
\begin{equation}
    \Delta \sim \frac{1}{\nu_{ii}}\frac{1}{B}\frac{\dd B}{\dd \bar{t}} \sim -\frac{1}{a\nu_{ii} \tau}\sim -10^{-11} \left(\frac{T}{0.3\,\mathrm{eV}}\right)^{1/2},
\end{equation}where $\bar{t}$ is cosmic time [defined below Eq.~\eqref{metric}]. Naturally, the value of $\beta_i$ at any given $T$ depends on the evolution of the magnetic field. A lower bound on the value of $\Bt$ at any given time for a given initial condition is the one that would develop from a decay on the kinetic reconnection timescale, ${\tau \sim 10 \lambda_B/\tilde{v}_A}$ [Eq.~\eqref{kinetic_rec_maintext}]. Solving Eqs.~\eqref{b4l5},~\eqref{kinetic_rec_maintext},~\eqref{tau=t_numerical} and~\eqref{vA} simultaneously, we find that this is
\begin{multline}
     \Bt(t) \sim 10^{-13}\,\mathrm{G} \left(\frac{T}{0.3\,\mathrm{eV}}\right)^{5/18}\\ \times  \left[\frac{\lambda_B(t_*)}{10^{-12} \,\mathrm{Mpc}}\right]^{5/9}\left[\frac{\Bt(t_*)}{10^{-5.5} \,\mathrm{G}}\right]^{4/9}\label{kinetic_B_pred_T}.
\end{multline}Using this lower bound on $\Bt$, we can obtain an upper limit on $|\beta_i \Delta|$:
\begin{multline}
     |\beta_i \Delta| \lesssim 10^{-6} \left(\frac{T}{0.3\,\mathrm{eV}}\right)^{-1/18}\\ \times  \left[\frac{\lambda_B(t_*)}{10^{-12} \,\mathrm{Mpc}}\right]^{-10/9}\left[\frac{\Bt(t_*)}{10^{-5.5} \,\mathrm{G}}\right]^{-8/9}. \label{beta_Delta}
\end{multline}Eq.~\eqref{beta_Delta} suggests that the threshold for instability \eqref{instability_condition} is never met, unless $\lambda_B(t_*)$ and/or $\Bt(t_*)$ are so small as to be inconsistent with the observational constraint~\eqref{constraint_delay}.

\noindent\textbf{Numerical simulation.} The numerical simulations visualised in Fig.~\ref{fig:slices} and described in the Supplementary information were conducted using the spectral MHD code Snoopy~\cite{Lesur15}. The code solves the equations of incompressible MHD in Minkowski spacetime with hyper-viscosity and hyper-resistivity both of order~$n$, \textit{viz}.,
\begin{align}
    \frac{\p \bu}{\p t}+\bu\bcdot \bnabla \bu & = -\bnabla p + (\bnabla \times \bB)\times \bB \blue{-(-1)^{n/2} \nu_n \nabla^{n}\bu}, \\
    \frac{\p \bB}{\p t} & = \bnabla \times (\bu \times \bB) \blue{-(-1)^{n/2} \eta_{n}\nabla^n\bB},\label{mhd}
\end{align}where $p$, the thermal pressure, is determined via the incompressibility condition
\begin{equation}
    \bnabla \bcdot \bu =0.
\end{equation}The code uses a pseudo-spectral algorithm in a periodic box of size $2\pi$, with a $2/3$ dealiasing rule. Snoopy performs time integration of non-dissipative terms using a low-storage, third-order, Runge-Kutta scheme, whereas dissipative terms are solved using an implicit method that preserves the overall third-order accuracy of the numerical scheme. In all runs presented here, we employ $\nu_n = \eta_n =  10^{-12}$, ${n=6}$ and use a resolution of $512^3$.
\vspace{-3mm}
\normalsize
\acknowledgements

We are grateful to {R.~Blandford}, {B.~Chandran}, {F.~Rincon} and {D.~Uzdensky} for stimulating discussions, and to {K.~Subramanian} for posing a question that led us to write Section~B of the Supplementary Information. D.N.H. was supported by a UK STFC studentship. The work of A.A.S. was supported in part by the UK EPSRC grant EP/R034737/1. This work used the ARCHER UK National Supercomputing Service (http://www.archer.ac.uk).

\vspace{-2mm}
\section*{Author contributions}

D.N.H. conducted the study and wrote the manuscript, A.A.S. provided conceptual advice and comments on the manuscript.

\vspace{-3mm}
\section*{Data availability}

The data that support the findings of this study are available from the corresponding author upon reasonable request.

\onecolumngrid
\newpage

\setcounter{page}{1}

\section*{Supplementary information}
\setstretch{1.25}

\subsection{Decay of helical PMFs \label{helical}}

\begin{figure*}[t]
\centering
    \includegraphics[width=0.8\textwidth]{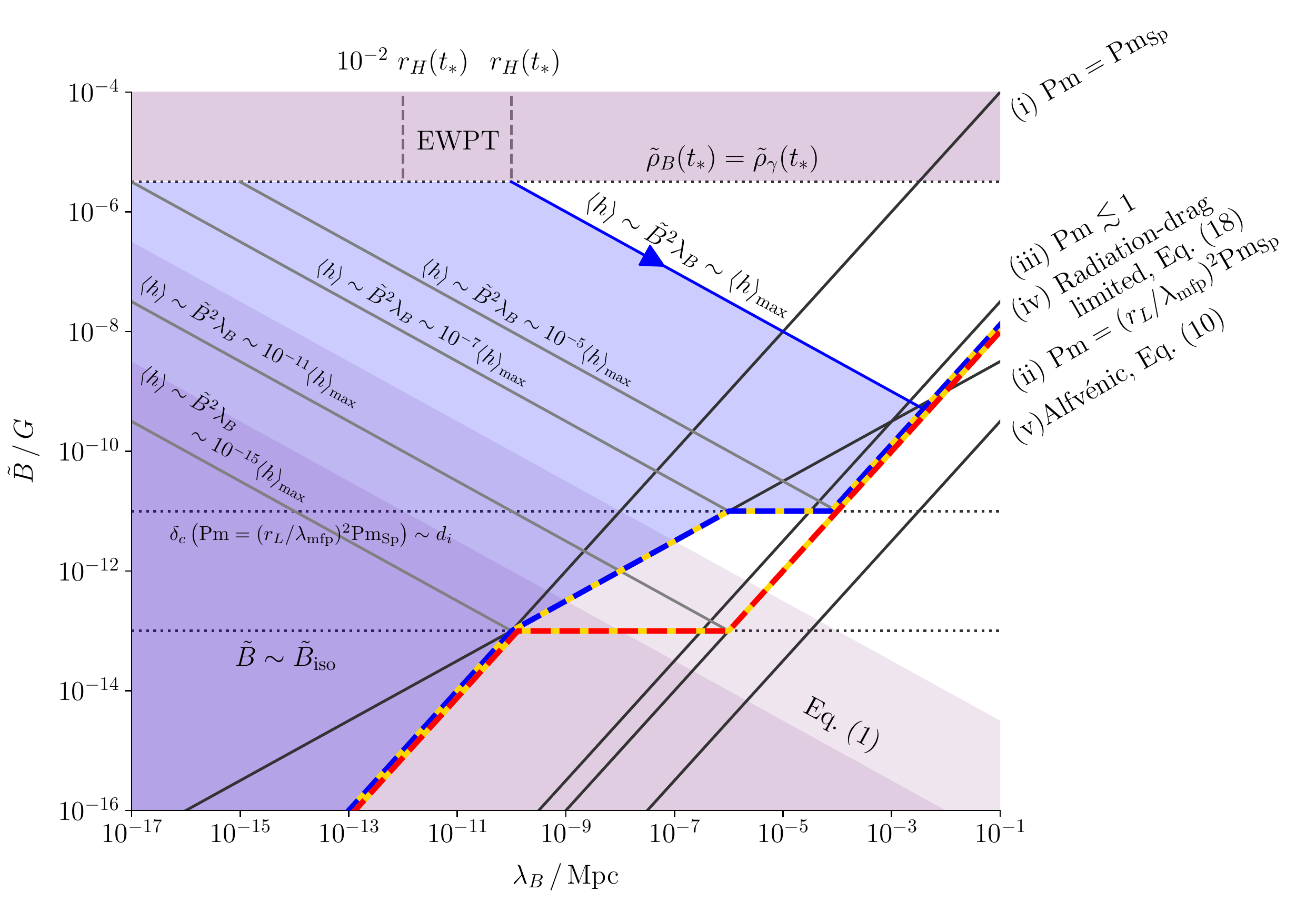}
    \caption{\textbf{Evolution of a maximally helical PMF.}\\
    % The Coulomb-gauge ($\bnabla \bcdot \bAt = 0$) magnetic-helicity density $\tilde{h}=\tilde{\bA}\bcdot \tilde{\bB}$,  , taken from a direct numerical simulation of non-helical MHD turbulence decaying from a magnetically dominated state.
    Analogue of Fig.~\ref{fig:b4l5}, showing the decay of a maximally helical magnetic field generated at the EWPT.
    }
    \label{fig:helical}
\end{figure*}

For completeness, here we provide the results for maximally helical fields that correspond to those presented in the main text for non-helical fields; these results are relevant for magnetogenesis mechanisms that are capable of parity violation. The decay of such fields conserves the net magnetic helicity, resulting in the self-similar scaling
\begin{equation}
    \langle h\rangle = \Bt^2 \lambda_B \sim \const.\label{helical_scaling}
\end{equation}As in the non-helical case, the decay proceeds on reconnection timescales~\cite{HoskingSchekochihin20decay}; the possible decay regimes are the same as those described in the main text. Under Eq.~\eqref{helical_scaling}, the PMF evolution in the $(\Bt, \lambda_B)$ plane is parallel to Eq.~\eqref{constraint_delay} [see Fig.~\ref{fig:helical}]. Thus, any field satisfying
\begin{equation}
    \Bt(t_*) \gtrsim 10^{-17} \,\mathrm{G}\, \left[\frac{\lambda_B(t_*)}{1\,\mathrm{Mpc}}\right]^{-1/2}
\end{equation}will satisfy the observational constraint~\eqref{constraint_delay} at recombination, as is well known~(see, e.g., \cite{DurrerNeronov13, Subramanian16}). 

The locus of present-day PMF states for decays that occur on the reconnection timescales explained in the main text is shown by the blue-gold line in Fig.~\eqref{fig:helical}. Analogously to $I_{\bL,\,\mathrm{max}}$ and $I_{H,\,\mathrm{max}}$ in the main text, we denote the largest value of the mean magnetic-helicity density $\langle h \rangle$ that is consistent with EWPT magnetogenesis by $\langle h \rangle_{\mathrm{max}}$ [this corresponds to $\rhot_{B}(t_*)=\rhot_{\gamma}(t_*)$ and $\lambda_B(t_*)= r_H(t_*)$]. For $\langle h \rangle \lesssim 10^{-15}\langle h \rangle_{\mathrm{max}}$, the decay of PMFs terminates on line~(i) in Fig.~\ref{fig:helical}, which corresponds to Eq.~\eqref{MHD_rec_maintext} of the main text with $\mathrm{Pm}\sim\PmSp$ [Eq.~\eqref{Pm}]. For ${10^{-15}\langle h \rangle_{\mathrm{max}}\lesssim \langle h \rangle \lesssim 10^{-11}\langle h \rangle_{\mathrm{max}}}$, the decay of PMFs terminates on line~(ii) [Eq.~\eqref{line(ii)} in Methods], which corresponds to Eq.~\eqref{MHD_rec_maintext} of the main text with $\mathrm{Pm}=(r_L/\lambda_{\mathrm{mfp}})^2 \PmSp$. For $10^{-7}\langle h \rangle_{\mathrm{max}}\lesssim \langle h \rangle \lesssim 10^{-5}\langle h \rangle_{\mathrm{max}}$, decays terminate at $\Bt\sim 10^{-11},\mathrm{G}$, which corresponds to $\delta_c\sim \lambda_{\mathrm{mfp}}$, as explained in the main text. Finally, decays are radiation-drag limited for $\langle h \rangle \gtrsim 10^{-5}\langle h \rangle_{\mathrm{max}}$, and therefore terminate on line~(iv) [Eq.~\eqref{MHD_processing_alpha} in Methods]. We note that, for $\langle h \rangle\lesssim 10^{-5} \langle h \rangle_{\mathrm{max}}$, the role of magnetic reconnection in determining the decay timescale implies significantly stronger relic fields than would be expected under the decay physics envisaged by~\cite{BanerjeeJedamzik04}, i.e., Alfv\'{e}nic [Eq.~\eqref{Alfvenic}; line~(v)] or radiation-drag-limited [line~(iv)] decay.

As in the main text, we also indicate by a red-gold line the locus of present-day PMF states if $\mathrm{Pm}\lesssim 1$ (due to plasma microinstabilities) for $\Bt\gtrsim \Bt_{\mathrm{iso}}$.

\subsection{Decay of non-helical magnetic fields with $I_H=0$ \label{individually_nonhelical}}

As explained in the main text, the invariance of $I_H$ follows from the conservation of the fluctuation level of magnetic helicity. While we view fluctuations in magnetic helicity to be a generic feature of real MHD turbulence\footnote{It should also be noted that all extant numerical work on this subject~\cite{BiskampMuller99, BiskampMuller00, Christensson01, BanerjeeJedamzik04, FrickStepanov10, BereraLinkmann14, Brandenburg15, Brandenburg17, ReppinBanerjee17, Bhat21} has exclusively employed initial conditions \textit{with} helicity fluctuations.}, it is nonetheless possible to construct artificial field configurations for which the helicity of each magnetic structure vanishes --- this will be the case if they have no twists and do not interlink. Strictly, therefore, the possibility that PMFs might have been generated without helicity fluctuations cannot be ruled out. 

\textit{A priori}, it appears that this kind of field might relax in a fundamentally different manner to the one described in the main text. This was the view that we expressed in Ref.~\cite{HoskingSchekochihin20decay}: there, we suggested that fields with $I_H=0$ might decay subject to the conservation of invariants associated with the velocity, rather than the magnetic, field. This is because individually non-helical structures (unlike helical ones) can relax under entirely flux-frozen dynamics, by driving flows with $\ut \sim \Bt$ (a process sometimes called kinetic diffusion~\cite{DurrerNeronov13}). Plausibly, the decay of those flows would respect the invariance of the hydrodynamic Loitsyansky integral,
\begin{equation}
    I_{\bL}\equiv-\int\dd^3 \br \,r^2 \langle \but(\bx)\bcdot\but(\bx+\br) \rangle,\label{Loitsyansky}
\end{equation}which encodes the conservation of angular momentum $\bL = \bx \times \bu$~\cite{LandauLifshitzFluids} (in the same fluctuating manner as $I_H$ encodes helicity conservation~\cite{HoskingSchekochihin20decay}).\footnote{$I_{\bL}$ is related to the small-$k$ asymptotic of the kinetic-energy spectrum, $\mcE_K(k)$, of isotropic turbulence without long-range spatial correlations by $\mcE_{K}(k\to 0) = I_{\bL} k^4/24 \pi^2 + O(k^6)$~\cite{Davidson15}.} Denoting the characteristic size and scale of the velocity field by $\ut$ and $\lambda_u$ respectively, $I_{\bL}\sim \ut^2 \lambda_u^5$. Conservation of $I_{\bL}$ therefore implies $\ut^2 \lambda_u^5\sim\const$. This suggests that $\Bt^2 \lambda_B^5\sim \const$ also, if $\ut \sim \Bt$ and $\lambda_B \sim \lambda_u$, which seems reasonable for, e.g., a magnetic field maintained by the dynamo effect\footnote{We note that the dynamo effect in a decaying velocity field has been studied by~\cite{Brandenburg17}.}. This returns us to~Eq.~\eqref{b2l5_const}, i.e., to the same prediction that was shown to be inconsistent with the observational constraints by Ref.~\cite{WagstaffBanerjee16}. 

On the other hand, if the magnetic field \textit{were} maintained by dynamo, then it seems unlikely that $I_H=0$ would be maintained. This is because random helicity fluctuations could be generated freely at resistive scales (as the Lundquist number is order unity there), where dynamo primarily generates magnetic field (at least in its kinematic stage)~\cite{Rincon19}. Thus, $I_H$ could become non-zero, although it would not need to be conserved if magnetic energy remained concentrated at the resistive scales. If, however, the dynamo-replenished magnetic fields later transferred to larger scales and saturated with $\lambda_B \sim \lambda_u$, as supposed above, while still having helicity fluctuations, then $I_H$ would become invariant, because the integral scale of the magnetic field would be much larger than the resistive scale. This would push us back to the scaling $I_H\sim \Bt^4 \lambda_B^5\sim\const$ [Eq.~\eqref{b4l5}]. Moreover, we conjecture that the size of the conserved product $\Bt^4 \lambda_B^5$ would be of the same order as its value for the initial field with $I_H=0$, because memory of its $\Bt$ and $\lambda_B$ would be retained by the velocity field.

\begin{figure*}
    \centering
    \includegraphics[height=0.85\textheight]{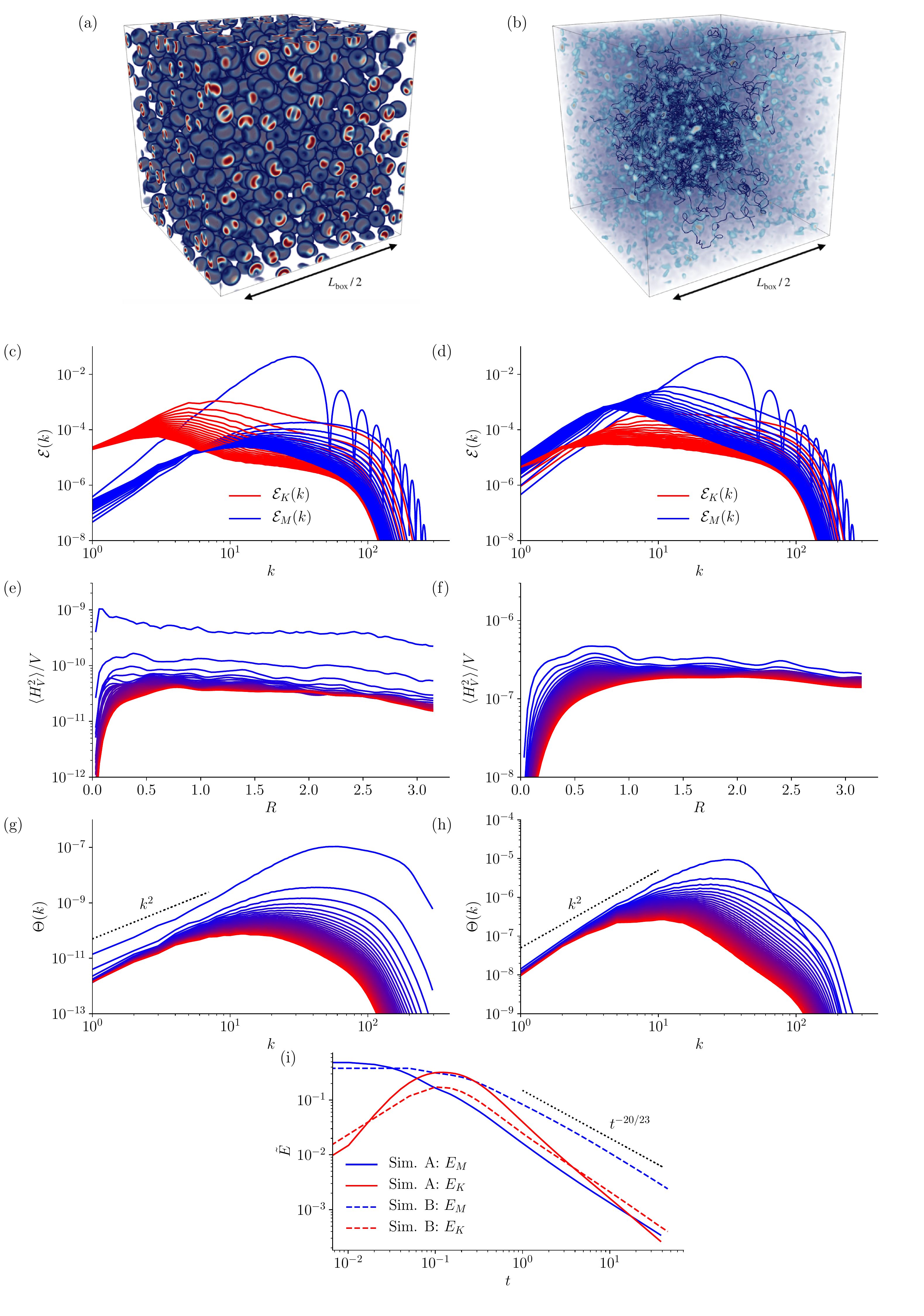}
    \caption{\textbf{Simulations of MHD turbulence decaying from a magnetically dominated state.}
    \\
    Left-hand plots are for Simulation A, which had $I_H=0$ initially; right-hand plots are for Simulation B, which had $I_H\neq0$ initially. Panels (a-b) show a 3D plot of the magnetic-energy distribution at the initial time, in a volume $1/8$ the size of the simulation domain; panels (c-d) show energy spectra (kinetic in red, magnetic in blue), plotted at intervals of $2.0$ between $t=0$ and $t=38.0$ (time is measured in code units based on normalising the box size and the mean-square magnetic field to $2\pi$ and $1$, respectively, so that one time unit is equal to the initial Alfv\'{e}n crossing time of the box); panels (e-f) show $\langle H_V^2 \rangle/V$ (computed as an average over many spheres of radius $R$ and volume $V$, distributed throughout the simulation domain) vs. $R$, plotted at intervals of $1.0$ between $t=0.25$ (blue) and $t=38.25$ (red); panels (g-h) show the helicity-variance spectrum $\Theta(k)$, at the same times as for (e-f); panel (i) shows the evolution of magnetic energy $E_M$ and kinetic energy $E_K$ for each simulation as functions of time, with the theoretical prediction for the decay on the slow-reconnection timescale, $E_M \propto t^{-20/23}$, given for reference [this follows from Eq.~\eqref{MHD_rec_maintext}, generalised appropriately for hyper-dissipation, with $S<S_c$ --- see~\cite{HoskingSchekochihin20decay}].}
    \label{fig:simulations}
\end{figure*}
In Fig.~\ref{fig:simulations}, we present results from a numerical simulation (Simulation A) designed to assess these arguments. We initialise a large number of untwisted, non-interlinking magnetic-flux loops (otherwise distributed in a random, statistically isotropic way) in a periodic simulation domain [see Fig.~\ref{fig:simulations}(a)]. At $t=0$, $I_H=0$ because the loops each have zero magnetic helicity. For the purpose of comparison, we also present a second simulation (Simulation B) with the same setup but without the non-interlinking condition, instead starting with many loops superimposed on top of each other. This field has complex initial topology, as Fig.~\ref{fig:simulations}(b) indicates, but no net helicity, in the sense that $\langle H_V\rangle/V\ll \tilde{B}^2\lambda_B$ for all $V$. However, unlike the field in Simulation A, it has $I_H\neq 0$, because superimposing loops creates linkages in the magnetic field.

Figs.~\ref{fig:simulations}(c-d)
show the evolution of energy spectra for the two simulations. In Simulation A, unlike in Simulation B, there is an immediate and rapid decay of the magnetic energy as the loops contract and drive flows. There is a corresponding decay of the large-scale (low-$k$) spectral tail, which demonstrates the non-invariance of $I_{\bL_M}$. The newly generated kinetic energy is comparable in magnitude to the initial magnetic energy [see Fig.~\ref{fig:simulations}(i)], and its spectrum peaks close to the initial peak of the magnetic-energy spectrum.\footnote{At larger scales, it exhibits a power law close to $\mcE(k)\propto k^2$, suggesting that it is a `Saffman turbulence' --- roughly speaking, eddies are translational rather than rotational~\cite{Davidson15}. However, we note that each flux tube must individually relax in a momentum-conserving manner, so it is unlikely that the relaxation could generate true Saffman turbulence, which has a stochastic momentum distribution. Instead, it is likely that the momentum distribution is ``quasi-random'', in the sense described by Ref.~\cite{HoskingSchekochihin21k2} --- in an arbitrarily large simulation domain, one would find that the $\mcE_K(k)\propto k^2$ spectrum transitions to $\mcE_K(k)\propto k^4$ at sufficiently large scales. Similarly, the large-scale spectrum of the magnetic energy appears to be somewhat shallower than $\mcE_M(k)\propto k^4$ --- we think that this too is an effect of the finite size of the simulation domain.} On the other hand, the contraction of the loops leaves magnetic energy concentrated at small (resistive) scales, where it can be refuelled by the dynamo effect associated with the newly generated flows. As we anticipated above, this resistive-scale magnetic field has random fluctuations in magnetic helicity: this is shown explicitly in Fig.~\ref{fig:simulations}(e) where, for volumes $V$ taken to be spheres of radius $R$, we plot $\langle H_V^2\rangle/V$ vs. $R$ at regular intervals in time (the average is taken over a large sample of spheres with centres throughout the simulation box).
While $\langle H_V^2\rangle \propto V^{2/3}$ at $t=0$ (not shown) because $H_V$ is dominated by random surface contributions at this time, this scaling is replaced by $\langle H_V^2\rangle \propto V$ as soon as turbulence develops, indicating $I_H\neq 0$ [see Eq.~\eqref{randomwalk} of the main text]. Though $I_H$, which is the value of $\langle H_V^2\rangle/V$ in the flat part of the curves in Figs.~\ref{fig:simulations}(e-f), decays by around an order of magnitude during the first few eddy-turnover times, Fig.~\ref{fig:simulations}(e) shows that its decay ceases after that.
This is consistent with our suggestion above that $I_H$ should become constant when the dynamo saturates, due to migration of the helicity-containing scale towards the flow scale $\lambda_u$.\footnote{We identify the migration as dynamo-induced because it occurs under conditions of dominant kinetic energy [see Fig.~\ref{fig:simulations}(i)]. An alternative explanation is that it occurs because of the non-helical inverse-transfer effect described in the main text. The connection between the two phenomena, and the role that the invariance of $I_H$ might have in constraining the nonlinear dynamo's evolution, are topics to which we plan to return in future work.} This interpretation is supported by the evolution of the helicity-variance spectrum $\Theta(k)$ [see Fig.~\ref{fig:simulations}(g)], which encodes the characteristic size of helicity fluctuations at each scale.\footnote{Note that $\Theta(k)$ is not the same as the helicity spectrum, which is close to zero for all $k$ for both simulations, as the field is non-helical at all scales.} In Simulation A, $\Theta(k)$ is concentrated around the dissipation scale at early times (though after the decay of the magnetic loops), but later moves to larger scales. $I_H$ [which is proportional to the coefficient of $k^2$ in the $\Theta(k\to 0) \propto k^2$ asymptotic~\cite{HoskingSchekochihin20decay}] ceases to decay once the peak of $\Theta(k)$ is moderately separated from the dissipation scales.

The value of $I_H$ ultimately attained by the magnetic field in Simulation A is smaller than the one in Simulation B by a factor of around $10^4$. This appears to contradict our conjecture that dynamo should generate $I_H$ of the same size as $\Bt^4 \lambda_B^5$ at the initial time. On the other hand, we note that (i) this factor may well be smaller for a simulation at larger resolution and larger Prandtl number (recent work has shown that extremely large resolutions are required to probe the asymptotic nature of the large-$\mathrm{Pm}$ dynamo~\cite{Galishnikova22}), and that (ii) the strong scaling of $I_H$ with $\Bt$ and $\lambda_B$ means that even a factor-$10^4$ reduction in $I_H$ corresponds only to a factor-$10$ reduction in $\Bt$ or $\lambda_B$. This means that a PMF generated with $I_H=0$ would migrate only a relatively short distance on the $(\Bt, \lambda_B)$ plane (Fig.~\ref{fig:b4l5}) before settling to decay with $\Bt^4\lambda_B^5\sim \const$.

To summarise, there appear to be both theoretical and numerical reasons to believe that a PMF generated with $I_H(t_*)=0$ at the initial time $t_*$ would, via an initial period of rapid decay and subsequent regeneration via dynamo, develop ${I_H\sim \Bt(t_*)^4\lambda_B(t_*)^5\sim \const}$. At later times, a magnetically dominated state would likely be re-established because the flows will drive Alfv\'{e}nic turbulence, which cascades to small scales and is dissipated by viscosity (which may be large, if associated with neutrinos or photons), while background ``quasi-force-free'' magnetic fields persist, decaying only on the magnetic-reconnection timescale, as described in the main text. There is some evidence of this in Fig.~\ref{fig:simulations}(i), which shows that magnetic energy becomes larger than kinetic in Simulation A at late times.

\subsection{The effect of the large-scale spectral slope: coexistence of flux and helicity invariants}

In this paper, we have contrasted our theory of $I_H$-conserving PMF decay with the previously accepted theory based on ``selective decay of small-scale structure'', i.e., the invariance of the large-scale asymptotic of the magnetic-energy spectrum. One success of our theory is that it explains the inverse-transfer effect observed in simulations of magnetic fields initialised with ${\mcE_M(k \to 0)\propto k^4}$ (\cite{Zrake14, Brandenburg15}; see main text); this effect is manifestly not compatible with selective decay. On the other hand, Ref.~\cite{ReppinBanerjee17} observe that inverse transfer is \textit{not} present in simulations that are initialised with sufficiently shallow large-scale spectra [namely, with ${\mcE_M(k \to 0, t=0)\propto k^n}$, where $n<3$]. Instead, they find that the $k\to 0$ asymptotic of $\mcE_M(k)$ is preserved. This result raises questions of whether a ``selective-decay-like'' principle might be at work in such decays, and what its effect might be on the laws for the decay of energy and growth of the integral scale. In this Section, we explain the invariance of this $k^n$ asymptotic as a consequence of the conservation of magnetic flux, but also argue that, beyond an initial transient, flux conservation does not affect the decay laws if $n>3/2$ (as is the case in all models of EWPT magnetogenesis of which we are aware). It is therefore not necessary to know the precise value of $n$ to compute the present-day properties of EGMFs under the relic-field hypothesis --- the theory presented in the main text is valid independently of it.

\subsubsection{Invariance of the large-scale spectral asymptotic for $n\leq 3$}

In general, the large-scale spectral asymptotic is frozen in time when the coefficient of $k^n$ in $\mcE_M(k\to 0)$ is proportional to some statistical invariant. As explained in the main text, this is not the case when correlations in $\bB$ decay rapidly with distance, because then $\mcE_M(k \to 0)\propto I_{\bL_M}k^4$ [Eq.~\eqref{magnetic_Loitsyansky}] where $I_{\bL_M} \neq \const$. However, for $n\leq 3$, it turns out that the coefficient of $k^n$ is proportional to an invariant that is related to the conservation of magnetic flux. Physically, this invariant encodes the fact that, over sufficiently large volumes, local fluctuations in magnetic flux may sum to a non-zero net fluctuation level, which must be conserved as the field decays. Spatial correlations must be long (and hence spectra must be shallow) for the fluctuation level to be non-zero, because $\bnabla \bcdot \bB =0$ means that magnetic structures without sufficiently strong far-field components have net zero flux. The relevant measure of correlation strength is the large-$r$ asymptotic of the magnetic field's longitudinal correlation function, $\chi_B(r)\equiv \langle B_r(\bx)B_r(\bx+\br)\rangle /\langle B_r^2 \rangle$, where $B_r = \bB \bcdot \br/r$. The argument is particularly transparent if ${\chi_B(r\to\infty)\propto r^{-3}}$ (as, for example, would be the case for a superposition of many randomly positioned and oriented magnetic dipoles), as then it can be shown that
\begin{equation}
    \mcE_M(k\to 0) = \frac{I_{\bB}k^2}{4\pi^2},
\end{equation}where
\begin{equation}
    I_{\bB}\equiv \int\dd^3 \br \langle \bBt(\bx)\bcdot\bBt(\bx+\br) \rangle = \lim_{V\to\infty} \frac{1}{V} \left\langle \left(\int_V \dd^3 \bx\,\bBt \right)^2\right\rangle \equiv  \lim_{V\to\infty} \frac{\langle \Bt_V^2\rangle}{V}\label{magnetic_Saffman}
\end{equation}is the Saffman flux invariant~\cite{HoskingSchekochihin20decay}. The invariance of $I_{\bB}$ encodes conservation of the fluctuation level of magnetic flux in the same manner as the invariance of $I_H$ does for magnetic helicity. More generally, if $\mcE_M(k\to 0) = C k^n$ with $n>-1$, then it can be shown that $\chi_B$ satisfies
\begin{equation}
    \chi_B(r\to \infty) \begin{dcases*}
                    \leq O(r^{-1-n}) & if  $n=2m$, $m=2,3,4,\dots$;\\
                    = f_n C r^{-1-n} & otherwise, \end{dcases*}\label{asymptotic_result}
\end{equation}where $f_n$ is a numerical coefficient that depends only on $n$. Furthermore, it can also be shown that
\begin{align}
    \lim_{R\to\infty}\langle \Bt_V^2\rangle \begin{dcases*}
            \propto R^2 & if  $n>3$, \\
            = g_n C R^2 \ln R & if  $n=3$, \\
            = g_n C R^{5-n} & if  $-1<n<3$, \\
                 \end{dcases*}\label{P2_asymptotic2}
\end{align}where $g_n$ is a different numerical coefficient dependent only on $n$, and $R$ is the radius of a spherical control volume $V$. These results are straightforward analogues of ones that we derived for the kinetic-energy spectrum of hydrodynamic turbulence in Ref.~\cite{HoskingSchekochihin21k2}. The rate of change of $\langle \Bt_V^2\rangle$ due to the advection of flux through the surface of $V$ scales as
\begin{equation}
    \frac{\dd}{\dd t} \langle \Bt_V^2\rangle \propto V^{2/3}\propto R^2 \implies \frac{\dd}{\dd t} \log \langle \Bt_V^2\rangle \propto \begin{dcases*}
            1 & if  $n>3$, \\
            1/\ln R & if  $n=3$, \\
            R^{n-3} & if  $-1<n<3$, \\
                 \end{dcases*}\label{logBv}
\end{equation}so the timescale associated with changes in $\langle \Bt_V^2\rangle$ is an increasing function of $R$ for $n\leq 3$. This means that the decay is constrained by the conservation of magnetic flux via
\begin{equation}
    \lim_{R\to\infty}\frac{\langle \Bt_V^2\rangle}{R^{5-n}}= \const =  g_n C. \label{B_V_const}
\end{equation}Eq.~\eqref{B_V_const} shows that $C=\const$ for $n \leq 3$, which explains the invariance of $\mcE_M(k\to0)$ observed by Ref.~\cite{ReppinBanerjee17}.

\subsubsection{Conservation of magnetic flux does not affect the decay laws for $\Bt$ and $\lambda_B$}

We now turn to the effect that the need to satisfy the new constraint~\eqref{B_V_const} has on the decay laws. A fully self-similar decay satisfying Eq.~\eqref{B_V_const} would have
\begin{equation}
    \lim_{R\to\infty} \frac{\langle\Bt_V^2\rangle}{R^{5-n}}\sim \Bt^2 \lambda_B^{1+n}\sim \const, \label{selective_decay}
\end{equation}which is the selective-decay scaling considered by~\cite{BanerjeeJedamzik04}. However, Eq.~\eqref{selective_decay} cannot describe the true evolution as it is inconsistent with the invariance of $I_H$, as we now explain. While, in principle, $I_H$ can be small compared to $\Bt^4 \lambda_B^5$ (see Section~\ref{individually_nonhelical} of the Supplementary Information), it cannot be much larger than this: $I_H\sim \Bt^4 \lambda_B^5$ corresponds to magnetic fields that are locally maximally helical.\footnote{We expect that $\langle H_V^2\rangle\propto R^3$ even in the presence of slowly decaying correlations in $\bB$. This is because spatial correlations in the magnetic helicity decay faster than those in the magnetic field. To see why, it is convenient to imagine a turbulence consisting of a superposition of uncorrelated magnetic structures. For $n<4$, the far-field component of $\bB$ associated with any given structure must scale as $r^{-1-n}$ [Eq.~\eqref{asymptotic_result}], so the far-field component of the vector potential $\bA$ due to that structure is proportional to $r^{-n}$, and hence the far-field component of $h$ is proportional to $r^{-2n-1}$. $I_H$ diverges only if the helicity correlation function $\langle h(\bx)h(\bx+\br)\rangle \geq O(r^{-3})$ as $r\to\infty$ [see Eq.~\eqref{I_H} of the main text], which occurs if $n<1$. Thus, $\langle H_V^2\rangle\propto R^3$ provided that $n>1$. As explained in the main text, this scaling implies $I_H = \lim_{R\to\infty}\langle H_V^2\rangle/R^{3}\sim \Bt^4 \lambda_B^{5}\sim \const$.} Therefore, adopting Eq.~\eqref{selective_decay}, we can write
\begin{equation}
    I_H \lesssim B^4 \lambda_B^5 \sim \Bt^{2(2n-3)/(n+1)}.\label{decay_I_H}
\end{equation}Assuming that $n>3/2$, Eq.~\eqref{decay_I_H} requires $I_H$ to be smaller than a decreasing function of time, which contradicts its invariance. 

On the other hand, the scaling $ I_H\sim \Bt^4 \lambda_B^{5} \sim \const$ [Eq.~\eqref{b4l5} of the main text] is \textit{not} incompatible with Eq.~\eqref{B_V_const}, as, under this scaling,
\begin{equation}
    \lim_{R\to\infty}\frac{\langle \Bt_V^2\rangle}{R^{5-n}}\lesssim \Bt^2 \lambda_B^{1+n} \sim \Bt^{2(3-2n)/5}. \label{B_V_inc}
\end{equation}Again assuming that $n>3/2$, Eq.~\eqref{B_V_inc} only requires ${\lim_{R\to\infty}\langle \Bt_V^2\rangle/R^{5-n}}$ to be smaller than an increasing function of time, which does not contradict its conservation. We conclude that while the selective-decay scaling \eqref{selective_decay} is ruled out by $I_H$ conservation, the converse is not true: Eq.~\eqref{b4l5} is compatible with the conservation of the magnetic-flux fluctuation level, and thus with the invariance of the large-scale spectral asymptotic. We therefore expect Eq.~\eqref{b4l5} of the main text to hold regardless of the value of $n$ (although for $n < 4$, some transient order-unity variation in $\Bt^4 \lambda_B^5$ should be expected as a result of departures from self-similarity; see below).

That conservation of $I_H$ should provide the relevant constraint even in the presence of magnetic-flux fluctuations is also reasonable physically. Under Eq.~\eqref{b4l5} of the main text, Eqs.~\eqref{B_V_const} and~\eqref{B_V_inc} imply that 
% $\lim_{R\to\infty}\langle B_V^2\rangle/R^{5-n}$ decays relative to the dimensional scaling $\Bt^2 \lambda_B^{1+n}$; in other words, 
the expectation value of the squared magnetic flux contained within the volume $V$ decreases relative to its ``maximal'' value of $\Bt^2 \lambda_B^{1+n} R^{5-n}$. This makes sense: while there is a dynamical tendency for magnetic fields to favour locally maximally helical states~\cite{Taylor74, Servidio08} (meaning that we expect $I_H\sim \Bt^4 \lambda_B^5$), there is no physical reason that that they should maintain states of maximal magnetic flux (in the sense that ${\lim_{R\to\infty}\langle \Bt_V^2\rangle/R^{5-n}}\sim \Bt^2 \lambda_B^{1+n}$).

\begin{figure*}
    \centering
    \includegraphics[width=0.6\textwidth]{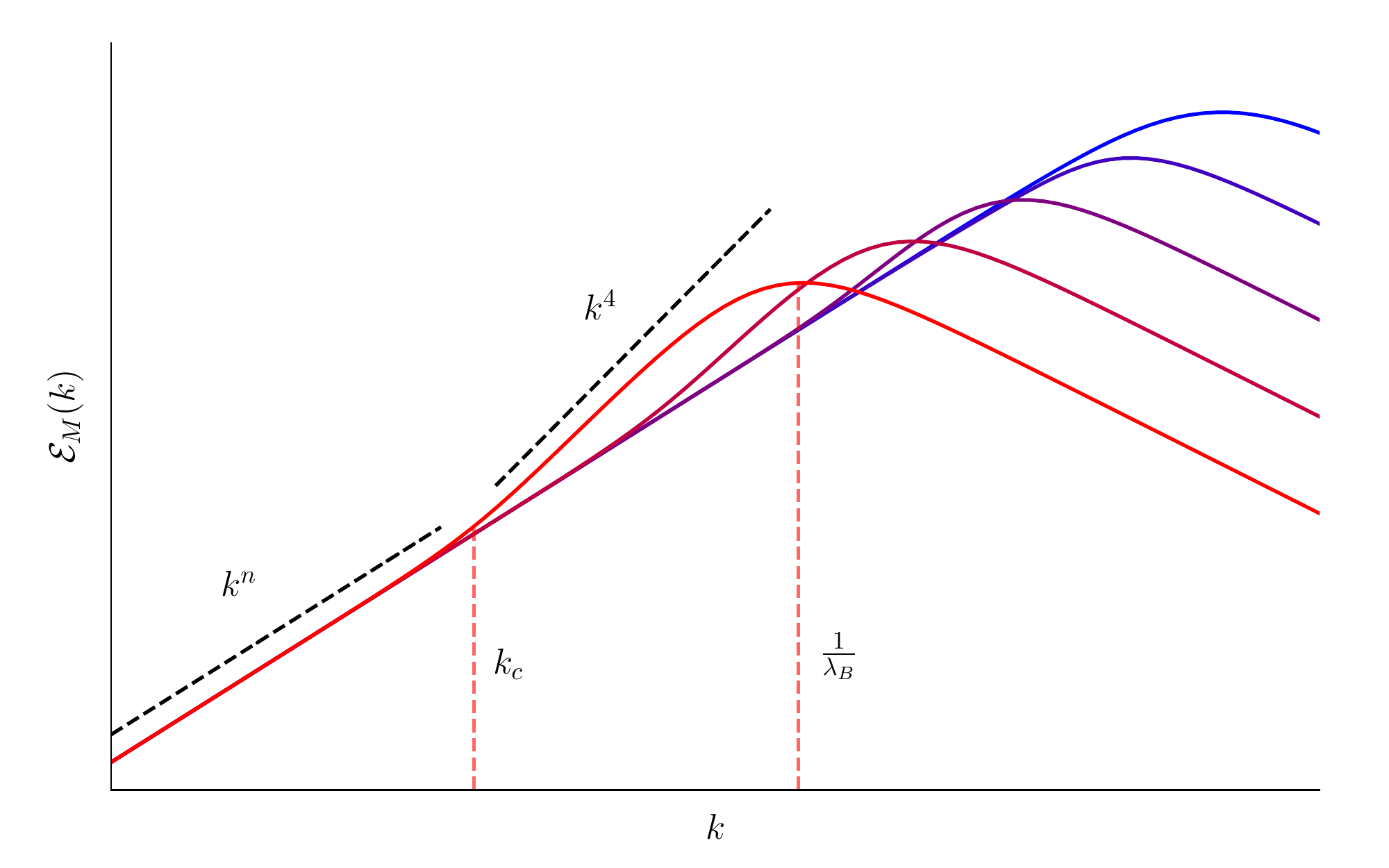}
    \caption{\textbf{Schematic of the evolution of $\mcE_M(t,k)$ for $\mcE_M(t=0,k\to0)\propto k^n$.}\\
    For $n\leq 3$, the $k\to 0$ asymptotic of $\mcE_M(t,k)$ is preserved as the turbulence decays; this is a consequence of magnetic-flux conservation. Nonetheless, the inverse-transfer effect persists, though it only occurs for $k>k_c(t)$, where $k_c$ is given by Eq.~\eqref{k_c}. The progression of time is from blue to red in this figure.}
    \label{fig:inverse_transfer}
\end{figure*}

A schematic of the evolution of the magnetic-energy spectrum decaying in a manner that satisfies both \eqref{B_V_const} and Eq.~\eqref{b4l5} of the main text is shown in Fig.~\ref{fig:inverse_transfer}. Under Eq.~\eqref{b4l5} of the main text, the spectral peak at $k\sim 1/\lambda_B$ grows relative to the position it would occupy under selective decay --- i.e., there is an inverse transfer --- nonetheless, the $k\to 0$ asymptotic is preserved. This leads to the development of a spectral knee at $k=k_c$, where $k_c^{-1}$ is the minimal scale for the applicability of $\langle \Bt_V^2 \rangle \propto R^{5-n} $. Because coalescence of structures via magnetic reconnection is a local process, we argue that it should not generate correlations on scales much larger than $\lambda_B$. This means that the spectrum between $k_c$ and $\lambda_B^{-1}$ should be proportional to $k^4$ [cf. Eq.~\eqref{EMexpansion} of the main text; see \cite{HoskingSchekochihin21k2} for discussion of the correspondence between long-range correlations and broken-power-law spectra].

The size of $k_c$ can be estimated by equating the invariant asymptotic with the growing $k^4$ component of the spectrum, which on dimensional grounds is of size $\sim \Bt^2 \lambda_B^5 k^4$:
\begin{equation}
    B_0^2 L_0 (k L_0)^n \sim \Bt^2 \lambda_B^5 k^4.  
\end{equation}From $\Bt^4 \lambda_B^5\sim \const$, we have
\begin{equation}
    k_c \sim \frac{1}{\lambda_B}\left[\frac{\lambda_B}{\lambda_B(0)}\right]^{-\frac{2n-3}{2(4-n)}} \sim \frac{1}{\lambda_B}\left[\frac{\Bt}{\Bt(0)}\right]^{\frac{2(2n-3)}{5(4-n)}}, \label{k_c}
\end{equation}a decreasing function of time.

The evolution of the magnetic-energy spectrum depicted in Fig.~\ref{fig:inverse_transfer} is manifestly non-self-similar. As a result, transient order-unity changes in $\Bt^4 \lambda_B^5$ and the decay timescale as a function of $\Bt$ and $\lambda_B$ should be expected at early times. On the other hand, the decay does become approximately self-similar at late times, when $k_c \ll 1/\lambda_B$, so any deviation from the theory proposed in the main text becomes small as $t$ becomes large. Finally, we acknowledge that, while we expect the evolution depicted in Fig.~\ref{fig:inverse_transfer} to be valid for any initial spectrum with $n>3/2$, Ref.~\cite{ReppinBanerjee17} do not observe the formation of a ``$k^4$ bulge'' in their simulation with $n=2$. We believe this to be a result of insufficient scale separation in that simulation: with $n=2$, Eq.~\eqref{k_c} implies $k_c \lambda_B\sim [\lambda_B/\lambda_B(0)]^{-1/4}$, so with $\lambda_B/\lambda_B(0)\simeq 10$ (see Fig.~16 of Ref.~\cite{ReppinBanerjee17}), $k_c \lambda_B\simeq 0.6$. It is therefore not surprising that these scales cannot be distinguished.

% \clearpage
\bibliography{decay_mod_prx}% Produces the bibliography via BibTeX.

\begin{thebibliography}{83}%
\makeatletter
\providecommand \@ifxundefined [1]{%
 \@ifx{#1\undefined}
}%
\providecommand \@ifnum [1]{%
 \ifnum #1\expandafter \@firstoftwo
 \else \expandafter \@secondoftwo
 \fi
}%
\providecommand \@ifx [1]{%
 \ifx #1\expandafter \@firstoftwo
 \else \expandafter \@secondoftwo
 \fi
}%
\providecommand \natexlab [1]{#1}%
\providecommand \enquote  [1]{``#1''}%
\providecommand \bibnamefont  [1]{#1}%
\providecommand \bibfnamefont [1]{#1}%
\providecommand \citenamefont [1]{#1}%
\providecommand \href@noop [0]{\@secondoftwo}%
\providecommand \href [0]{\begingroup \@sanitize@url \@href}%
\providecommand \@href[1]{\@@startlink{#1}\@@href}%
\providecommand \@@href[1]{\endgroup#1\@@endlink}%
\providecommand \@sanitize@url [0]{\catcode `\\12\catcode `\$12\catcode
  `\&12\catcode `\#12\catcode `\^12\catcode `\_12\catcode `\%12\relax}%
\providecommand \@@startlink[1]{}%
\providecommand \@@endlink[0]{}%
\providecommand \url  [0]{\begingroup\@sanitize@url \@url }%
\providecommand \@url [1]{\endgroup\@href {#1}{\urlprefix }}%
\providecommand \urlprefix  [0]{URL }%
\providecommand \Eprint [0]{\href }%
\providecommand \doibase [0]{https://doi.org/}%
\providecommand \selectlanguage [0]{\@gobble}%
\providecommand \bibinfo  [0]{\@secondoftwo}%
\providecommand \bibfield  [0]{\@secondoftwo}%
\providecommand \translation [1]{[#1]}%
\providecommand \BibitemOpen [0]{}%
\providecommand \bibitemStop [0]{}%
\providecommand \bibitemNoStop [0]{.\EOS\space}%
\providecommand \EOS [0]{\spacefactor3000\relax}%
\providecommand \BibitemShut  [1]{\csname bibitem#1\endcsname}%
\let\auto@bib@innerbib\@empty
%</preamble>
\bibitem [{\citenamefont {{Neronov}}\ and\ \citenamefont
  {{Vovk}}(2010)}]{NeronovVovk10}%
  \BibitemOpen
  \bibfield  {author} {\bibinfo {author} {\bibfnamefont {A.}~\bibnamefont
  {{Neronov}}}\ and\ \bibinfo {author} {\bibfnamefont {I.}~\bibnamefont
  {{Vovk}}},\ }\bibfield  {title} {\bibinfo {title} {{Evidence for strong
  extragalactic magnetic fields from Fermi observations of TeV blazars}},\
  }\href {https://doi.org/10.1126/science.1184192} {\bibfield  {journal}
  {\bibinfo  {journal} {Science}\ }\textbf {\bibinfo {volume} {328}},\ \bibinfo
  {pages} {73} (\bibinfo {year} {2010})}\BibitemShut {NoStop}%
\bibitem [{\citenamefont {{Tavecchio}}\ \emph {et~al.}(2010)\citenamefont
  {{Tavecchio}}, \citenamefont {{Ghisellini}}, \citenamefont {{Foschini}},
  \citenamefont {{Bonnoli}}, \citenamefont {{Ghirlanda}},\ and\ \citenamefont
  {{Coppi}}}]{Tavecchio10}%
  \BibitemOpen
  \bibfield  {author} {\bibinfo {author} {\bibfnamefont {F.}~\bibnamefont
  {{Tavecchio}}}, \bibinfo {author} {\bibfnamefont {G.}~\bibnamefont
  {{Ghisellini}}}, \bibinfo {author} {\bibfnamefont {L.}~\bibnamefont
  {{Foschini}}}, \bibinfo {author} {\bibfnamefont {G.}~\bibnamefont
  {{Bonnoli}}}, \bibinfo {author} {\bibfnamefont {G.}~\bibnamefont
  {{Ghirlanda}}},\ and\ \bibinfo {author} {\bibfnamefont {P.}~\bibnamefont
  {{Coppi}}},\ }\bibfield  {title} {\bibinfo {title} {{The intergalactic
  magnetic field constrained by Fermi/Large Area Telescope observations of the
  TeV blazar 1ES0229+200}},\ }\href
  {https://doi.org/10.1111/j.1745-3933.2010.00884.x} {\bibfield  {journal}
  {\bibinfo  {journal} {Mon. Not. R. Astron. Soc.}\ }\textbf {\bibinfo {volume}
  {406}},\ \bibinfo {pages} {L70} (\bibinfo {year} {2010})}\BibitemShut
  {NoStop}%
\bibitem [{\citenamefont {{Taylor}}\ \emph {et~al.}(2011)\citenamefont
  {{Taylor}}, \citenamefont {{Vovk}},\ and\ \citenamefont
  {{Neronov}}}]{Taylor11}%
  \BibitemOpen
  \bibfield  {author} {\bibinfo {author} {\bibfnamefont {A.~M.}\ \bibnamefont
  {{Taylor}}}, \bibinfo {author} {\bibfnamefont {I.}~\bibnamefont {{Vovk}}},\
  and\ \bibinfo {author} {\bibfnamefont {A.}~\bibnamefont {{Neronov}}},\
  }\bibfield  {title} {\bibinfo {title} {{Extragalactic magnetic fields
  constraints from simultaneous GeV-TeV observations of blazars}},\ }\href
  {https://doi.org/10.1051/0004-6361/201116441} {\bibfield  {journal} {\bibinfo
   {journal} {Astron. Astrophys.}\ }\textbf {\bibinfo {volume} {529}},\
  \bibinfo {eid} {A144} (\bibinfo {year} {2011})}\BibitemShut {NoStop}%
\bibitem [{\citenamefont {{Dermer}}\ \emph {et~al.}(2011)\citenamefont
  {{Dermer}}, \citenamefont {{Cavadini}}, \citenamefont {{Razzaque}},
  \citenamefont {{Finke}}, \citenamefont {{Chiang}},\ and\ \citenamefont
  {{Lott}}}]{Dermer11}%
  \BibitemOpen
  \bibfield  {author} {\bibinfo {author} {\bibfnamefont {C.~D.}\ \bibnamefont
  {{Dermer}}}, \bibinfo {author} {\bibfnamefont {M.}~\bibnamefont
  {{Cavadini}}}, \bibinfo {author} {\bibfnamefont {S.}~\bibnamefont
  {{Razzaque}}}, \bibinfo {author} {\bibfnamefont {J.~D.}\ \bibnamefont
  {{Finke}}}, \bibinfo {author} {\bibfnamefont {J.}~\bibnamefont {{Chiang}}},\
  and\ \bibinfo {author} {\bibfnamefont {B.}~\bibnamefont {{Lott}}},\
  }\bibfield  {title} {\bibinfo {title} {{Time delay of cascade radiation for
  TeV blazars and the measurement of the intergalactic magnetic field}},\
  }\href {https://doi.org/10.1088/2041-8205/733/2/L21} {\bibfield  {journal}
  {\bibinfo  {journal} {Astrophys. J. Lett.}\ }\textbf {\bibinfo {volume}
  {733}},\ \bibinfo {pages} {L21} (\bibinfo {year} {2011})}\BibitemShut
  {NoStop}%
\bibitem [{\citenamefont {{Dolag}}\ \emph {et~al.}(2011)\citenamefont
  {{Dolag}}, \citenamefont {{Kachelriess}}, \citenamefont {{Ostapchenko}},\
  and\ \citenamefont {{Tom{\`a}s}}}]{Dolag11}%
  \BibitemOpen
  \bibfield  {author} {\bibinfo {author} {\bibfnamefont {K.}~\bibnamefont
  {{Dolag}}}, \bibinfo {author} {\bibfnamefont {M.}~\bibnamefont
  {{Kachelriess}}}, \bibinfo {author} {\bibfnamefont {S.}~\bibnamefont
  {{Ostapchenko}}},\ and\ \bibinfo {author} {\bibfnamefont {R.}~\bibnamefont
  {{Tom{\`a}s}}},\ }\bibfield  {title} {\bibinfo {title} {{Lower limit on the
  strength and filling factor of extragalactic magnetic fields}},\ }\href
  {https://doi.org/10.1088/2041-8205/727/1/L4} {\bibfield  {journal} {\bibinfo
  {journal} {Astrophys. J. Lett.}\ }\textbf {\bibinfo {volume} {727}},\
  \bibinfo {eid} {L4} (\bibinfo {year} {2011})}\BibitemShut {NoStop}%
\bibitem [{\citenamefont {{Essey}}\ \emph {et~al.}(2011)\citenamefont
  {{Essey}}, \citenamefont {{Ando}},\ and\ \citenamefont
  {{Kusenko}}}]{Essey11}%
  \BibitemOpen
  \bibfield  {author} {\bibinfo {author} {\bibfnamefont {W.}~\bibnamefont
  {{Essey}}}, \bibinfo {author} {\bibfnamefont {S.}~\bibnamefont {{Ando}}},\
  and\ \bibinfo {author} {\bibfnamefont {A.}~\bibnamefont {{Kusenko}}},\
  }\bibfield  {title} {\bibinfo {title} {{Determination of intergalactic
  magnetic fields from gamma ray data}},\ }\href
  {https://doi.org/10.1016/j.astropartphys.2011.06.010} {\bibfield  {journal}
  {\bibinfo  {journal} {Astroparticle Physics}\ }\textbf {\bibinfo {volume}
  {35}},\ \bibinfo {pages} {135} (\bibinfo {year} {2011})}\BibitemShut
  {NoStop}%
\bibitem [{\citenamefont {{Huan}}\ \emph {et~al.}(2011)\citenamefont {{Huan}},
  \citenamefont {{Weisgarber}}, \citenamefont {{Arlen}},\ and\ \citenamefont
  {{Wakely}}}]{Huan11}%
  \BibitemOpen
  \bibfield  {author} {\bibinfo {author} {\bibfnamefont {H.}~\bibnamefont
  {{Huan}}}, \bibinfo {author} {\bibfnamefont {T.}~\bibnamefont
  {{Weisgarber}}}, \bibinfo {author} {\bibfnamefont {T.}~\bibnamefont
  {{Arlen}}},\ and\ \bibinfo {author} {\bibfnamefont {S.~P.}\ \bibnamefont
  {{Wakely}}},\ }\bibfield  {title} {\bibinfo {title} {{A new model for
  gamma-ray cascades in extragalactic magnetic fields}},\ }\href
  {https://doi.org/10.1088/2041-8205/735/2/L28} {\bibfield  {journal} {\bibinfo
   {journal} {Astrophys. J. Lett.}\ }\textbf {\bibinfo {volume} {735}},\
  \bibinfo {eid} {L28} (\bibinfo {year} {2011})}\BibitemShut {NoStop}%
\bibitem [{\citenamefont {{Tavecchio}}\ \emph {et~al.}(2011)\citenamefont
  {{Tavecchio}}, \citenamefont {{Ghisellini}}, \citenamefont {{Bonnoli}},\ and\
  \citenamefont {{Foschini}}}]{Tavecchio11}%
  \BibitemOpen
  \bibfield  {author} {\bibinfo {author} {\bibfnamefont {F.}~\bibnamefont
  {{Tavecchio}}}, \bibinfo {author} {\bibfnamefont {G.}~\bibnamefont
  {{Ghisellini}}}, \bibinfo {author} {\bibfnamefont {G.}~\bibnamefont
  {{Bonnoli}}},\ and\ \bibinfo {author} {\bibfnamefont {L.}~\bibnamefont
  {{Foschini}}},\ }\bibfield  {title} {\bibinfo {title} {{Extreme TeV blazars
  and the intergalactic magnetic field}},\ }\href
  {https://doi.org/10.1111/j.1365-2966.2011.18657.x} {\bibfield  {journal}
  {\bibinfo  {journal} {Mon. Not. R. Astron. Soc.}\ }\textbf {\bibinfo {volume}
  {414}},\ \bibinfo {pages} {3566} (\bibinfo {year} {2011})}\BibitemShut
  {NoStop}%
\bibitem [{\citenamefont {{Takahashi}}\ \emph {et~al.}(2012)\citenamefont
  {{Takahashi}}, \citenamefont {{Mori}}, \citenamefont {{Ichiki}},\ and\
  \citenamefont {{Inoue}}}]{Takahashi12}%
  \BibitemOpen
  \bibfield  {author} {\bibinfo {author} {\bibfnamefont {K.}~\bibnamefont
  {{Takahashi}}}, \bibinfo {author} {\bibfnamefont {M.}~\bibnamefont {{Mori}}},
  \bibinfo {author} {\bibfnamefont {K.}~\bibnamefont {{Ichiki}}},\ and\
  \bibinfo {author} {\bibfnamefont {S.}~\bibnamefont {{Inoue}}},\ }\bibfield
  {title} {\bibinfo {title} {{Lower bounds on intergalactic magnetic fields
  from simultaneously observed GeV-TeV light curves of the blazar Mrk 501}},\
  }\href {https://doi.org/10.1088/2041-8205/744/1/L7} {\bibfield  {journal}
  {\bibinfo  {journal} {Astrophys. J. Lett.}\ }\textbf {\bibinfo {volume}
  {744}},\ \bibinfo {eid} {L7} (\bibinfo {year} {2012})}\BibitemShut {NoStop}%
\bibitem [{\citenamefont {{Arlen}}\ \emph {et~al.}(2014)\citenamefont
  {{Arlen}}, \citenamefont {{Vassilev}}, \citenamefont {{Weisgarber}},
  \citenamefont {{Wakely}},\ and\ \citenamefont {{Yusef Shafi}}}]{Arlen14}%
  \BibitemOpen
  \bibfield  {author} {\bibinfo {author} {\bibfnamefont {T.~C.}\ \bibnamefont
  {{Arlen}}}, \bibinfo {author} {\bibfnamefont {V.~V.}\ \bibnamefont
  {{Vassilev}}}, \bibinfo {author} {\bibfnamefont {T.}~\bibnamefont
  {{Weisgarber}}}, \bibinfo {author} {\bibfnamefont {S.~P.}\ \bibnamefont
  {{Wakely}}},\ and\ \bibinfo {author} {\bibfnamefont {S.}~\bibnamefont {{Yusef
  Shafi}}},\ }\bibfield  {title} {\bibinfo {title} {{Intergalactic magnetic
  fields and gamma-ray observations of extreme TeV blazars}},\ }\href
  {https://doi.org/10.1088/0004-637X/796/1/18} {\bibfield  {journal} {\bibinfo
  {journal} {Astrophys. J.}\ }\textbf {\bibinfo {volume} {796}},\ \bibinfo
  {eid} {18} (\bibinfo {year} {2014})}\BibitemShut {NoStop}%
\bibitem [{\citenamefont {{Finke}}\ \emph {et~al.}(2015)\citenamefont
  {{Finke}}, \citenamefont {{Reyes}}, \citenamefont {{Georganopoulos}},
  \citenamefont {{Reynolds}}, \citenamefont {{Ajello}}, \citenamefont
  {{Fegan}},\ and\ \citenamefont {{McCann}}}]{Finke15}%
  \BibitemOpen
  \bibfield  {author} {\bibinfo {author} {\bibfnamefont {J.~D.}\ \bibnamefont
  {{Finke}}}, \bibinfo {author} {\bibfnamefont {L.~C.}\ \bibnamefont
  {{Reyes}}}, \bibinfo {author} {\bibfnamefont {M.}~\bibnamefont
  {{Georganopoulos}}}, \bibinfo {author} {\bibfnamefont {K.}~\bibnamefont
  {{Reynolds}}}, \bibinfo {author} {\bibfnamefont {M.}~\bibnamefont
  {{Ajello}}}, \bibinfo {author} {\bibfnamefont {S.~J.}\ \bibnamefont
  {{Fegan}}},\ and\ \bibinfo {author} {\bibfnamefont {K.}~\bibnamefont
  {{McCann}}},\ }\bibfield  {title} {\bibinfo {title} {{Constraints on the
  intergalactic magnetic field with gamma-ray observations of blazars}},\
  }\href {https://doi.org/10.1088/0004-637X/814/1/20} {\bibfield  {journal}
  {\bibinfo  {journal} {Astrophys. J.}\ }\textbf {\bibinfo {volume} {814}},\
  \bibinfo {eid} {20} (\bibinfo {year} {2015})}\BibitemShut {NoStop}%
\bibitem [{\citenamefont {{Archambault}}\ \emph {et~al.}(2017)\citenamefont
  {{Archambault}}, \citenamefont {{Archer}}, \citenamefont {{Benbow}},
  \citenamefont {{Buchovecky}}, \citenamefont {{Bugaev}}, \citenamefont
  {{Cerruti}}, \citenamefont {{Connolly}}, \citenamefont {{Cui}}, \citenamefont
  {{Falcone}}, \citenamefont {{Fern{\'a}ndez Alonso}} \emph
  {et~al.}}]{Archambault17}%
  \BibitemOpen
  \bibfield  {author} {\bibinfo {author} {\bibfnamefont {S.}~\bibnamefont
  {{Archambault}}}, \bibinfo {author} {\bibfnamefont {A.}~\bibnamefont
  {{Archer}}}, \bibinfo {author} {\bibfnamefont {W.}~\bibnamefont {{Benbow}}},
  \bibinfo {author} {\bibfnamefont {M.}~\bibnamefont {{Buchovecky}}}, \bibinfo
  {author} {\bibfnamefont {V.}~\bibnamefont {{Bugaev}}}, \bibinfo {author}
  {\bibfnamefont {M.}~\bibnamefont {{Cerruti}}}, \bibinfo {author}
  {\bibfnamefont {M.~P.}\ \bibnamefont {{Connolly}}}, \bibinfo {author}
  {\bibfnamefont {W.}~\bibnamefont {{Cui}}}, \bibinfo {author} {\bibfnamefont
  {A.}~\bibnamefont {{Falcone}}}, \bibinfo {author} {\bibfnamefont
  {M.}~\bibnamefont {{Fern{\'a}ndez Alonso}}}, \emph {et~al.},\ }\bibfield
  {title} {\bibinfo {title} {{Search for magnetically broadened cascade
  emission from blazars with VERITAS}},\ }\href
  {https://doi.org/10.3847/1538-4357/835/2/288} {\bibfield  {journal} {\bibinfo
   {journal} {Astrophys. J.}\ }\textbf {\bibinfo {volume} {835}},\ \bibinfo
  {eid} {288} (\bibinfo {year} {2017})}\BibitemShut {NoStop}%
\bibitem [{\citenamefont {{Durrer}}\ and\ \citenamefont
  {{Neronov}}(2013)}]{DurrerNeronov13}%
  \BibitemOpen
  \bibfield  {author} {\bibinfo {author} {\bibfnamefont {R.}~\bibnamefont
  {{Durrer}}}\ and\ \bibinfo {author} {\bibfnamefont {A.}~\bibnamefont
  {{Neronov}}},\ }\bibfield  {title} {\bibinfo {title} {{Cosmological magnetic
  fields: their generation, evolution and observation}},\ }\href
  {https://doi.org/10.1007/s00159-013-0062-7} {\bibfield  {journal} {\bibinfo
  {journal} {Astron. Astrophys. Rev.}\ }\textbf {\bibinfo {volume} {21}},\
  \bibinfo {eid} {62} (\bibinfo {year} {2013})}\BibitemShut {NoStop}%
\bibitem [{\citenamefont {{Subramanian}}(2016)}]{Subramanian16}%
  \BibitemOpen
  \bibfield  {author} {\bibinfo {author} {\bibfnamefont {K.}~\bibnamefont
  {{Subramanian}}},\ }\bibfield  {title} {\bibinfo {title} {{The origin,
  evolution and signatures of primordial magnetic fields}},\ }\href
  {https://doi.org/10.1088/0034-4885/79/7/076901} {\bibfield  {journal}
  {\bibinfo  {journal} {Rep. Prog. Phys.}\ }\textbf {\bibinfo {volume} {79}},\
  \bibinfo {eid} {076901} (\bibinfo {year} {2016})}\BibitemShut {NoStop}%
\bibitem [{\citenamefont {{Vachaspati}}(2021)}]{Vachaspati21}%
  \BibitemOpen
  \bibfield  {author} {\bibinfo {author} {\bibfnamefont {T.}~\bibnamefont
  {{Vachaspati}}},\ }\bibfield  {title} {\bibinfo {title} {{Progress on
  cosmological magnetic fields}},\ }\href
  {https://doi.org/10.1088/1361-6633/ac03a9} {\bibfield  {journal} {\bibinfo
  {journal} {Rep. Prog. Phys.}\ }\textbf {\bibinfo {volume} {84}},\ \bibinfo
  {eid} {074901} (\bibinfo {year} {2021})}\BibitemShut {NoStop}%
\bibitem [{\citenamefont {{Ackermann}}\ \emph {et~al.}(2018)\citenamefont
  {{Ackermann}}, \citenamefont {{Ajello}}, \citenamefont {{Baldini}},
  \citenamefont {{Ballet}}, \citenamefont {{Barbiellini}}, \citenamefont
  {{Bastieri}}, \citenamefont {{Bellazzini}}, \citenamefont {{Bissaldi}},
  \citenamefont {{Blandford}} \emph {et~al.}}]{Ackermann18}%
  \BibitemOpen
  \bibfield  {author} {\bibinfo {author} {\bibfnamefont {M.}~\bibnamefont
  {{Ackermann}}}, \bibinfo {author} {\bibfnamefont {M.}~\bibnamefont
  {{Ajello}}}, \bibinfo {author} {\bibfnamefont {L.}~\bibnamefont {{Baldini}}},
  \bibinfo {author} {\bibfnamefont {J.}~\bibnamefont {{Ballet}}}, \bibinfo
  {author} {\bibfnamefont {G.}~\bibnamefont {{Barbiellini}}}, \bibinfo {author}
  {\bibfnamefont {D.}~\bibnamefont {{Bastieri}}}, \bibinfo {author}
  {\bibfnamefont {R.}~\bibnamefont {{Bellazzini}}}, \bibinfo {author}
  {\bibfnamefont {E.}~\bibnamefont {{Bissaldi}}}, \bibinfo {author}
  {\bibfnamefont {R.~D.}\ \bibnamefont {{Blandford}}}, \emph {et~al.},\
  }\bibfield  {title} {\bibinfo {title} {{The search for spatial extension in
  high-latitude sources detected by the Fermi Large Area Telescope}},\ }\href
  {https://doi.org/10.3847/1538-4365/aacdf7} {\bibfield  {journal} {\bibinfo
  {journal} {Astrophys. J. Suppl.}\ }\textbf {\bibinfo {volume} {237}},\
  \bibinfo {eid} {32} (\bibinfo {year} {2018})}\BibitemShut {NoStop}%
\bibitem [{\citenamefont {{Broderick}}\ \emph {et~al.}(2012)\citenamefont
  {{Broderick}}, \citenamefont {{Chang}},\ and\ \citenamefont
  {{Pfrommer}}}]{Broderick12}%
  \BibitemOpen
  \bibfield  {author} {\bibinfo {author} {\bibfnamefont {A.~E.}\ \bibnamefont
  {{Broderick}}}, \bibinfo {author} {\bibfnamefont {P.}~\bibnamefont
  {{Chang}}},\ and\ \bibinfo {author} {\bibfnamefont {C.}~\bibnamefont
  {{Pfrommer}}},\ }\bibfield  {title} {\bibinfo {title} {{The cosmological
  impact of luminous TeV blazars. I. Implications of plasma instabilities for
  the intergalactic magnetic field and extragalactic gamma-ray background}},\
  }\href {https://doi.org/10.1088/0004-637X/752/1/22} {\bibfield  {journal}
  {\bibinfo  {journal} {Astrophys. J.}\ }\textbf {\bibinfo {volume} {752}},\
  \bibinfo {eid} {22} (\bibinfo {year} {2012})}\BibitemShut {NoStop}%
\bibitem [{\citenamefont {{Broderick}}\ \emph {et~al.}(2018)\citenamefont
  {{Broderick}}, \citenamefont {{Tiede}}, \citenamefont {{Chang}},
  \citenamefont {{Lamberts}}, \citenamefont {{Pfrommer}}, \citenamefont
  {{Puchwein}}, \citenamefont {{Shalaby}},\ and\ \citenamefont
  {{Werhahn}}}]{Broderick18}%
  \BibitemOpen
  \bibfield  {author} {\bibinfo {author} {\bibfnamefont {A.~E.}\ \bibnamefont
  {{Broderick}}}, \bibinfo {author} {\bibfnamefont {P.}~\bibnamefont
  {{Tiede}}}, \bibinfo {author} {\bibfnamefont {P.}~\bibnamefont {{Chang}}},
  \bibinfo {author} {\bibfnamefont {A.}~\bibnamefont {{Lamberts}}}, \bibinfo
  {author} {\bibfnamefont {C.}~\bibnamefont {{Pfrommer}}}, \bibinfo {author}
  {\bibfnamefont {E.}~\bibnamefont {{Puchwein}}}, \bibinfo {author}
  {\bibfnamefont {M.}~\bibnamefont {{Shalaby}}},\ and\ \bibinfo {author}
  {\bibfnamefont {M.}~\bibnamefont {{Werhahn}}},\ }\bibfield  {title} {\bibinfo
  {title} {{Missing gamma-ray halos and the need for new physics in the
  gamma-ray sky}},\ }\href {https://doi.org/10.3847/1538-4357/aae5f2}
  {\bibfield  {journal} {\bibinfo  {journal} {Astrophys. J.}\ }\textbf
  {\bibinfo {volume} {868}},\ \bibinfo {eid} {87} (\bibinfo {year}
  {2018})}\BibitemShut {NoStop}%
\bibitem [{\citenamefont {{Alves Batista}}\ \emph {et~al.}(2019)\citenamefont
  {{Alves Batista}}, \citenamefont {{Saveliev}},\ and\ \citenamefont {{de
  Gouveia Dal Pino}}}]{Batista19}%
  \BibitemOpen
  \bibfield  {author} {\bibinfo {author} {\bibfnamefont {R.}~\bibnamefont
  {{Alves Batista}}}, \bibinfo {author} {\bibfnamefont {A.}~\bibnamefont
  {{Saveliev}}},\ and\ \bibinfo {author} {\bibfnamefont {E.~M.}\ \bibnamefont
  {{de Gouveia Dal Pino}}},\ }\bibfield  {title} {\bibinfo {title} {{The impact
  of plasma instabilities on the spectra of TeV blazars}},\ }\href
  {https://doi.org/10.1093/mnras/stz2389} {\bibfield  {journal} {\bibinfo
  {journal} {Mon. Not. R. Astron. Soc.}\ }\textbf {\bibinfo {volume} {489}},\
  \bibinfo {pages} {3836} (\bibinfo {year} {2019})}\BibitemShut {NoStop}%
\bibitem [{\citenamefont {{Perry}}\ and\ \citenamefont
  {{Lyubarsky}}(2021)}]{PerryLyubarsky21}%
  \BibitemOpen
  \bibfield  {author} {\bibinfo {author} {\bibfnamefont {R.}~\bibnamefont
  {{Perry}}}\ and\ \bibinfo {author} {\bibfnamefont {Y.}~\bibnamefont
  {{Lyubarsky}}},\ }\bibfield  {title} {\bibinfo {title} {{The role of resonant
  plasma instabilities in the evolution of blazar-induced pair beams}},\ }\href
  {https://doi.org/10.1093/mnras/stab324} {\bibfield  {journal} {\bibinfo
  {journal} {Mon. Not. R. Astron. Soc.}\ }\textbf {\bibinfo {volume} {503}},\
  \bibinfo {pages} {2215} (\bibinfo {year} {2021})}\BibitemShut {NoStop}%
\bibitem [{\citenamefont {{Addazi}}\ \emph {et~al.}(2022)\citenamefont
  {{Addazi}}, \citenamefont {{Alvarez-Muniz}}, \citenamefont {{Alves Batista}},
  \citenamefont {{Amelino-Camelia}}, \citenamefont {{Antonelli}}, \citenamefont
  {{Arzano}}, \citenamefont {{Asorey}}, \citenamefont {{Atteia}}, \citenamefont
  {{Bahamonde}}, \citenamefont {{Bajardi}} \emph {et~al.}}]{Addazi22}%
  \BibitemOpen
  \bibfield  {author} {\bibinfo {author} {\bibfnamefont {A.}~\bibnamefont
  {{Addazi}}}, \bibinfo {author} {\bibfnamefont {J.}~\bibnamefont
  {{Alvarez-Muniz}}}, \bibinfo {author} {\bibfnamefont {R.}~\bibnamefont
  {{Alves Batista}}}, \bibinfo {author} {\bibfnamefont {G.}~\bibnamefont
  {{Amelino-Camelia}}}, \bibinfo {author} {\bibfnamefont {V.}~\bibnamefont
  {{Antonelli}}}, \bibinfo {author} {\bibfnamefont {M.}~\bibnamefont
  {{Arzano}}}, \bibinfo {author} {\bibfnamefont {M.}~\bibnamefont {{Asorey}}},
  \bibinfo {author} {\bibfnamefont {J.~L.}\ \bibnamefont {{Atteia}}}, \bibinfo
  {author} {\bibfnamefont {S.}~\bibnamefont {{Bahamonde}}}, \bibinfo {author}
  {\bibfnamefont {F.}~\bibnamefont {{Bajardi}}}, \emph {et~al.},\ }\bibfield
  {title} {\bibinfo {title} {{Quantum gravity phenomenology at the dawn of the
  multi-messenger era --- A review}},\ }\href
  {https://doi.org/10.1016/j.ppnp.2022.103948} {\bibfield  {journal} {\bibinfo
  {journal} {Progress in Particle and Nuclear Physics}\ }\textbf {\bibinfo
  {volume} {125}},\ \bibinfo {eid} {103948} (\bibinfo {year}
  {2022})}\BibitemShut {NoStop}%
\bibitem [{\citenamefont {{Alves Batista}}\ and\ \citenamefont
  {{Saveliev}}(2021)}]{BatistaSaveliev21}%
  \BibitemOpen
  \bibfield  {author} {\bibinfo {author} {\bibfnamefont {R.}~\bibnamefont
  {{Alves Batista}}}\ and\ \bibinfo {author} {\bibfnamefont {A.}~\bibnamefont
  {{Saveliev}}},\ }\bibfield  {title} {\bibinfo {title} {{The gamma-ray window
  to intergalactic magnetism}},\ }\href
  {https://doi.org/10.3390/universe7070223} {\bibfield  {journal} {\bibinfo
  {journal} {Universe}\ }\textbf {\bibinfo {volume} {7}},\ \bibinfo {pages}
  {223} (\bibinfo {year} {2021})}\BibitemShut {NoStop}%
\bibitem [{\citenamefont {{Beck}}\ \emph {et~al.}(2013)\citenamefont {{Beck}},
  \citenamefont {{Hanasz}}, \citenamefont {{Lesch}}, \citenamefont {{Remus}},\
  and\ \citenamefont {{Stasyszyn}}}]{Beck13}%
  \BibitemOpen
  \bibfield  {author} {\bibinfo {author} {\bibfnamefont {A.~M.}\ \bibnamefont
  {{Beck}}}, \bibinfo {author} {\bibfnamefont {M.}~\bibnamefont {{Hanasz}}},
  \bibinfo {author} {\bibfnamefont {H.}~\bibnamefont {{Lesch}}}, \bibinfo
  {author} {\bibfnamefont {R.~S.}\ \bibnamefont {{Remus}}},\ and\ \bibinfo
  {author} {\bibfnamefont {F.~A.}\ \bibnamefont {{Stasyszyn}}},\ }\bibfield
  {title} {\bibinfo {title} {{On the magnetic fields in voids.}},\ }\href
  {https://doi.org/10.1093/mnrasl/sls026} {\bibfield  {journal} {\bibinfo
  {journal} {Mon. Not. R. Astron. Soc.}\ }\textbf {\bibinfo {volume} {429}},\
  \bibinfo {pages} {L60} (\bibinfo {year} {2013})}\BibitemShut {NoStop}%
\bibitem [{\citenamefont {{Banerjee}}\ and\ \citenamefont
  {{Jedamzik}}(2004)}]{BanerjeeJedamzik04}%
  \BibitemOpen
  \bibfield  {author} {\bibinfo {author} {\bibfnamefont {R.}~\bibnamefont
  {{Banerjee}}}\ and\ \bibinfo {author} {\bibfnamefont {K.}~\bibnamefont
  {{Jedamzik}}},\ }\bibfield  {title} {\bibinfo {title} {{Evolution of cosmic
  magnetic fields: From the very early Universe, to recombination, to the
  present}},\ }\href {https://doi.org/10.1103/PhysRevD.70.123003} {\bibfield
  {journal} {\bibinfo  {journal} {Phys. Rev. D}\ }\textbf {\bibinfo {volume}
  {70}},\ \bibinfo {eid} {123003} (\bibinfo {year} {2004})}\BibitemShut
  {NoStop}%
\bibitem [{\citenamefont {{Vachaspati}}(1991)}]{Vachaspati91}%
  \BibitemOpen
  \bibfield  {author} {\bibinfo {author} {\bibfnamefont {T.}~\bibnamefont
  {{Vachaspati}}},\ }\bibfield  {title} {\bibinfo {title} {{Magnetic fields
  from cosmological phase transitions}},\ }\href
  {https://doi.org/10.1016/0370-2693(91)90051-Q} {\bibfield  {journal}
  {\bibinfo  {journal} {Phys. Lett. B}\ }\textbf {\bibinfo {volume} {265}},\
  \bibinfo {pages} {258} (\bibinfo {year} {1991})}\BibitemShut {NoStop}%
\bibitem [{\citenamefont {{Wagstaff}}\ and\ \citenamefont
  {{Banerjee}}(2016)}]{WagstaffBanerjee16}%
  \BibitemOpen
  \bibfield  {author} {\bibinfo {author} {\bibfnamefont {J.~M.}\ \bibnamefont
  {{Wagstaff}}}\ and\ \bibinfo {author} {\bibfnamefont {R.}~\bibnamefont
  {{Banerjee}}},\ }\bibfield  {title} {\bibinfo {title} {{Extragalactic
  magnetic fields unlikely generated at the electroweak phase transition}},\
  }\href {https://doi.org/10.1088/1475-7516/2016/01/002} {\bibfield  {journal}
  {\bibinfo  {journal} {J. Cosmol. Astropart. Phys.}\ }\textbf {\bibinfo
  {volume} {2016}},\ \bibinfo {eid} {002}}\BibitemShut {NoStop}%
\bibitem [{\citenamefont {{Taylor}}(1986)}]{Taylor86}%
  \BibitemOpen
  \bibfield  {author} {\bibinfo {author} {\bibfnamefont {J.~B.}\ \bibnamefont
  {{Taylor}}},\ }\bibfield  {title} {\bibinfo {title} {{Relaxation and magnetic
  reconnection in plasmas}},\ }\href
  {https://doi.org/10.1103/RevModPhys.58.741} {\bibfield  {journal} {\bibinfo
  {journal} {Rev. Mod. Phys.}\ }\textbf {\bibinfo {volume} {58}},\ \bibinfo
  {pages} {741} (\bibinfo {year} {1986})}\BibitemShut {NoStop}%
\bibitem [{\citenamefont {{Vachaspati}}(2001)}]{Vachaspati01}%
  \BibitemOpen
  \bibfield  {author} {\bibinfo {author} {\bibfnamefont {T.}~\bibnamefont
  {{Vachaspati}}},\ }\bibfield  {title} {\bibinfo {title} {{Estimate of the
  primordial magnetic field helicity}},\ }\href
  {https://doi.org/10.1103/PhysRevLett.87.251302} {\bibfield  {journal}
  {\bibinfo  {journal} {Phys. Rev. Lett.}\ }\textbf {\bibinfo {volume} {87}},\
  \bibinfo {eid} {251302} (\bibinfo {year} {2001})}\BibitemShut {NoStop}%
\bibitem [{\citenamefont {{Boyarsky}}\ \emph {et~al.}(2021)\citenamefont
  {{Boyarsky}}, \citenamefont {{Cheianov}}, \citenamefont {{Ruchayskiy}},\ and\
  \citenamefont {{Sobol}}}]{Boyarksy21}%
  \BibitemOpen
  \bibfield  {author} {\bibinfo {author} {\bibfnamefont {A.}~\bibnamefont
  {{Boyarsky}}}, \bibinfo {author} {\bibfnamefont {V.}~\bibnamefont
  {{Cheianov}}}, \bibinfo {author} {\bibfnamefont {O.}~\bibnamefont
  {{Ruchayskiy}}},\ and\ \bibinfo {author} {\bibfnamefont {O.}~\bibnamefont
  {{Sobol}}},\ }\bibfield  {title} {\bibinfo {title} {{Equilibration of the
  chiral asymmetry due to finite electron mass in electron-positron plasma}},\
  }\href {https://doi.org/10.1103/PhysRevD.103.013003} {\bibfield  {journal}
  {\bibinfo  {journal} {Phys. Rev. D}\ }\textbf {\bibinfo {volume} {103}},\
  \bibinfo {eid} {013003} (\bibinfo {year} {2021})}\BibitemShut {NoStop}%
\bibitem [{\citenamefont {{Brandenburg}}\ \emph
  {et~al.}(2017{\natexlab{a}})\citenamefont {{Brandenburg}}, \citenamefont
  {{Kahniashvili}}, \citenamefont {{Mandal}}, \citenamefont {{Pol}},
  \citenamefont {{Tevzadze}},\ and\ \citenamefont
  {{Vachaspati}}}]{Brandenburg17_cosmic}%
  \BibitemOpen
  \bibfield  {author} {\bibinfo {author} {\bibfnamefont {A.}~\bibnamefont
  {{Brandenburg}}}, \bibinfo {author} {\bibfnamefont {T.}~\bibnamefont
  {{Kahniashvili}}}, \bibinfo {author} {\bibfnamefont {S.}~\bibnamefont
  {{Mandal}}}, \bibinfo {author} {\bibfnamefont {A.~R.}\ \bibnamefont {{Pol}}},
  \bibinfo {author} {\bibfnamefont {A.~G.}\ \bibnamefont {{Tevzadze}}},\ and\
  \bibinfo {author} {\bibfnamefont {T.}~\bibnamefont {{Vachaspati}}},\
  }\bibfield  {title} {\bibinfo {title} {{Evolution of hydromagnetic turbulence
  from the electroweak phase transition}},\ }\href
  {https://doi.org/10.1103/PhysRevD.96.123528} {\bibfield  {journal} {\bibinfo
  {journal} {Phys. Rev. D}\ }\textbf {\bibinfo {volume} {96}},\ \bibinfo {eid}
  {123528} (\bibinfo {year} {2017}{\natexlab{a}})}\BibitemShut {NoStop}%
\bibitem [{\citenamefont {{Brandenburg}}\ \emph
  {et~al.}(2017{\natexlab{b}})\citenamefont {{Brandenburg}}, \citenamefont
  {{Schober}}, \citenamefont {{Rogachevskii}}, \citenamefont {{Kahniashvili}},
  \citenamefont {{Boyarsky}}, \citenamefont {{Fr{\"o}hlich}}, \citenamefont
  {{Ruchayskiy}},\ and\ \citenamefont {{Kleeorin}}}]{Brandenburg17_chiral}%
  \BibitemOpen
  \bibfield  {author} {\bibinfo {author} {\bibfnamefont {A.}~\bibnamefont
  {{Brandenburg}}}, \bibinfo {author} {\bibfnamefont {J.}~\bibnamefont
  {{Schober}}}, \bibinfo {author} {\bibfnamefont {I.}~\bibnamefont
  {{Rogachevskii}}}, \bibinfo {author} {\bibfnamefont {T.}~\bibnamefont
  {{Kahniashvili}}}, \bibinfo {author} {\bibfnamefont {A.}~\bibnamefont
  {{Boyarsky}}}, \bibinfo {author} {\bibfnamefont {J.}~\bibnamefont
  {{Fr{\"o}hlich}}}, \bibinfo {author} {\bibfnamefont {O.}~\bibnamefont
  {{Ruchayskiy}}},\ and\ \bibinfo {author} {\bibfnamefont {N.}~\bibnamefont
  {{Kleeorin}}},\ }\bibfield  {title} {\bibinfo {title} {{The turbulent chiral
  magnetic cascade in the early Universe}},\ }\href
  {https://doi.org/10.3847/2041-8213/aa855d} {\bibfield  {journal} {\bibinfo
  {journal} {Astrophys. J. Lett.}\ }\textbf {\bibinfo {volume} {845}},\
  \bibinfo {eid} {L21} (\bibinfo {year} {2017}{\natexlab{b}})}\BibitemShut
  {NoStop}%
\bibitem [{\citenamefont {{Zrake}}(2014)}]{Zrake14}%
  \BibitemOpen
  \bibfield  {author} {\bibinfo {author} {\bibfnamefont {J.}~\bibnamefont
  {{Zrake}}},\ }\bibfield  {title} {\bibinfo {title} {{Inverse cascade of
  nonhelical magnetic turbulence in a relativistic fluid}},\ }\href
  {https://doi.org/10.1088/2041-8205/794/2/L26} {\bibfield  {journal} {\bibinfo
   {journal} {Astrophys. J. Lett.}\ }\textbf {\bibinfo {volume} {794}},\
  \bibinfo {eid} {L26} (\bibinfo {year} {2014})}\BibitemShut {NoStop}%
\bibitem [{\citenamefont {{Brandenburg}}\ \emph {et~al.}(2015)\citenamefont
  {{Brandenburg}}, \citenamefont {{Kahniashvili}},\ and\ \citenamefont
  {{Tevzadze}}}]{Brandenburg15}%
  \BibitemOpen
  \bibfield  {author} {\bibinfo {author} {\bibfnamefont {A.}~\bibnamefont
  {{Brandenburg}}}, \bibinfo {author} {\bibfnamefont {T.}~\bibnamefont
  {{Kahniashvili}}},\ and\ \bibinfo {author} {\bibfnamefont {A.~G.}\
  \bibnamefont {{Tevzadze}}},\ }\bibfield  {title} {\bibinfo {title}
  {{Nonhelical inverse transfer of a decaying turbulent magnetic field}},\
  }\href {https://doi.org/10.1103/PhysRevLett.114.075001} {\bibfield  {journal}
  {\bibinfo  {journal} {Phys. Rev. Lett.}\ }\textbf {\bibinfo {volume} {114}},\
  \bibinfo {eid} {075001} (\bibinfo {year} {2015})}\BibitemShut {NoStop}%
\bibitem [{\citenamefont {{Kahniashvili}}\ \emph {et~al.}(2013)\citenamefont
  {{Kahniashvili}}, \citenamefont {{Tevzadze}}, \citenamefont {{Brandenburg}},\
  and\ \citenamefont {{Neronov}}}]{Kahniashvili13}%
  \BibitemOpen
  \bibfield  {author} {\bibinfo {author} {\bibfnamefont {T.}~\bibnamefont
  {{Kahniashvili}}}, \bibinfo {author} {\bibfnamefont {A.~G.}\ \bibnamefont
  {{Tevzadze}}}, \bibinfo {author} {\bibfnamefont {A.}~\bibnamefont
  {{Brandenburg}}},\ and\ \bibinfo {author} {\bibfnamefont {A.}~\bibnamefont
  {{Neronov}}},\ }\bibfield  {title} {\bibinfo {title} {{Evolution of
  primordial magnetic fields from phase transitions}},\ }\href
  {https://doi.org/10.1103/PhysRevD.87.083007} {\bibfield  {journal} {\bibinfo
  {journal} {Phys. Rev. D}\ }\textbf {\bibinfo {volume} {87}},\ \bibinfo {eid}
  {083007} (\bibinfo {year} {2013})}\BibitemShut {NoStop}%
\bibitem [{\citenamefont {{Ellis}}\ \emph {et~al.}(2019)\citenamefont
  {{Ellis}}, \citenamefont {{Fairbairn}}, \citenamefont {{Lewicki}},
  \citenamefont {{Vaskonen}},\ and\ \citenamefont {{Wickens}}}]{Ellis19}%
  \BibitemOpen
  \bibfield  {author} {\bibinfo {author} {\bibfnamefont {J.}~\bibnamefont
  {{Ellis}}}, \bibinfo {author} {\bibfnamefont {M.}~\bibnamefont
  {{Fairbairn}}}, \bibinfo {author} {\bibfnamefont {M.}~\bibnamefont
  {{Lewicki}}}, \bibinfo {author} {\bibfnamefont {V.}~\bibnamefont
  {{Vaskonen}}},\ and\ \bibinfo {author} {\bibfnamefont {A.}~\bibnamefont
  {{Wickens}}},\ }\bibfield  {title} {\bibinfo {title} {{Intergalactic magnetic
  fields from first-order phase transitions}},\ }\href
  {https://doi.org/10.1088/1475-7516/2019/09/019} {\bibfield  {journal}
  {\bibinfo  {journal} {J. Cosmol. Astropart. Phys.}\ }\textbf {\bibinfo
  {volume} {2019}},\ \bibinfo {eid} {019}}\BibitemShut {NoStop}%
\bibitem [{\citenamefont {{Mtchedlidze}}\ \emph {et~al.}(2022)\citenamefont
  {{Mtchedlidze}}, \citenamefont {{Dom{\'\i}nguez-Fern{\'a}ndez}},
  \citenamefont {{Du}}, \citenamefont {{Brandenburg}}, \citenamefont
  {{Kahniashvili}}, \citenamefont {{O'Sullivan}}, \citenamefont {{Schmidt}},\
  and\ \citenamefont {{Br{\"u}ggen}}}]{Mtchedlidze22}%
  \BibitemOpen
  \bibfield  {author} {\bibinfo {author} {\bibfnamefont {S.}~\bibnamefont
  {{Mtchedlidze}}}, \bibinfo {author} {\bibfnamefont {P.}~\bibnamefont
  {{Dom{\'\i}nguez-Fern{\'a}ndez}}}, \bibinfo {author} {\bibfnamefont
  {X.}~\bibnamefont {{Du}}}, \bibinfo {author} {\bibfnamefont {A.}~\bibnamefont
  {{Brandenburg}}}, \bibinfo {author} {\bibfnamefont {T.}~\bibnamefont
  {{Kahniashvili}}}, \bibinfo {author} {\bibfnamefont {S.}~\bibnamefont
  {{O'Sullivan}}}, \bibinfo {author} {\bibfnamefont {W.}~\bibnamefont
  {{Schmidt}}},\ and\ \bibinfo {author} {\bibfnamefont {M.}~\bibnamefont
  {{Br{\"u}ggen}}},\ }\bibfield  {title} {\bibinfo {title} {{Evolution of
  primordial magnetic fields during large-scale structure formation}},\ }\href
  {https://doi.org/10.3847/1538-4357/ac5960} {\bibfield  {journal} {\bibinfo
  {journal} {Astrophys. J.}\ }\textbf {\bibinfo {volume} {929}},\ \bibinfo
  {eid} {127} (\bibinfo {year} {2022})}\BibitemShut {NoStop}%
\bibitem [{\citenamefont {{Hosking}}\ and\ \citenamefont
  {{Schekochihin}}(2021)}]{HoskingSchekochihin20decay}%
  \BibitemOpen
  \bibfield  {author} {\bibinfo {author} {\bibfnamefont {D.~N.}\ \bibnamefont
  {{Hosking}}}\ and\ \bibinfo {author} {\bibfnamefont {A.~A.}\ \bibnamefont
  {{Schekochihin}}},\ }\bibfield  {title} {\bibinfo {title}
  {Reconnection-controlled decay of magnetohydrodynamic turbulence and the role
  of invariants},\ }\href {https://doi.org/10.1103/PhysRevX.11.041005}
  {\bibfield  {journal} {\bibinfo  {journal} {Phys. Rev. X}\ }\textbf {\bibinfo
  {volume} {11}},\ \bibinfo {pages} {041005} (\bibinfo {year}
  {2021})}\BibitemShut {NoStop}%
\bibitem [{\citenamefont {{Zhou}}\ \emph {et~al.}(2019)\citenamefont {{Zhou}},
  \citenamefont {{Bhat}}, \citenamefont {{Loureiro}},\ and\ \citenamefont
  {{Uzdensky}}}]{Zhou19}%
  \BibitemOpen
  \bibfield  {author} {\bibinfo {author} {\bibfnamefont {M.}~\bibnamefont
  {{Zhou}}}, \bibinfo {author} {\bibfnamefont {P.}~\bibnamefont {{Bhat}}},
  \bibinfo {author} {\bibfnamefont {N.~F.}\ \bibnamefont {{Loureiro}}},\ and\
  \bibinfo {author} {\bibfnamefont {D.~A.}\ \bibnamefont {{Uzdensky}}},\
  }\bibfield  {title} {\bibinfo {title} {{Magnetic island merger as a mechanism
  for inverse magnetic energy transfer}},\ }\href
  {https://doi.org/10.1103/PhysRevResearch.1.012004} {\bibfield  {journal}
  {\bibinfo  {journal} {Phys. Rev. Res.}\ }\textbf {\bibinfo {volume} {1}},\
  \bibinfo {eid} {012004} (\bibinfo {year} {2019})}\BibitemShut {NoStop}%
\bibitem [{\citenamefont {{Zhou}}\ \emph {et~al.}(2020)\citenamefont {{Zhou}},
  \citenamefont {{Loureiro}},\ and\ \citenamefont {{Uzdensky}}}]{Zhou20}%
  \BibitemOpen
  \bibfield  {author} {\bibinfo {author} {\bibfnamefont {M.}~\bibnamefont
  {{Zhou}}}, \bibinfo {author} {\bibfnamefont {N.~F.}\ \bibnamefont
  {{Loureiro}}},\ and\ \bibinfo {author} {\bibfnamefont {D.~A.}\ \bibnamefont
  {{Uzdensky}}},\ }\bibfield  {title} {\bibinfo {title} {{Multi-scale dynamics
  of magnetic flux tubes and inverse magnetic energy transfer}},\ }\href
  {https://doi.org/10.1017/S0022377820000641} {\bibfield  {journal} {\bibinfo
  {journal} {J. Plasma Phys.}\ }\textbf {\bibinfo {volume} {86}},\ \bibinfo
  {eid} {535860401} (\bibinfo {year} {2020})}\BibitemShut {NoStop}%
\bibitem [{\citenamefont {{Bhat}}\ \emph {et~al.}(2021)\citenamefont {{Bhat}},
  \citenamefont {{Zhou}},\ and\ \citenamefont {{Loureiro}}}]{Bhat21}%
  \BibitemOpen
  \bibfield  {author} {\bibinfo {author} {\bibfnamefont {P.}~\bibnamefont
  {{Bhat}}}, \bibinfo {author} {\bibfnamefont {M.}~\bibnamefont {{Zhou}}},\
  and\ \bibinfo {author} {\bibfnamefont {N.~F.}\ \bibnamefont {{Loureiro}}},\
  }\bibfield  {title} {\bibinfo {title} {{Inverse energy transfer in decaying,
  three-dimensional, non-helical magnetic turbulence due to magnetic
  reconnection}},\ }\href {https://doi.org/10.1093/mnras/staa3849} {\bibfield
  {journal} {\bibinfo  {journal} {Mon. Not. R. Astron. Soc.}\ }\textbf
  {\bibinfo {volume} {501}},\ \bibinfo {pages} {3074} (\bibinfo {year}
  {2021})}\BibitemShut {NoStop}%
\bibitem [{\citenamefont {{Jedamzik}}\ and\ \citenamefont
  {{Pogosian}}(2020)}]{JedamzikPogosian20}%
  \BibitemOpen
  \bibfield  {author} {\bibinfo {author} {\bibfnamefont {K.}~\bibnamefont
  {{Jedamzik}}}\ and\ \bibinfo {author} {\bibfnamefont {L.}~\bibnamefont
  {{Pogosian}}},\ }\bibfield  {title} {\bibinfo {title} {{Relieving the Hubble
  tension with primordial magnetic fields}},\ }\href
  {https://doi.org/10.1103/PhysRevLett.125.181302} {\bibfield  {journal}
  {\bibinfo  {journal} {Phys. Rev. Lett.}\ }\textbf {\bibinfo {volume} {125}},\
  \bibinfo {eid} {181302} (\bibinfo {year} {2020})}\BibitemShut {NoStop}%
\bibitem [{\citenamefont {{Galli}}\ \emph {et~al.}(2022)\citenamefont
  {{Galli}}, \citenamefont {{Pogosian}}, \citenamefont {{Jedamzik}},\ and\
  \citenamefont {{Balkenhol}}}]{Galli22}%
  \BibitemOpen
  \bibfield  {author} {\bibinfo {author} {\bibfnamefont {S.}~\bibnamefont
  {{Galli}}}, \bibinfo {author} {\bibfnamefont {L.}~\bibnamefont {{Pogosian}}},
  \bibinfo {author} {\bibfnamefont {K.}~\bibnamefont {{Jedamzik}}},\ and\
  \bibinfo {author} {\bibfnamefont {L.}~\bibnamefont {{Balkenhol}}},\
  }\bibfield  {title} {\bibinfo {title} {{Consistency of Planck, ACT, and SPT
  constraints on magnetically assisted recombination and forecasts for future
  experiments}},\ }\href {https://doi.org/10.1103/PhysRevD.105.023513}
  {\bibfield  {journal} {\bibinfo  {journal} {Phys. Rev. D}\ }\textbf {\bibinfo
  {volume} {105}},\ \bibinfo {eid} {023513} (\bibinfo {year}
  {2022})}\BibitemShut {NoStop}%
\bibitem [{\citenamefont {{Banerjee}}\ and\ \citenamefont
  {{Jedamzik}}(2003)}]{BanerjeeJedamzik03}%
  \BibitemOpen
  \bibfield  {author} {\bibinfo {author} {\bibfnamefont {R.}~\bibnamefont
  {{Banerjee}}}\ and\ \bibinfo {author} {\bibfnamefont {K.}~\bibnamefont
  {{Jedamzik}}},\ }\bibfield  {title} {\bibinfo {title} {{Are cluster magnetic
  fields primordial?}},\ }\href {https://doi.org/10.1103/PhysRevLett.91.251301}
  {\bibfield  {journal} {\bibinfo  {journal} {Phys. Rev. Lett.}\ }\textbf
  {\bibinfo {volume} {91}},\ \bibinfo {eid} {251301} (\bibinfo {year}
  {2003})}\BibitemShut {NoStop}%
\bibitem [{\citenamefont {{Brandenburg}}\ \emph {et~al.}(1996)\citenamefont
  {{Brandenburg}}, \citenamefont {{Enqvist}},\ and\ \citenamefont
  {{Olesen}}}]{Brandenburg96}%
  \BibitemOpen
  \bibfield  {author} {\bibinfo {author} {\bibfnamefont {A.}~\bibnamefont
  {{Brandenburg}}}, \bibinfo {author} {\bibfnamefont {K.}~\bibnamefont
  {{Enqvist}}},\ and\ \bibinfo {author} {\bibfnamefont {P.}~\bibnamefont
  {{Olesen}}},\ }\bibfield  {title} {\bibinfo {title} {{Large-scale magnetic
  fields from hydromagnetic turbulence in the very early Universe}},\ }\href
  {https://doi.org/10.1103/PhysRevD.54.1291} {\bibfield  {journal} {\bibinfo
  {journal} {Phys. Rev. D}\ }\textbf {\bibinfo {volume} {54}},\ \bibinfo
  {pages} {1291} (\bibinfo {year} {1996})}\BibitemShut {NoStop}%
\bibitem [{\citenamefont {{Jedamzik}}\ and\ \citenamefont
  {{Saveliev}}(2019)}]{JedamzikSaveliev19}%
  \BibitemOpen
  \bibfield  {author} {\bibinfo {author} {\bibfnamefont {K.}~\bibnamefont
  {{Jedamzik}}}\ and\ \bibinfo {author} {\bibfnamefont {A.}~\bibnamefont
  {{Saveliev}}},\ }\bibfield  {title} {\bibinfo {title} {{Stringent limit on
  primordial magnetic fields from the cosmic microwave background radiation}},\
  }\href {https://doi.org/10.1103/PhysRevLett.123.021301} {\bibfield  {journal}
  {\bibinfo  {journal} {Phys. Rev. Lett.}\ }\textbf {\bibinfo {volume} {123}},\
  \bibinfo {eid} {021301} (\bibinfo {year} {2019})}\BibitemShut {NoStop}%
\bibitem [{\citenamefont {{Turok}}(1992)}]{Turok92}%
  \BibitemOpen
  \bibfield  {author} {\bibinfo {author} {\bibfnamefont {N.}~\bibnamefont
  {{Turok}}},\ }\bibfield  {title} {\bibinfo {title} {{Electroweak bubbles:
  nucleation and growth}},\ }\href
  {https://doi.org/10.1103/PhysRevLett.68.1803} {\bibfield  {journal} {\bibinfo
   {journal} {Phys. Rev. Lett.}\ }\textbf {\bibinfo {volume} {68}},\ \bibinfo
  {pages} {1803} (\bibinfo {year} {1992})}\BibitemShut {NoStop}%
\bibitem [{\citenamefont {Davidson}(2015)}]{Davidson15}%
  \BibitemOpen
  \bibfield  {author} {\bibinfo {author} {\bibfnamefont {P.~A.}\ \bibnamefont
  {Davidson}},\ }\href@noop {} {\emph {\bibinfo {title} {{Turbulence: an
  Introduction for Scientists and Engineers}}}}\ (\bibinfo  {publisher} {Oxford
  University Press},\ \bibinfo {year} {2015})\BibitemShut {NoStop}%
\bibitem [{\citenamefont {{Landau}}\ and\ \citenamefont
  {{Lifshitz}}(1959)}]{LandauLifshitzFluids}%
  \BibitemOpen
  \bibfield  {author} {\bibinfo {author} {\bibfnamefont {L.~D.}\ \bibnamefont
  {{Landau}}}\ and\ \bibinfo {author} {\bibfnamefont {E.~M.}\ \bibnamefont
  {{Lifshitz}}},\ }\href@noop {} {\emph {\bibinfo {title} {{Fluid
  Mechanics}}}}\ (\bibinfo  {publisher} {{Pergamon Press}},\ \bibinfo {year}
  {1959})\BibitemShut {NoStop}%
\bibitem [{\citenamefont {{Reppin}}\ and\ \citenamefont
  {{Banerjee}}(2017)}]{ReppinBanerjee17}%
  \BibitemOpen
  \bibfield  {author} {\bibinfo {author} {\bibfnamefont {J.}~\bibnamefont
  {{Reppin}}}\ and\ \bibinfo {author} {\bibfnamefont {R.}~\bibnamefont
  {{Banerjee}}},\ }\bibfield  {title} {\bibinfo {title} {{Nonhelical turbulence
  and the inverse transfer of energy: a parameter study}},\ }\href
  {https://doi.org/10.1103/PhysRevE.96.053105} {\bibfield  {journal} {\bibinfo
  {journal} {Phys. Rev. E}\ }\textbf {\bibinfo {volume} {96}},\ \bibinfo {eid}
  {053105} (\bibinfo {year} {2017})}\BibitemShut {NoStop}%
\bibitem [{\citenamefont {{Zhou}}\ \emph {et~al.}(2022)\citenamefont {{Zhou}},
  \citenamefont {{Sharma}},\ and\ \citenamefont
  {{Brandenburg}}}]{Zhou22_Hosking}%
  \BibitemOpen
  \bibfield  {author} {\bibinfo {author} {\bibfnamefont {H.}~\bibnamefont
  {{Zhou}}}, \bibinfo {author} {\bibfnamefont {R.}~\bibnamefont {{Sharma}}},\
  and\ \bibinfo {author} {\bibfnamefont {A.}~\bibnamefont {{Brandenburg}}},\
  }\bibfield  {title} {\bibinfo {title} {{Scaling of the Hosking integral in
  decaying magnetically dominated turbulence}},\ }\href
  {https://doi.org/10.1017/S002237782200109X} {\bibfield  {journal} {\bibinfo
  {journal} {J. Plasma Phys.}\ }\textbf {\bibinfo {volume} {88}},\ \bibinfo
  {eid} {905880602} (\bibinfo {year} {2022})}\BibitemShut {NoStop}%
\bibitem [{\citenamefont {{Brandenburg}}(2022)}]{Brandenburg22_Hosking}%
  \BibitemOpen
  \bibfield  {author} {\bibinfo {author} {\bibfnamefont {A.}~\bibnamefont
  {{Brandenburg}}},\ }\bibfield  {title} {\bibinfo {title} {{Hosking integral
  in nonhelical Hall cascade}},\ }\href@noop {} {\bibfield  {journal} {\bibinfo
   {journal} {arXiv e-prints}\ ,\ \bibinfo {eid} {arXiv:2211.14197}} (\bibinfo
  {year} {2022})}\BibitemShut {NoStop}%
\bibitem [{\citenamefont {{Kahniashvili}}\ \emph {et~al.}(2010)\citenamefont
  {{Kahniashvili}}, \citenamefont {{Brandenburg}}, \citenamefont {{Tevzadze}},\
  and\ \citenamefont {{Ratra}}}]{Kahniashvili10}%
  \BibitemOpen
  \bibfield  {author} {\bibinfo {author} {\bibfnamefont {T.}~\bibnamefont
  {{Kahniashvili}}}, \bibinfo {author} {\bibfnamefont {A.}~\bibnamefont
  {{Brandenburg}}}, \bibinfo {author} {\bibfnamefont {A.~G.}\ \bibnamefont
  {{Tevzadze}}},\ and\ \bibinfo {author} {\bibfnamefont {B.}~\bibnamefont
  {{Ratra}}},\ }\bibfield  {title} {\bibinfo {title} {{Numerical simulations of
  the decay of primordial magnetic turbulence}},\ }\href
  {https://doi.org/10.1103/PhysRevD.81.123002} {\bibfield  {journal} {\bibinfo
  {journal} {Phys. Rev. D}\ }\textbf {\bibinfo {volume} {81}},\ \bibinfo {eid}
  {123002} (\bibinfo {year} {2010})}\BibitemShut {NoStop}%
\bibitem [{\citenamefont {{Loureiro}}\ \emph {et~al.}(2007)\citenamefont
  {{Loureiro}}, \citenamefont {{Schekochihin}},\ and\ \citenamefont
  {{Cowley}}}]{Loureiro07}%
  \BibitemOpen
  \bibfield  {author} {\bibinfo {author} {\bibfnamefont {N.~F.}\ \bibnamefont
  {{Loureiro}}}, \bibinfo {author} {\bibfnamefont {A.~A.}\ \bibnamefont
  {{Schekochihin}}},\ and\ \bibinfo {author} {\bibfnamefont {S.~C.}\
  \bibnamefont {{Cowley}}},\ }\bibfield  {title} {\bibinfo {title}
  {{Instability of current sheets and formation of plasmoid chains}},\ }\href
  {https://doi.org/10.1063/1.2783986} {\bibfield  {journal} {\bibinfo
  {journal} {Phys. Plasmas}\ }\textbf {\bibinfo {volume} {14}},\ \bibinfo
  {pages} {100703} (\bibinfo {year} {2007})}\BibitemShut {NoStop}%
\bibitem [{\citenamefont {{Uzdensky}}\ \emph {et~al.}(2010)\citenamefont
  {{Uzdensky}}, \citenamefont {{Loureiro}},\ and\ \citenamefont
  {{Schekochihin}}}]{Uzdensky10}%
  \BibitemOpen
  \bibfield  {author} {\bibinfo {author} {\bibfnamefont {D.~A.}\ \bibnamefont
  {{Uzdensky}}}, \bibinfo {author} {\bibfnamefont {N.~F.}\ \bibnamefont
  {{Loureiro}}},\ and\ \bibinfo {author} {\bibfnamefont {A.~A.}\ \bibnamefont
  {{Schekochihin}}},\ }\bibfield  {title} {\bibinfo {title} {{Fast magnetic
  reconnection in the plasmoid-dominated regime}},\ }\href
  {https://doi.org/10.1103/PhysRevLett.105.235002} {\bibfield  {journal}
  {\bibinfo  {journal} {Phys. Rev. Lett.}\ }\textbf {\bibinfo {volume} {105}},\
  \bibinfo {eid} {235002} (\bibinfo {year} {2010})}\BibitemShut {NoStop}%
\bibitem [{\citenamefont {{Bhattacharjee}}\ \emph {et~al.}(2009)\citenamefont
  {{Bhattacharjee}}, \citenamefont {{Huang}}, \citenamefont {{Yang}},\ and\
  \citenamefont {{Rogers}}}]{Bhattacharjee09}%
  \BibitemOpen
  \bibfield  {author} {\bibinfo {author} {\bibfnamefont {A.}~\bibnamefont
  {{Bhattacharjee}}}, \bibinfo {author} {\bibfnamefont {Y.-M.}\ \bibnamefont
  {{Huang}}}, \bibinfo {author} {\bibfnamefont {H.}~\bibnamefont {{Yang}}},\
  and\ \bibinfo {author} {\bibfnamefont {B.}~\bibnamefont {{Rogers}}},\
  }\bibfield  {title} {\bibinfo {title} {{Fast reconnection in
  high-Lundquist-number plasmas due to the plasmoid Instability}},\ }\href
  {https://doi.org/10.1063/1.3264103} {\bibfield  {journal} {\bibinfo
  {journal} {Phys. Plasmas}\ }\textbf {\bibinfo {volume} {16}},\ \bibinfo {eid}
  {112102} (\bibinfo {year} {2009})}\BibitemShut {NoStop}%
\bibitem [{\citenamefont {{Schekochihin}}(2020)}]{Schekochihin20}%
  \BibitemOpen
  \bibfield  {author} {\bibinfo {author} {\bibfnamefont {A.~A.}\ \bibnamefont
  {{Schekochihin}}},\ }\bibfield  {title} {\bibinfo {title} {{MHD turbulence: a
  biased review}},\ }\Eprint {https://arxiv.org/abs/2010.00699}
  {arXiv:2010.00699}  (\bibinfo {year} {2020})\BibitemShut {NoStop}%
\bibitem [{\citenamefont {{Loureiro}}\ \emph {et~al.}(2012)\citenamefont
  {{Loureiro}}, \citenamefont {{Samtaney}}, \citenamefont {{Schekochihin}},\
  and\ \citenamefont {{Uzdensky}}}]{Loureiro12}%
  \BibitemOpen
  \bibfield  {author} {\bibinfo {author} {\bibfnamefont {N.~F.}\ \bibnamefont
  {{Loureiro}}}, \bibinfo {author} {\bibfnamefont {R.}~\bibnamefont
  {{Samtaney}}}, \bibinfo {author} {\bibfnamefont {A.~A.}\ \bibnamefont
  {{Schekochihin}}},\ and\ \bibinfo {author} {\bibfnamefont {D.~A.}\
  \bibnamefont {{Uzdensky}}},\ }\bibfield  {title} {\bibinfo {title} {{Magnetic
  reconnection and stochastic plasmoid chains in high-Lundquist-number
  plasmas}},\ }\href {https://doi.org/10.1063/1.3703318} {\bibfield  {journal}
  {\bibinfo  {journal} {Phys. Plasmas}\ }\textbf {\bibinfo {volume} {19}},\
  \bibinfo {pages} {042303} (\bibinfo {year} {2012})}\BibitemShut {NoStop}%
\bibitem [{\citenamefont {{Spitzer}}(1956)}]{Spitzer56}%
  \BibitemOpen
  \bibfield  {author} {\bibinfo {author} {\bibfnamefont {L.}~\bibnamefont
  {{Spitzer}}},\ }\href@noop {} {\emph {\bibinfo {title} {{Physics of Fully
  Ionized Gases}}}}\ (\bibinfo  {publisher} {Interscience Publishers},\
  \bibinfo {year} {1956})\BibitemShut {NoStop}%
\bibitem [{\citenamefont {{Braginskii}}(1965)}]{Braginskii65}%
  \BibitemOpen
  \bibfield  {author} {\bibinfo {author} {\bibfnamefont {S.~I.}\ \bibnamefont
  {{Braginskii}}},\ }\bibfield  {title} {\bibinfo {title} {{Transport processes
  in a plasma}},\ }\href@noop {} {\bibfield  {journal} {\bibinfo  {journal}
  {Rev. Plasma Phys.}\ }\textbf {\bibinfo {volume} {1}},\ \bibinfo {pages}
  {205} (\bibinfo {year} {1965})}\BibitemShut {NoStop}%
\bibitem [{\citenamefont {{Ji}}\ \emph {et~al.}(2022)\citenamefont {{Ji}},
  \citenamefont {{Daughton}}, \citenamefont {{Jara-Almonte}}, \citenamefont
  {{Le}}, \citenamefont {{Stanier}},\ and\ \citenamefont {{Yoo}}}]{Ji22}%
  \BibitemOpen
  \bibfield  {author} {\bibinfo {author} {\bibfnamefont {H.}~\bibnamefont
  {{Ji}}}, \bibinfo {author} {\bibfnamefont {W.}~\bibnamefont {{Daughton}}},
  \bibinfo {author} {\bibfnamefont {J.}~\bibnamefont {{Jara-Almonte}}},
  \bibinfo {author} {\bibfnamefont {A.}~\bibnamefont {{Le}}}, \bibinfo {author}
  {\bibfnamefont {A.}~\bibnamefont {{Stanier}}},\ and\ \bibinfo {author}
  {\bibfnamefont {J.}~\bibnamefont {{Yoo}}},\ }\bibfield  {title} {\bibinfo
  {title} {{Magnetic reconnection in the era of exascale computing and
  multiscale experiments}},\ }\href@noop {} {\bibfield  {journal} {\bibinfo
  {journal} {arXiv e-prints}\ ,\ \bibinfo {eid} {arXiv:2202.09004}} (\bibinfo
  {year} {2022})}\BibitemShut {NoStop}%
\bibitem [{\citenamefont {{Liu}}\ \emph {et~al.}(2022)\citenamefont {{Liu}},
  \citenamefont {{Cassak}}, \citenamefont {{Li}}, \citenamefont {{Hesse}},
  \citenamefont {{Lin}},\ and\ \citenamefont {{Genestreti}}}]{Liu22}%
  \BibitemOpen
  \bibfield  {author} {\bibinfo {author} {\bibfnamefont {Y.-H.}\ \bibnamefont
  {{Liu}}}, \bibinfo {author} {\bibfnamefont {P.}~\bibnamefont {{Cassak}}},
  \bibinfo {author} {\bibfnamefont {X.}~\bibnamefont {{Li}}}, \bibinfo {author}
  {\bibfnamefont {M.}~\bibnamefont {{Hesse}}}, \bibinfo {author} {\bibfnamefont
  {S.-C.}\ \bibnamefont {{Lin}}},\ and\ \bibinfo {author} {\bibfnamefont
  {K.}~\bibnamefont {{Genestreti}}},\ }\bibfield  {title} {\bibinfo {title}
  {{First-principles theory of the rate of magnetic reconnection in
  magnetospheric and solar plasmas}},\ }\href
  {https://doi.org/10.1038/s42005-022-00854-x} {\bibfield  {journal} {\bibinfo
  {journal} {Communications Physics}\ }\textbf {\bibinfo {volume} {5}},\
  \bibinfo {eid} {97} (\bibinfo {year} {2022})}\BibitemShut {NoStop}%
\bibitem [{\citenamefont {{Comisso}}\ and\ \citenamefont
  {{Bhattacharjee}}(2016)}]{ComissoBhattacharjee16}%
  \BibitemOpen
  \bibfield  {author} {\bibinfo {author} {\bibfnamefont {L.}~\bibnamefont
  {{Comisso}}}\ and\ \bibinfo {author} {\bibfnamefont {A.}~\bibnamefont
  {{Bhattacharjee}}},\ }\bibfield  {title} {\bibinfo {title} {{On the value of
  the reconnection rate}},\ }\href {https://doi.org/10.1017/S002237781600101X}
  {\bibfield  {journal} {\bibinfo  {journal} {J. Plasma Phys.}\ }\textbf
  {\bibinfo {volume} {82}},\ \bibinfo {eid} {595820601} (\bibinfo {year}
  {2016})}\BibitemShut {NoStop}%
\bibitem [{\citenamefont {{Cassak}}\ \emph {et~al.}(2017)\citenamefont
  {{Cassak}}, \citenamefont {{Liu}},\ and\ \citenamefont {{Shay}}}]{Cassak17}%
  \BibitemOpen
  \bibfield  {author} {\bibinfo {author} {\bibfnamefont {P.~A.}\ \bibnamefont
  {{Cassak}}}, \bibinfo {author} {\bibfnamefont {Y.~H.}\ \bibnamefont
  {{Liu}}},\ and\ \bibinfo {author} {\bibfnamefont {M.~A.}\ \bibnamefont
  {{Shay}}},\ }\bibfield  {title} {\bibinfo {title} {{A review of the 0.1
  reconnection rate problem}},\ }\href
  {https://doi.org/10.1017/S0022377817000666} {\bibfield  {journal} {\bibinfo
  {journal} {J. Plasma Phys.}\ }\textbf {\bibinfo {volume} {83}},\ \bibinfo
  {eid} {715830501} (\bibinfo {year} {2017})}\BibitemShut {NoStop}%
\bibitem [{\citenamefont {{Schekochihin}}\ \emph {et~al.}(2010)\citenamefont
  {{Schekochihin}}, \citenamefont {{Cowley}}, \citenamefont {{Rincon}},\ and\
  \citenamefont {{Rosin}}}]{Schekochihin10}%
  \BibitemOpen
  \bibfield  {author} {\bibinfo {author} {\bibfnamefont {A.~A.}\ \bibnamefont
  {{Schekochihin}}}, \bibinfo {author} {\bibfnamefont {S.~C.}\ \bibnamefont
  {{Cowley}}}, \bibinfo {author} {\bibfnamefont {F.}~\bibnamefont {{Rincon}}},\
  and\ \bibinfo {author} {\bibfnamefont {M.~S.}\ \bibnamefont {{Rosin}}},\
  }\bibfield  {title} {\bibinfo {title} {{Magnetofluid dynamics of magnetized
  cosmic plasma: firehose and gyrothermal instabilities}},\ }\href
  {https://doi.org/10.1111/j.1365-2966.2010.16493.x} {\bibfield  {journal}
  {\bibinfo  {journal} {Mon. Not. R. Astron. Soc.}\ }\textbf {\bibinfo {volume}
  {405}},\ \bibinfo {pages} {291} (\bibinfo {year} {2010})}\BibitemShut
  {NoStop}%
\bibitem [{\citenamefont {{Schekochihin}}\ \emph {et~al.}(2005)\citenamefont
  {{Schekochihin}}, \citenamefont {{Cowley}}, \citenamefont {{Kulsrud}},
  \citenamefont {{Hammett}},\ and\ \citenamefont {{Sharma}}}]{Schekochihin05}%
  \BibitemOpen
  \bibfield  {author} {\bibinfo {author} {\bibfnamefont {A.~A.}\ \bibnamefont
  {{Schekochihin}}}, \bibinfo {author} {\bibfnamefont {S.~C.}\ \bibnamefont
  {{Cowley}}}, \bibinfo {author} {\bibfnamefont {R.~M.}\ \bibnamefont
  {{Kulsrud}}}, \bibinfo {author} {\bibfnamefont {G.~W.}\ \bibnamefont
  {{Hammett}}},\ and\ \bibinfo {author} {\bibfnamefont {P.}~\bibnamefont
  {{Sharma}}},\ }\bibfield  {title} {\bibinfo {title} {{Plasma instabilities
  and magnetic field growth in clusters of galaxies}},\ }\href
  {https://doi.org/10.1086/431202} {\bibfield  {journal} {\bibinfo  {journal}
  {Astrophys. J.}\ }\textbf {\bibinfo {volume} {629}},\ \bibinfo {pages} {139}
  (\bibinfo {year} {2005})}\BibitemShut {NoStop}%
\bibitem [{\citenamefont {{Winarto}}\ and\ \citenamefont
  {{Kunz}}(2022)}]{Winarto22}%
  \BibitemOpen
  \bibfield  {author} {\bibinfo {author} {\bibfnamefont {H.~W.}\ \bibnamefont
  {{Winarto}}}\ and\ \bibinfo {author} {\bibfnamefont {M.~W.}\ \bibnamefont
  {{Kunz}}},\ }\bibfield  {title} {\bibinfo {title} {{Triggering tearing in a
  forming current sheet with the mirror instability}},\ }\href
  {https://doi.org/10.1017/S0022377822000150} {\bibfield  {journal} {\bibinfo
  {journal} {J. Plasma Phys.}\ }\textbf {\bibinfo {volume} {88}},\ \bibinfo
  {eid} {905880210} (\bibinfo {year} {2022})}\BibitemShut {NoStop}%
\bibitem [{\citenamefont {{St-Onge}}\ and\ \citenamefont
  {{Kunz}}(2018)}]{St-OngeKunz18}%
  \BibitemOpen
  \bibfield  {author} {\bibinfo {author} {\bibfnamefont {D.~A.}\ \bibnamefont
  {{St-Onge}}}\ and\ \bibinfo {author} {\bibfnamefont {M.~W.}\ \bibnamefont
  {{Kunz}}},\ }\bibfield  {title} {\bibinfo {title} {{Fluctuation dynamo in a
  collisionless, weakly magnetized plasma}},\ }\href
  {https://doi.org/10.3847/2041-8213/aad638} {\bibfield  {journal} {\bibinfo
  {journal} {Astrophys. J. Lett.}\ }\textbf {\bibinfo {volume} {863}},\
  \bibinfo {eid} {L25} (\bibinfo {year} {2018})}\BibitemShut {NoStop}%
\bibitem [{\citenamefont {{Kunz}}\ \emph {et~al.}(2016)\citenamefont {{Kunz}},
  \citenamefont {{Stone}},\ and\ \citenamefont {{Quataert}}}]{Kunz16}%
  \BibitemOpen
  \bibfield  {author} {\bibinfo {author} {\bibfnamefont {M.~W.}\ \bibnamefont
  {{Kunz}}}, \bibinfo {author} {\bibfnamefont {J.~M.}\ \bibnamefont
  {{Stone}}},\ and\ \bibinfo {author} {\bibfnamefont {E.}~\bibnamefont
  {{Quataert}}},\ }\bibfield  {title} {\bibinfo {title} {{Magnetorotational
  turbulence and dynamo in a collisionless plasma}},\ }\href
  {https://doi.org/10.1103/PhysRevLett.117.235101} {\bibfield  {journal}
  {\bibinfo  {journal} {Phys. Rev. Lett.}\ }\textbf {\bibinfo {volume} {117}},\
  \bibinfo {eid} {235101} (\bibinfo {year} {2016})}\BibitemShut {NoStop}%
\bibitem [{\citenamefont {{Bennett}}\ \emph {et~al.}(2003)\citenamefont
  {{Bennett}}, \citenamefont {{Halpern}}, \citenamefont {{Hinshaw}},
  \citenamefont {{Jarosik}}, \citenamefont {{Kogut}}, \citenamefont {{Limon}},
  \citenamefont {{Meyer}}, \citenamefont {{Page}}, \citenamefont {{Spergel}},
  \citenamefont {{Tucker}} \emph {et~al.}}]{Bennett03}%
  \BibitemOpen
  \bibfield  {author} {\bibinfo {author} {\bibfnamefont {C.~L.}\ \bibnamefont
  {{Bennett}}}, \bibinfo {author} {\bibfnamefont {M.}~\bibnamefont
  {{Halpern}}}, \bibinfo {author} {\bibfnamefont {G.}~\bibnamefont
  {{Hinshaw}}}, \bibinfo {author} {\bibfnamefont {N.}~\bibnamefont
  {{Jarosik}}}, \bibinfo {author} {\bibfnamefont {A.}~\bibnamefont {{Kogut}}},
  \bibinfo {author} {\bibfnamefont {M.}~\bibnamefont {{Limon}}}, \bibinfo
  {author} {\bibfnamefont {S.~S.}\ \bibnamefont {{Meyer}}}, \bibinfo {author}
  {\bibfnamefont {L.}~\bibnamefont {{Page}}}, \bibinfo {author} {\bibfnamefont
  {D.~N.}\ \bibnamefont {{Spergel}}}, \bibinfo {author} {\bibfnamefont {G.~S.}\
  \bibnamefont {{Tucker}}}, \emph {et~al.},\ }\bibfield  {title} {\bibinfo
  {title} {{First-year Wilkinson Microwave Anisotropy Probe (WMAP)
  observations: preliminary maps and basic results}},\ }\href
  {https://doi.org/10.1086/377253} {\bibfield  {journal} {\bibinfo  {journal}
  {Astrophys. J. Suppl.}\ }\textbf {\bibinfo {volume} {148}},\ \bibinfo {pages}
  {1} (\bibinfo {year} {2003})}\BibitemShut {NoStop}%
\bibitem [{\citenamefont {{Parra}}(2019)}]{Parra19}%
  \BibitemOpen
  \bibfield  {author} {\bibinfo {author} {\bibfnamefont {F.~I.}\ \bibnamefont
  {{Parra}}},\ }\href@noop {} {\bibinfo {title} {Collisional plasma physics.
  \textit{Lecture Notes for an Oxford MMathPhys course}}},\ \bibinfo
  {howpublished}
  {\url{http://www-thphys.physics.ox.ac.uk/people/FelixParra/CollisionalPlasmaPhysics/CollisionalPlasmaPhysics.html}}
  (\bibinfo {year} {2019})\BibitemShut {NoStop}%
\bibitem [{\citenamefont {{Melville}}\ \emph {et~al.}(2016)\citenamefont
  {{Melville}}, \citenamefont {{Schekochihin}},\ and\ \citenamefont
  {{Kunz}}}]{Melville16}%
  \BibitemOpen
  \bibfield  {author} {\bibinfo {author} {\bibfnamefont {S.}~\bibnamefont
  {{Melville}}}, \bibinfo {author} {\bibfnamefont {A.~A.}\ \bibnamefont
  {{Schekochihin}}},\ and\ \bibinfo {author} {\bibfnamefont {M.~W.}\
  \bibnamefont {{Kunz}}},\ }\bibfield  {title} {\bibinfo {title}
  {{Pressure-anisotropy-driven microturbulence and magnetic-field evolution in
  shearing, collisionless plasma}},\ }\href
  {https://doi.org/10.1093/mnras/stw793} {\bibfield  {journal} {\bibinfo
  {journal} {Mon. Not. R. Astron. Soc.}\ }\textbf {\bibinfo {volume} {459}},\
  \bibinfo {pages} {2701} (\bibinfo {year} {2016})}\BibitemShut {NoStop}%
\bibitem [{\citenamefont {{Lesur}}(2015)}]{Lesur15}%
  \BibitemOpen
  \bibfield  {author} {\bibinfo {author} {\bibfnamefont {G.}~\bibnamefont
  {{Lesur}}},\ }\href@noop {} {\bibinfo {title} {{Snoopy: general purpose
  spectral solver, Astrophysics Source Code Library (ascl:1505.022)}}}
  (\bibinfo {year} {2015})\BibitemShut {NoStop}%
\bibitem [{\citenamefont {{Biskamp}}\ and\ \citenamefont
  {{M{\"u}ller}}(1999)}]{BiskampMuller99}%
  \BibitemOpen
  \bibfield  {author} {\bibinfo {author} {\bibfnamefont {D.}~\bibnamefont
  {{Biskamp}}}\ and\ \bibinfo {author} {\bibfnamefont {W.-C.}\ \bibnamefont
  {{M{\"u}ller}}},\ }\bibfield  {title} {\bibinfo {title} {{Decay laws for
  three-dimensional magnetohydrodynamic turbulence}},\ }\href
  {https://doi.org/10.1103/PhysRevLett.83.2195} {\bibfield  {journal} {\bibinfo
   {journal} {Phys. Rev. Lett.}\ }\textbf {\bibinfo {volume} {83}},\ \bibinfo
  {pages} {2195} (\bibinfo {year} {1999})}\BibitemShut {NoStop}%
\bibitem [{\citenamefont {{M{\"u}ller}}\ and\ \citenamefont
  {{Biskamp}}(2000)}]{BiskampMuller00}%
  \BibitemOpen
  \bibfield  {author} {\bibinfo {author} {\bibfnamefont {W.-C.}\ \bibnamefont
  {{M{\"u}ller}}}\ and\ \bibinfo {author} {\bibfnamefont {D.}~\bibnamefont
  {{Biskamp}}},\ }\bibfield  {title} {\bibinfo {title} {{Scaling properties of
  three-dimensional magnetohydrodynamic turbulence}},\ }\href
  {https://doi.org/10.1103/PhysRevLett.84.475} {\bibfield  {journal} {\bibinfo
  {journal} {Phys. Rev. Lett.}\ }\textbf {\bibinfo {volume} {84}},\ \bibinfo
  {pages} {475} (\bibinfo {year} {2000})}\BibitemShut {NoStop}%
\bibitem [{\citenamefont {{Christensson}}\ \emph {et~al.}(2001)\citenamefont
  {{Christensson}}, \citenamefont {{Hindmarsh}},\ and\ \citenamefont
  {{Brandenburg}}}]{Christensson01}%
  \BibitemOpen
  \bibfield  {author} {\bibinfo {author} {\bibfnamefont {M.}~\bibnamefont
  {{Christensson}}}, \bibinfo {author} {\bibfnamefont {M.}~\bibnamefont
  {{Hindmarsh}}},\ and\ \bibinfo {author} {\bibfnamefont {A.}~\bibnamefont
  {{Brandenburg}}},\ }\bibfield  {title} {\bibinfo {title} {{Inverse cascade in
  decaying three-dimensional magnetohydrodynamic turbulence}},\ }\href
  {https://doi.org/10.1103/PhysRevE.64.056405} {\bibfield  {journal} {\bibinfo
  {journal} {Phys. Rev. E}\ }\textbf {\bibinfo {volume} {64}},\ \bibinfo
  {pages} {056405} (\bibinfo {year} {2001})}\BibitemShut {NoStop}%
\bibitem [{\citenamefont {{Frick}}\ and\ \citenamefont
  {{Stepanov}}(2010)}]{FrickStepanov10}%
  \BibitemOpen
  \bibfield  {author} {\bibinfo {author} {\bibfnamefont {P.}~\bibnamefont
  {{Frick}}}\ and\ \bibinfo {author} {\bibfnamefont {R.}~\bibnamefont
  {{Stepanov}}},\ }\bibfield  {title} {\bibinfo {title} {{Long-term free decay
  of MHD turbulence}},\ }\href {https://doi.org/10.1209/0295-5075/92/34007}
  {\bibfield  {journal} {\bibinfo  {journal} {Europhys. Lett.}\ }\textbf
  {\bibinfo {volume} {92}},\ \bibinfo {pages} {34007} (\bibinfo {year}
  {2010})}\BibitemShut {NoStop}%
\bibitem [{\citenamefont {{Berera}}\ and\ \citenamefont
  {{Linkmann}}(2014)}]{BereraLinkmann14}%
  \BibitemOpen
  \bibfield  {author} {\bibinfo {author} {\bibfnamefont {A.}~\bibnamefont
  {{Berera}}}\ and\ \bibinfo {author} {\bibfnamefont {M.}~\bibnamefont
  {{Linkmann}}},\ }\bibfield  {title} {\bibinfo {title} {{Magnetic helicity and
  the evolution of decaying magnetohydrodynamic turbulence}},\ }\href
  {https://doi.org/10.1103/PhysRevE.90.041003} {\bibfield  {journal} {\bibinfo
  {journal} {Phys. Rev. E}\ }\textbf {\bibinfo {volume} {90}},\ \bibinfo {eid}
  {041003} (\bibinfo {year} {2014})}\BibitemShut {NoStop}%
\bibitem [{\citenamefont {{Brandenburg}}\ and\ \citenamefont
  {{Kahniashvili}}(2017)}]{Brandenburg17}%
  \BibitemOpen
  \bibfield  {author} {\bibinfo {author} {\bibfnamefont {A.}~\bibnamefont
  {{Brandenburg}}}\ and\ \bibinfo {author} {\bibfnamefont {T.}~\bibnamefont
  {{Kahniashvili}}},\ }\bibfield  {title} {\bibinfo {title} {{Classes of
  hydrodynamic and magnetohydrodynamic turbulent decay}},\ }\href
  {https://doi.org/10.1103/PhysRevLett.118.055102} {\bibfield  {journal}
  {\bibinfo  {journal} {Phys. Rev. Lett.}\ }\textbf {\bibinfo {volume} {118}},\
  \bibinfo {eid} {055102} (\bibinfo {year} {2017})}\BibitemShut {NoStop}%
\bibitem [{\citenamefont {{Rincon}}(2019)}]{Rincon19}%
  \BibitemOpen
  \bibfield  {author} {\bibinfo {author} {\bibfnamefont {F.}~\bibnamefont
  {{Rincon}}},\ }\bibfield  {title} {\bibinfo {title} {{Dynamo theories}},\
  }\href {https://doi.org/10.1017/S0022377819000539} {\bibfield  {journal}
  {\bibinfo  {journal} {J. Plasma Phys.}\ }\textbf {\bibinfo {volume} {85}},\
  \bibinfo {eid} {205850401} (\bibinfo {year} {2019})}\BibitemShut {NoStop}%
\bibitem [{\citenamefont {{Hosking}}\ and\ \citenamefont
  {{Schekochihin}}(2022)}]{HoskingSchekochihin21k2}%
  \BibitemOpen
  \bibfield  {author} {\bibinfo {author} {\bibfnamefont {D.~N.}\ \bibnamefont
  {{Hosking}}}\ and\ \bibinfo {author} {\bibfnamefont {A.~A.}\ \bibnamefont
  {{Schekochihin}}},\ }\bibfield  {title} {\bibinfo {title} {{Emergence of
  long-range correlations and thermal spectra in forced turbulence}},\ }\Eprint
  {https://arxiv.org/abs/2202.00462} {arXiv:2202.00462}  (\bibinfo {year}
  {2022})\BibitemShut {NoStop}%
\bibitem [{\citenamefont {{Galishnikova}}\ \emph {et~al.}(2022)\citenamefont
  {{Galishnikova}}, \citenamefont {{Kunz}},\ and\ \citenamefont
  {{Schekochihin}}}]{Galishnikova22}%
  \BibitemOpen
  \bibfield  {author} {\bibinfo {author} {\bibfnamefont {A.~K.}\ \bibnamefont
  {{Galishnikova}}}, \bibinfo {author} {\bibfnamefont {M.~W.}\ \bibnamefont
  {{Kunz}}},\ and\ \bibinfo {author} {\bibfnamefont {A.~A.}\ \bibnamefont
  {{Schekochihin}}},\ }\bibfield  {title} {\bibinfo {title} {{Tearing
  instability and current-sheet disruption in the turbulent dynamo}},\ }\Eprint
  {https://arxiv.org/abs/2201.07757} {arXiv:2201.07757}  (\bibinfo {year}
  {2022})\BibitemShut {NoStop}%
\bibitem [{\citenamefont {{Taylor}}(1974)}]{Taylor74}%
  \BibitemOpen
  \bibfield  {author} {\bibinfo {author} {\bibfnamefont {J.~B.}\ \bibnamefont
  {{Taylor}}},\ }\bibfield  {title} {\bibinfo {title} {{Relaxation of toroidal
  plasma and generation of reverse magnetic fields}},\ }\href
  {https://doi.org/10.1103/PhysRevLett.33.1139} {\bibfield  {journal} {\bibinfo
   {journal} {Phys. Rev. Lett.}\ }\textbf {\bibinfo {volume} {33}},\ \bibinfo
  {pages} {1139} (\bibinfo {year} {1974})}\BibitemShut {NoStop}%
\bibitem [{\citenamefont {{Servidio}}\ \emph {et~al.}(2008)\citenamefont
  {{Servidio}}, \citenamefont {{Matthaeus}},\ and\ \citenamefont
  {{Dmitruk}}}]{Servidio08}%
  \BibitemOpen
  \bibfield  {author} {\bibinfo {author} {\bibfnamefont {S.}~\bibnamefont
  {{Servidio}}}, \bibinfo {author} {\bibfnamefont {W.~H.}\ \bibnamefont
  {{Matthaeus}}},\ and\ \bibinfo {author} {\bibfnamefont {P.}~\bibnamefont
  {{Dmitruk}}},\ }\bibfield  {title} {\bibinfo {title} {{Depression of
  nonlinearity in decaying isotropic MHD turbulence}},\ }\href
  {https://doi.org/10.1103/PhysRevLett.100.095005} {\bibfield  {journal}
  {\bibinfo  {journal} {Phys. Rev. Lett.}\ }\textbf {\bibinfo {volume} {100}},\
  \bibinfo {eid} {095005} (\bibinfo {year} {2008})}\BibitemShut {NoStop}%
\end{thebibliography}%

\end{document}